%% file: BT.tex
\documentclass[%
acmsmall,fleqn,screen,nonacm]{acmart}
\setcopyright{none}

\usepackage[capitalise,noabbrev]{cleveref}
\crefformat{equation}{#2#1#3}

\usepackage{mathtools}
\usepackage{varwidth}
\usepackage{pifont}

\usepackage{listings}
\usepackage{mdframed}
\newenvironment{temp}{\begin{mdframed}[backgroundcolor=red!7, linewidth=0, skipabove=1ex, leftmargin=1ex, rightmargin=0, innerleftmargin=0, innerrightmargin=0, innertopmargin=0, innerbottommargin=0]\setlength{\abovedisplayskip}{0ex}\raisebox{-\height-3pt}[0pt][0pt]{\hspace{.965\textwidth}\color{red}\huge\ding{56}}}{\end{mdframed}}

\usepackage{wrapfig}

\usepackage{xifthen}
\newcommand{\varcitet}[3][]{\citeauthor{#2}#3~[\ifthenelse{\isempty{#1}}{\citeyear{#2}}{\citeyear[#1]{#2}}]}

\usepackage[inline]{enumitem} 
\newlist{inlineenum}{enumerate*}{1}
\setlist[inlineenum]{label=(\arabic*)}

\usepackage[obeyFinal,color=yellow,textsize=scriptsize%
,disable%
]{todonotes}

\newenvironment{aha}{\medskip}{\unskip\medskip} 
\makeatletter
\newcommand{\pause}{\medskip\centerline{$\ast\quad\ast\quad\ast$}\medskip\@afterindentfalse\@afterheading} 
\newcommand{\bigpause}{\medskip\centerline{$\ast\quad\ast\quad\ast\quad\ast\quad\ast\quad\ast\quad\ast\quad\ast\quad\ast\quad\ast\quad\ast$}\medskip\@afterindentfalse\@afterheading} 
\makeatother

\newcommand{\csp}{\hspace{.5em minus .1em}}
\newcommand{\equals}{\enskip=\enskip}

\usepackage{tikzit}
\input{string.tikzstyles}

\makeatletter
\@ifundefined{lhs2tex.lhs2tex.sty.read}%
  {\@namedef{lhs2tex.lhs2tex.sty.read}{}%
   \newcommand\SkipToFmtEnd{}%
   \newcommand\EndFmtInput{}%
   \long\def\SkipToFmtEnd#1\EndFmtInput{}%
  }\SkipToFmtEnd

\newcommand\ReadOnlyOnce[1]{\@ifundefined{#1}{\@namedef{#1}{}}\SkipToFmtEnd}
\usepackage{amstext}
\usepackage{amssymb}
\usepackage{stmaryrd}
\DeclareFontFamily{OT1}{cmtex}{}
\DeclareFontShape{OT1}{cmtex}{m}{n}
  {<5><6><7><8>cmtex8
   <9>cmtex9
   <10><10.95><12><14.4><17.28><20.74><24.88>cmtex10}{}
\DeclareFontShape{OT1}{cmtex}{m}{it}
  {<-> ssub * cmtt/m/it}{}

\DeclareFontShape{OT1}{cmtt}{bx}{n}
  {<5><6><7><8>cmtt8
   <9>cmbtt9
   <10><10.95><12><14.4><17.28><20.74><24.88>cmbtt10}{}
\DeclareFontShape{OT1}{cmtex}{bx}{n}
  {<-> ssub * cmtt/bx/n}{}

\newcommand{\Conid}[1]{\mathit{#1}}
\newcommand{\Varid}[1]{\mathit{#1}}
\newcommand{\anonymous}{\kern0.06em \vbox{\hrule\@width.5em}}

\renewcommand{\leq}{\leqslant}

\usepackage{polytable}

\@ifundefined{mathindent}%
  {\newdimen\mathindent\mathindent\leftmargini}%
  {}%

\def\resethooks{%
  \global\let\SaveRestoreHook\empty
  \global\let\ColumnHook\empty}
\newcommand*{\savecolumns}[1][default]%
  {\g@addto@macro\SaveRestoreHook{\savecolumns[#1]}}
\newcommand*{\restorecolumns}[1][default]%
  {\g@addto@macro\SaveRestoreHook{\restorecolumns[#1]}}
\newcommand*{\aligncolumn}[2]%
  {\g@addto@macro\ColumnHook{\column{#1}{#2}}}

\resethooks

\newcommand{\onelinecommentchars}{\quad-{}- }
\newcommand{\commentbeginchars}{\enskip\{-}
\newcommand{\commentendchars}{-\}\enskip}

\newcommand{\visiblecomments}{%
  \let\onelinecomment=\onelinecommentchars
  \let\commentbegin=\commentbeginchars
  \let\commentend=\commentendchars}

\newcommand{\invisiblecomments}{%
  \let\onelinecomment=\empty
  \let\commentbegin=\empty
  \let\commentend=\empty}

\visiblecomments

\newlength{\blanklineskip}
\newcommand{\hsindent}[1]{\quad}
\let\hspre\empty
\let\hspost\empty

\EndFmtInput
\makeatother
\ReadOnlyOnce{polycode.fmt}%
\makeatletter

\newcommand{\hsnewpar}[1]%
  {{\parskip=0pt\parindent=0pt\par\vskip #1\noindent}}

\newcommand{\hscodestyle}{}

\newcommand{\sethscode}[1]%
  {\expandafter\let\expandafter\hscode\csname #1\endcsname
   \expandafter\let\expandafter\endhscode\csname end#1\endcsname}

  {\par\noindent
   \advance\leftskip\mathindent
   \hscodestyle
   \let\\=\@normalcr
   \let\hspre\(\let\hspost\)%
   \pboxed}%
  {\endpboxed\)%
   \par\noindent
   \ignorespacesafterend}

  {\hsnewpar\abovedisplayskip
   \advance\leftskip\mathindent
   \hscodestyle
   \let\hspre\(\let\hspost\)%
   \pboxed}%
  {\endpboxed%
   \hsnewpar\belowdisplayskip
   \ignorespacesafterend}

  {\hsnewpar\abovedisplayskip
   \advance\leftskip\mathindent
   \hscodestyle
   \let\\=\@normalcr
   \(\pboxed}%
  {\endpboxed\)%
   \hsnewpar\belowdisplayskip
   \ignorespacesafterend}

\newcommand{\plainhs}{\sethscode{plainhscode}}

\plainhs

  {\hsnewpar\abovedisplayskip
   \advance\leftskip\mathindent
   \hscodestyle
   \let\\=\@normalcr
   \(\parray}%
  {\endparray\)%
   \hsnewpar\belowdisplayskip
   \ignorespacesafterend}

  {\parray}{\endparray}

  {\(\parray}{\endparray\)}

\def\codeframewidth{\arrayrulewidth}
\RequirePackage{calc}

  {\parskip=\abovedisplayskip\par\noindent
   \hscodestyle
   \arrayrulewidth=\codeframewidth
   \tabular{@{}|p{\linewidth-2\arraycolsep-2\arrayrulewidth-2pt}|@{}}%
   \hline\framedhslinecorrect\\{-1.5ex}%
   \let\endoflinesave=\\
   \let\\=\@normalcr
   \(\pboxed}%
  {\endpboxed\)%
   \framedhslinecorrect\endoflinesave{.5ex}\hline
   \endtabular
   \parskip=\belowdisplayskip\par\noindent
   \ignorespacesafterend}

\newcommand{\framedhslinecorrect}[2]%
  {#1[#2]}

  {\(\def\column##1##2{}%
   \let\>\undefined\let\<\undefined\let\\\undefined
   \newcommand\>[1][]{}\newcommand\<[1][]{}\newcommand\\[1][]{}%
   \def\fromto##1##2##3{##3}%
   }{\) }%

  {\let\orighscode=\hscode
   \let\origendhscode=\endhscode
   \def\endhscode{\def\hscode{\endgroup\def\@currenvir{hscode}\\}\begingroup}
   \orighscode\def\hscode{\endgroup\def\@currenvir{hscode}}}%
  {\origendhscode
   \global\let\hscode=\orighscode
   \global\let\endhscode=\origendhscode}%

\makeatother
\EndFmtInput
\ReadOnlyOnce{agda.fmt}%

\RequirePackage[T1]{fontenc}
\RequirePackage[utf8]{inputenc}
\RequirePackage{amsfonts}

\DeclareUnicodeCharacter{737}{\textsuperscript{l}}
\DeclareUnicodeCharacter{8718}{\ensuremath{\blacksquare}}
\DeclareUnicodeCharacter{8759}{::}
\DeclareUnicodeCharacter{9669}{\ensuremath{\triangleleft}}
\DeclareUnicodeCharacter{8799}{\ensuremath{\stackrel{\scriptscriptstyle ?}{=}}}
\DeclareUnicodeCharacter{10214}{\ensuremath{\llbracket}}
\DeclareUnicodeCharacter{10215}{\ensuremath{\rrbracket}}
\DeclareUnicodeCharacter{27E6}{\ensuremath{\llbracket}}
\DeclareUnicodeCharacter{27E7}{\ensuremath{\rrbracket}}
\DeclareUnicodeCharacter{2200}{\ensuremath{\forall}}

\DeclareUnicodeCharacter{2294}{\ensuremath{\sqcup}}
\DeclareUnicodeCharacter{2080}{\ensuremath{_0}}
\DeclareUnicodeCharacter{2081}{\ensuremath{_1}}
\DeclareUnicodeCharacter{2082}{\ensuremath{_2}}
\DeclareUnicodeCharacter{2083}{\ensuremath{_3}}

\DeclareUnicodeCharacter{2115}{\ensuremath{\mathbb{N}}}
\DeclareUnicodeCharacter{2236}{:}
\DeclareUnicodeCharacter{2237}{\ensuremath{\mathrel{::}}}
\DeclareUnicodeCharacter{03A3}{\ensuremath{\Sigma}}
\DeclareUnicodeCharacter{03BB}{\ensuremath{\lambda}}
\DeclareUnicodeCharacter{039B}{\ensuremath{\Lambda}}
\DeclareUnicodeCharacter{03C0}{\ensuremath{\pi}}
\DeclareUnicodeCharacter{03C3}{\ensuremath{\sigma}}
\DeclareUnicodeCharacter{03C9}{\ensuremath{\omega}}

\DeclareUnicodeCharacter{2032}{\ensuremath{\prime}}
\DeclareUnicodeCharacter{2113}{\ensuremath{\ell}}
\DeclareUnicodeCharacter{2207}{\ensuremath{\nabla}}
\DeclareUnicodeCharacter{220B}{\ensuremath{\ni}}
\DeclareUnicodeCharacter{2264}{\ensuremath{\leq}}
\DeclareUnicodeCharacter{21D2}{\ensuremath{\Rightarrow}}
\DeclareUnicodeCharacter{22A2}{\ensuremath{\vdash}}
\DeclareUnicodeCharacter{22A4}{\ensuremath{\top}}
\DeclareUnicodeCharacter{22A5}{\ensuremath{\bot}}

\DeclareUnicodeCharacter{1D57}{\ensuremath{^t}}

\renewcommand\Varid[1]{\mathord{\textsf{#1}}}
\let\Conid\Varid
\newcommand\Keyword[1]{\textsf{\textbf{#1}}}
\EndFmtInput

\definecolor{suppressed}{RGB}{225,225,225}
\definecolor{goal}{RGB}{209,243,205}
\newcommand{\highlight}[2]{\smash{\text{\colorbox{#1}{\kern-.1em\vphantom{\vrule height 1.35ex depth 0.16ex}\smash{\ensuremath{#2}}\kern-.1em}}}}

\newcommand{\ignorenext}[1]{}

\newcommand{\Con}[1]{\mathbf{#1}}

\newcommand{\Var}[1]{\mathit{#1}}

\begin{document}

\setlength{\mathindent}{2\parindent}

\author{Hsiang-Shang Ko}
\email{joshko@iis.sinica.edu.tw}
\orcid{0000-0002-2439-1048}
\author{Shin-Cheng Mu}
\email{scm@iis.sinica.edu.tw}
\orcid{0000-0002-4755-601X}
\affiliation{
  \institution{Academia Sinica}
  \department{Institute of Information Science}
  \streetaddress{128 Academia Road}
  \city{Taipei}
  \country{Taiwan}
  \postcode{115201}
}
\author{Jeremy Gibbons}
\email{jeremy.gibbons@cs.ox.ac.uk}
\orcid{0000-0002-8426-9917}
\affiliation{
  \institution{University of Oxford}
  \department{Department of Computer Science}
  \streetaddress{Wolfson Building, Parks Road}
  \city{Oxford}
  \country{UK}
  \postcode{OX1 3QD}
}

\title{Binomial Tabulation: A Short Story
}

\begin{abstract}
We reconstruct some of the development in Richard Bird's \citeyearpar{Bird-zippy-tabulations} paper \textit{Zippy Tabulations of Recursive Functions}, using dependent types and string diagrams rather than mere simple types.
This paper serves as an intuitive introduction to and demonstration of these concepts for the curious functional programmer, who ideally already has some exposure to dependent types and category theory, is not put off by basic concepts like indexed types and functors, and wants to see a more practical example.

The paper is presented in the form of a short story, narrated from the perspective of a functional programmer trying to follow the development in Bird's paper.
The first section recaps the original simply typed presentation.
The second section explores a series of refinements that can be made using dependent types.
The third section uses string diagrams to simplify arguments involving functors and naturality.
The short story ends there, but the paper concludes with a discussion and reflection in the afterword.
\end{abstract}

%
%

\maketitle



\section{Functions}
\label{sec:functions}

`What on earth is this function doing?'

\medskip

I stare at the late Richard Bird's~\citeyearpar{Bird-zippy-tabulations} paper \textit{Zippy Tabulations of Recursive Functions}, frowning.

\begin{lstlisting}
cd :: B a -> B (L a)
cd (Bin (Tip y) (Tip z)) = Tip [y,z]
cd (Bin (Tip y) u      ) = Tip (y : ys) where Tip ys = cd u
cd (Bin t       (Tip z)) = Bin (cd t) (mapB (: [z]) t)
cd (Bin t       u      ) = Bin (cd t) (zipBWith (:) t (cd u))
\end{lstlisting}
I know \lstinline{B} is this Haskell data type of binary trees:
\begin{lstlisting}
data B a = Tip a | Bin (B a) (B a)
\end{lstlisting}
Presumably \lstinline{mapB} and \lstinline{zipBWith} are the usual map and zip functions for these trees, and \lstinline{L}~is the standard list type.
But how did Richard come up with such an incomprehensible function definition?
He didn't bother to explain it in his paper.
It might have helped if he had provided some examples.

\subsection{Top-Down and Bottom-Up Algorithms}
\label{sec:algorithms}

From the explanations that \emph{are} in the paper, I suppose I can guess roughly what \lstinline{cd} should do.
Richard was studying the relationship between top-down and bottom-up algorithms that solve problems specified recursively on some input data structure.
When the input is a nonempty list, a generic top-down algorithm is defined by
\begin{lstlisting}
td :: (a -> s) -> (L s -> s) -> L a -> s
td f g [x] = f x
td f g xs  = g (map (td f g) (dc xs))
\end{lstlisting}
The input of \lstinline{td} is a nonempty list of \lstinline{a}'s, and the output is a solution of type \lstinline{s} for the input list.
Singleton lists form the base case, for which the given function~\lstinline{f} computes a solution directly.
If an input list \lstinline{xs} is not singleton, it is decomposed into shorter lists by some\csp\lstinline{dc :: L a -> L (L a)}.
Then \lstinline{td} recursively computes a subsolution for each shorter list.
Finally, \lstinline{g}~combines the subsolutions into a solution for \lstinline{xs}.
In fact, in order to cover a wider range of algorithms, Richard's definition is more abstract and general.
But I'm working with this simplified version as an easier starting point.

\begin{figure}[t]
\centering
\includegraphics[width=0.95\textwidth]{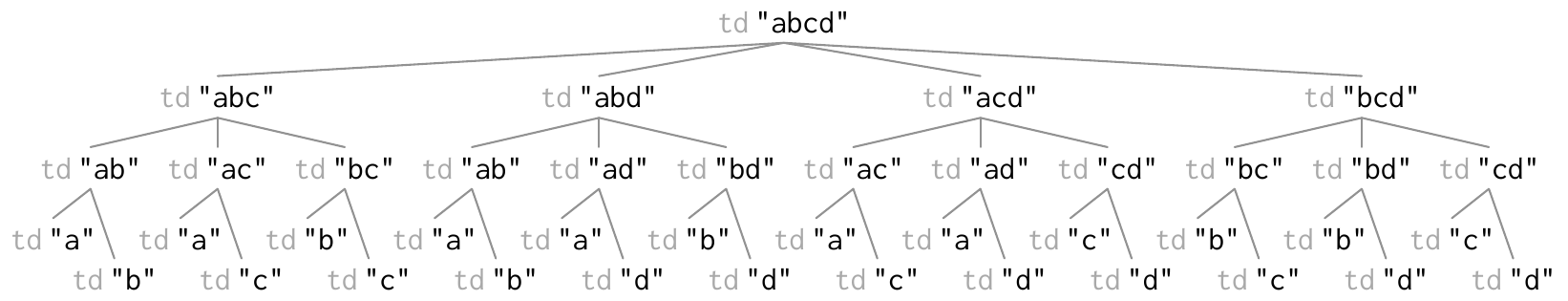}
\caption{Computing\csp\lstinline{td "abcd"}\csp top-down.}
\label{fig:td-call-tree}
\end{figure}

In the last example of the paper, \lstinline{dc} computes all the \emph{immediate sublists} of a given list --- all the lists with exactly one element missing:
\begin{lstlisting}
dc :: L a -> L (L a)
dc [x,y]       = [[x],[y]]
dc (xs ++ [x]) = [xs] ++ [ys ++ [x] | ys <- dc xs]
\end{lstlisting}
(Richard assumed that a list could be matched with a snoc pattern\csp\lstinline{xs ++ [x]}.)
For example, computing a solution for \lstinline{"abcd"} requires subsolutions for \lstinline{"abc"},\lstinline{"abd"}, \lstinline{"acd"}, and \lstinline{"bcd"}~(\cref{fig:td-call-tree}).
In turn, computing a solution for \lstinline{"abc"} requires subsolutions for \lstinline{"ab"}, \lstinline{"ac"}, and \lstinline{"bc"}, and so on.
When the problem decomposition reaches length-$2$ sublists ---~that's a bit of a mouthful, so let me just say `$2$-sublists' for short~--- it becomes evident that this \lstinline{dc} leads to \emph{overlapping subproblems}, and \lstinline{td} deals with that very inefficiently.
For example, a solution for \lstinline{"ab"} is required for solving the problem for \lstinline{"abc"} and \lstinline{"abd"}, and \lstinline{td} computes that solution twice.
And it gets worse further down: a solution for each $1$-sublist is computed $6$~times!

\begin{figure}[t]
\centering
\includegraphics[width=0.75\textwidth]{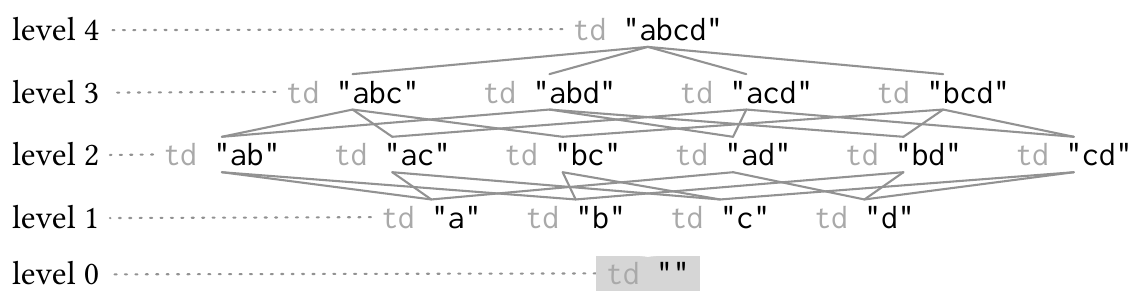}
\caption{Computing\csp\lstinline{td "abcd"}\csp bottom-up.}
\label{fig:sublists-lattice}
\end{figure}

It's better to proceed bottom-up instead, working upwards through a lattice of sublists~(\cref{fig:sublists-lattice}), level by level.
Level~$1$ consists of solutions for the $1$-sublists.
Then solutions for the $(k+1)$-sublists in level $k+1$ are computed from subsolutions in level~$k$.
Finally, the top level consists of a single solution, for the input list.
More specifically, if the levels were represented as lists, level~$2$ would be
\begin{lstlisting}
[td "ab", td "ac", td "bc", td "ad" ...]
\end{lstlisting}
One way to construct level~$3$ from level~$2$ would be using a function\csp\lstinline{cd' :: L a -> L (L a)}\csp that copies and rearranges the elements such that the subsolutions for the immediate sublists of the same list are brought together:
\begin{lstlisting}
[[td "ab", td "ac", td "bc"], [td "ab", td "ad", td "bd"] ...]
\end{lstlisting}
Then applying\csp\lstinline{map g}\csp to this list of lists would produce level~$3$:
\begin{lstlisting}
[td "abc", td "abd", td "acd", td "bcd" ...]
\end{lstlisting}
If such a function \lstinline{cd'} could be constructed, a bottom-up algorithm computing the same solution as \lstinline{td} would be given by
\begin{lstlisting}
bu' :: (a -> s) -> (L s -> s) -> L a -> s
bu' f g = head . loop . map f
  where loop [y] = [y]
        loop ys  = loop (map g (cd' ys))
\end{lstlisting}
Level~$1$ is obtained by applying \lstinline{f} to each element of the input list.
Then the \lstinline{loop} function keeps applying\csp\lstinline{map g . cd'}\csp to get the next level.
It stops at a level with a single element, which is a solution for the entire input list.
Like \lstinline{td}, this \lstinline{bu'} is a simplified version.
To cope with more general problems, Richard had to store something more complex in each level, but I don't think I need that.

\subsection{Rearranging Binary Trees}
\label{sec:bu}

That's what I understand about \lstinline{cd} so far.
But Richard must have realised at some point that it's difficult to define the \lstinline{cd} rearrangement using lists, and decided to represent each level using the \lstinline{B}~data type.
So\csp\lstinline{cd :: B a -> B (L a)}, and the actual bottom-up algorithm \lstinline{bu} is defined by
\begin{lstlisting}
bu :: (a -> s) -> (L s -> s) -> L a -> s
bu f g = unTip . loop . cvt . map f
  where loop (Tip x) = Tip x
        loop xs      = loop (mapB g (cd xs))
\end{lstlisting}
where \lstinline{unTip} extracts the element of a \lstinline{Tip} tree, and \lstinline{cvt} converts a list to a tree:
\begin{lstlisting}
unTip :: B a -> a            cvt :: L a -> B a
unTip (Tip x) = x            cvt [x]         = Tip x
                             cvt (xs ++ [x]) = Bin (cvt xs) (Tip x)
\end{lstlisting}

I wonder if I have to use~\lstinline{B} instead of lists in \lstinline{cd}.
If I'm given level~1 of the lattice~(\cref{fig:sublists-lattice}) as a 4-list, I know they are solutions for the four 1-sublists, and surely there's no problem rearranging them for level~2\ldots?

Oh wait, I don't actually know.
All I get is a 4-list.
This 4-list could be level~1, but it could well be level~3.
And I don't get any information from the elements --- the element type is parametrically quantified.
So there isn't enough context for me to decide whether I should produce a 6-list (level~2) or a 1-list (level~4).

Presumably, Richard's trees gave him more context.
I try to trace Richard's \lstinline{cd} to find out how it works.
Given input \lstinline{"abcd"}, the function\csp\lstinline{cvt . map f}\csp yields a tree slanted to the left as level~$1$:
\begin{lstlisting}
Bin (Bin (Bin (Tip (td "a")) (Tip (td "b"))) (Tip (td "c"))) (Tip (td "d"))
\end{lstlisting}
Following Richard's convention, I draw a\csp\lstinline{Tip x}\csp as \lstinline{x}, and\csp\lstinline{Bin t u}\csp as a dot with~\lstinline{t} to its left and \lstinline{u}~below~(\cref{fig:map_g_cd}).
Applying\csp\lstinline{mapB g . cd}\csp to this, I get level~$2$. 
For a closer look, I apply only \lstinline{cd} to level~$2$.
Indeed, with its clever mapping and zipping, \lstinline{cd} manages to bring together precisely the right elements, and produces a `level~$2{}^{1\kern-.2em}/_{\!2}$'. 
Then I reach level~$3$ by applying\csp\lstinline{mapB g}.
There's indeed more context for distinguishing levels $1$~and~$3$: their tree representations have the same size but different shapes.

The intuition about having enough context seems useful.
I was puzzled by why Richard started from singleton lists instead of the empty list.
The intuition helps to explain that too.
Level~$0$ would be a singleton list/tree, and I wouldn't know the number of values level~$1$ should contain.
That number is the length of the input list, and \lstinline{cd} doesn't get that information.
So there isn't enough context for going from level~$0$ to level~$1$, regardless of how levels are represented.

\subsection{Rearranging Binomial Trees}
\label{sec:binomial}

That still doesn't give me much insight into why \lstinline{cd} works though.
Presumably, \lstinline{cd} does something useful only for the trees built by\csp\lstinline{cvt . map f}\csp and \lstinline{cd} itself.
What are the constraints on these trees, and how does \lstinline{cd} exploit them?

\begin{figure}[t]
\centering
\includegraphics[width=0.85\textwidth]{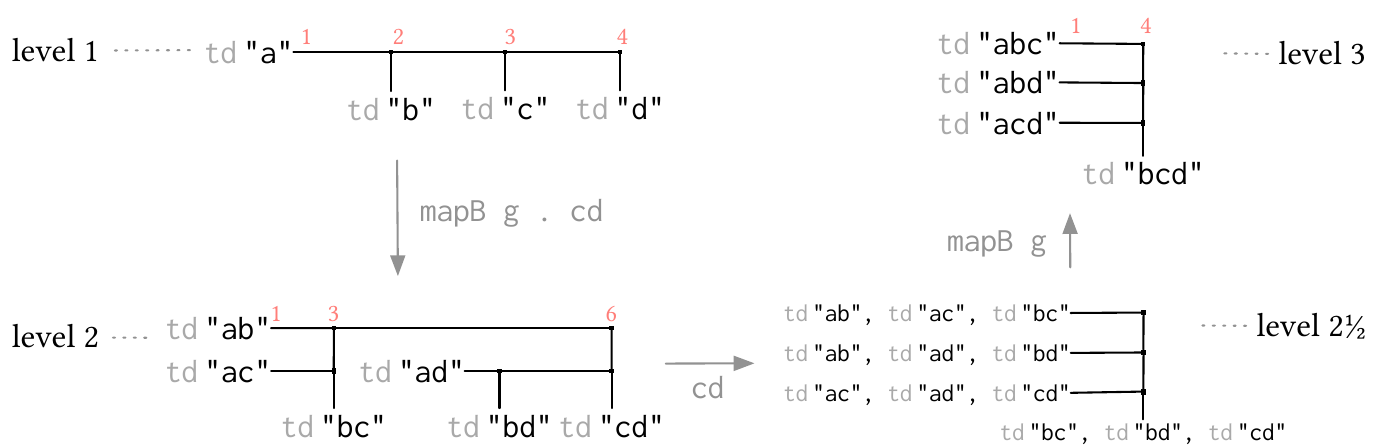}
\caption{How\csp\lstinline{mapB g . cd}\csp constructs a new level.}
\label{fig:map_g_cd}
\end{figure}

Richard did give a hint: the sizes $[1,2,3,4]$, $[1,3,6]$, and $[1,4]$ of subtrees along their left spines (the red numbers in Figure~\ref{fig:map_g_cd}) are the diagonals of Pascal's triangle --- the trees are related to \emph{binomial coefficients}!
The binomial coefficient~\ensuremath{C{}^{\Var n}_{\Var k}} is the number of ways of choosing $k$~elements from an $n$-list.
Indeed, each level~$k$ in the lattice~(\cref{fig:sublists-lattice}) contains values about $k$-sublists.
For example, level~$2$ has \ensuremath{C{}^{\Varid{4}}_{\mathrm 2}\;\unskip=\ignorenext\;\Varid{6}} values, and there are $6$~ways of choosing $2$~elements from a $4$-list.

Aha!
I can even see a pattern related to the choices in the tree representation of level~$2$
(\cref{fig:map_g_cd}): the right subtree is about all the 2-sublists that end with~\lstinline{'d'}, and the left subtree about the other 2-sublists not containing~\lstinline{'d'}.
To choose $2$~elements from \lstinline{"abcd"}, I can include the rightmost element~\lstinline{'d'} or not.
If \lstinline{'d'}~is included, there are \ensuremath{C{}^{\mathrm 3}_{\mathrm 1}} ways of choosing $1$~element from \lstinline{"abc"} to go with~\lstinline{'d'}.
If \lstinline{'d'}~is not included, there are \ensuremath{C{}^{\mathrm 3}_{\mathrm 2}} ways of choosing $2$~elements from \lstinline{"abc"}.
And the total number of $2$-sublists is \ensuremath{C{}^{\mathrm 3}_{\mathrm 2}\;\Varid{+}\;C{}^{\mathrm 3}_{\mathrm 1}\;\unskip=\ignorenext\;C{}^{\Varid{4}}_{\mathrm 2}}.
All the \lstinline{Bin} nodes fit this pattern.
I guess the trees are supposed to satisfy a binomial shape constraint (and the name of the~\lstinline{B} data type could refer to `binomial' as much as to `binary').

That's about as many clues as I can get from Richard's paper for now.
Given these clues, how do I prove that \lstinline{cd} indeed does the job --- bringing values about related immediate sublists together?
In fact, how do I even write that down as a formal specification?
And how does that help me to prove that \lstinline{td} equals \lstinline{bu}?

I'm worried that there will be many complex proofs waiting ahead for me.



\section{Types}

\subsection{Shapes in Types}
\label{sec:shape}

The binomial shape constraint seems to be the easiest next target, because there's a standard solution: capturing the tree shape in its type.
Shape-indexed data types are so common now that I'm tempted to say they should count as simple types.
They're common even in Haskell.
I prefer a proper dependently typed language though, so I open my favourite editor, and switch to Agda.

The classic example of shape-indexing is, of course, length-indexed lists (or `vectors'):
\begin{hscode}\SaveRestoreHook
\column{B}{@{}>{\hspre}l<{\hspost}@{}}%
\column{3}{@{}>{\hspre}l<{\hspost}@{}}%
\column{8}{@{}>{\hspre}l<{\hspost}@{}}%
\column{25}{@{}>{\hspre}l<{\hspost}@{}}%
\column{32}{@{}>{\hspre}l<{\hspost}@{}}%
\column{40}{@{}>{\hspre}l<{\hspost}@{}}%
\column{E}{@{}>{\hspre}l<{\hspost}@{}}%
\>[B]{}\Keyword{data}\;\Conid{Vec}\;\mathbin{:}\;\Conid{ℕ}\;\Varid{→}\;\Conid{Set}\;\Varid{→}\;\Conid{Set}\;\Keyword{where}{}\<[E]%
\\
\>[B]{}\hsindent{3}{}\<[3]%
\>[3]{}\Varid{[]}\;{}\<[8]%
\>[8]{}\mathbin{:}\;{}\<[25]%
\>[25]{}\Conid{Vec}\;{}\<[32]%
\>[32]{}\Con{zero}\;{}\<[40]%
\>[40]{}\Var a{}\<[E]%
\\
\>[B]{}\hsindent{3}{}\<[3]%
\>[3]{}\anonymous \kern.5pt\Varid{∷}\anonymous \;{}\<[8]%
\>[8]{}\mathbin{:}\;\Var a\;\Varid{→}\;\Conid{Vec}\;\Var n\;\Var a\;\Varid{→}\;{}\<[25]%
\>[25]{}\Conid{Vec}\;({}\<[32]%
\>[32]{}\Con{suc}\;\Var n)\;{}\<[40]%
\>[40]{}\Var a{}\<[E]%
\ColumnHook
\end{hscode}\resethooks
(I enjoy writing Agda types these days because I no longer have to quantify over each and every variable, such as \ensuremath{\Var a}~and~\ensuremath{\Var n} in the type of cons --- Agda supports implicit `generalisation of declared variables' now.)
The constructors used in a list of type \ensuremath{\Conid{Vec}\;\Var n\;\Var a} are completely determined by the length index~\ensuremath{\Var n}.
The data type definition could even be understood as if it were performing pattern matching on the index: if the index is \ensuremath{\Con{zero}}, then the list has to be nil; otherwise the index is a \ensuremath{\Con{suc}}, and the list has to start with a cons.
(\citet{Chapman-levitation} did develop a theory where data types can be defined by pattern matching on indices in this way.)

In the same vein, I write down a shape-indexed version of Richard's \lstinline{B}~data type~(\cref{sec:binomial}):
\begin{hscode}\SaveRestoreHook
\column{B}{@{}>{\hspre}l<{\hspost}@{}}%
\column{3}{@{}>{\hspre}l<{\hspost}@{}}%
\column{9}{@{}>{\hspre}l<{\hspost}@{}}%
\column{13}{@{}>{\hspre}l<{\hspost}@{}}%
\column{38}{@{}>{\hspre}l<{\hspost}@{}}%
\column{48}{@{}>{\hspre}l<{\hspost}@{}}%
\column{54}{@{}>{\hspre}l<{\hspost}@{}}%
\column{62}{@{}>{\hspre}l<{\hspost}@{}}%
\column{E}{@{}>{\hspre}l<{\hspost}@{}}%
\>[B]{}\Keyword{data}\;\Conid{B}\;\mathbin{:}\;\Conid{ℕ}\;\Varid{→}\;\Conid{ℕ}\;\Varid{→}\;\Conid{Set}\;\Varid{→}\;\Conid{Set}\;\Keyword{where}{}\<[E]%
\\
\>[B]{}\hsindent{3}{}\<[3]%
\>[3]{}\Con{tip_z}\;{}\<[9]%
\>[9]{}\mathbin{:}\;{}\<[13]%
\>[13]{}\Var a\;{}\<[38]%
\>[38]{}\Varid{→}\;\Conid{B}\;\Var n\;{}\<[54]%
\>[54]{}\Con{zero}\;{}\<[62]%
\>[62]{}\Var a{}\<[E]%
\\
\>[B]{}\hsindent{3}{}\<[3]%
\>[3]{}\Con{tip_s}\;{}\<[9]%
\>[9]{}\mathbin{:}\;{}\<[13]%
\>[13]{}\Var a\;{}\<[38]%
\>[38]{}\Varid{→}\;\Conid{B}\;(\Con{suc}\;{}\<[48]%
\>[48]{}\Var k)\;({}\<[54]%
\>[54]{}\Con{suc}\;\Var k)\;{}\<[62]%
\>[62]{}\Var a{}\<[E]%
\\
\>[B]{}\hsindent{3}{}\<[3]%
\>[3]{}\Con{bin}\;{}\<[9]%
\>[9]{}\mathbin{:}\;{}\<[13]%
\>[13]{}\Conid{B}\;\Var n\;(\Con{suc}\;\Var k)\;\Var a\;\Varid{→}\;\Conid{B}\;\Var n\;\Var k\;\Var a\;{}\<[38]%
\>[38]{}\Varid{→}\;\Conid{B}\;(\Con{suc}\;{}\<[48]%
\>[48]{}\Var n)\;({}\<[54]%
\>[54]{}\Con{suc}\;\Var k)\;{}\<[62]%
\>[62]{}\Var a{}\<[E]%
\ColumnHook
\end{hscode}\resethooks
The size of a tree of type \ensuremath{\Conid{B}\;\Var n\;\Var k\;\Var a} with $k \le n$ is precisely the binomial coefficient~\ensuremath{C{}^{\Var n}_{\Var k}}. Naturally, there are no inhabitants when $k > n$.
Like \ensuremath{\Conid{Vec}}, the indices $n, k$ determine the constructors.
If \ensuremath{\Var k}~is \ensuremath{\Con{zero}}, then the tree is a \ensuremath{\Con{tip_z}} with one element (\ensuremath{C{}^{\Var n}_{\mathrm 0}\;\unskip=\ignorenext\;\mathrm 1}).
If \ensuremath{\Var n}~is \ensuremath{\Con{suc}\;\Var k}, then the tree is a \ensuremath{\Con{tip_s}} also with one element (\ensuremath{C{}^{\mathrm 1\Varid{+}\Var k}_{\mathrm 1\Varid{+}\Var k}\;\unskip=\ignorenext\;\mathrm 1}).
Otherwise the tree is a \ensuremath{\Con{bin}}, and the sizes \ensuremath{C{}^{\Var n}_{\mathrm 1\Varid{+}\Var k}} and \ensuremath{C{}^{\Var n}_{\Var k}} of the two subtrees add up to the expected size \ensuremath{C{}^{\mathrm 1\Varid{+}\Var n}_{\mathrm 1\Varid{+}\Var k}} of the whole tree.
The trees are now truly \emph{binomial} rather than just binary, formalising Richard's hint about sizes.
I'll write \ensuremath{\Conid{B}\;\unskip^{\Var n}_{\Var k}} for \ensuremath{\Conid{B}\;\Var n\;\Var k}, by analogy with \ensuremath{C{}^{\Var n}_{\Var k}}.

And now I can give a more informative type to \lstinline{cd}~(\cref{sec:functions}):
\begin{hscode}\SaveRestoreHook
\column{B}{@{}>{\hspre}l<{\hspost}@{}}%
\column{E}{@{}>{\hspre}l<{\hspost}@{}}%
\>[B]{}\Varid{cd}\;\mathbin{:}\;\highlight{suppressed}{\mathrm 1\;\unskip\le\ignorenext\;\Var k}\;\Varid{→}\;\highlight{suppressed}{\Var k\;\unskip<\ignorenext\;\Var n}\;\Varid{→}\;\Conid{B}\;\unskip^{\Var n}_{\Var k}\;\Var a\;\Varid{→}\;\Conid{B}\;\unskip^{\Var n}_{\mathrm 1\Varid{+}\Var k}\;(\Conid{Vec}\;(\mathrm 1\;\Varid{+}\;\Var k)\;\Var a){}\<[E]%
\ColumnHook
\end{hscode}\resethooks
It takes as input the data for level~$k$ out of $n$~levels in the sublist lattice~(\cref{fig:sublists-lattice}), with $1 \le k < n$; these are the solutions for each of the \ensuremath{C{}^{\Var n}_{\Var k}} $k$-sublists of the original $n$-list.
And it returns as output the components for level \ensuremath{\mathrm 1\;\Varid{+}\;\Var k}; there are \ensuremath{C{}^{\Var n}_{\mathrm 1\Varid{+}\Var k}} of these, each a \ensuremath{(\mathrm 1\;\Varid{+}\;\Var k)}-list (to be fed into \lstinline{g} when used in \lstinline{bu}).
There are two (greyed out) `side conditions' \ensuremath{\mathrm 1\;\unskip\le\ignorenext\;\Var k} and \ensuremath{\Var k\;\unskip<\ignorenext\;\Var n}, which I don't want to bother with when I'm thinking at a higher level.
I'm going to check these conditions mentally and then ignore everything related to these conditions --- I'll call \ensuremath{\Varid{cd}} as if it were a function without the first two arguments, and skip over cases that are impossible due to the conditions.
Agda ensures that I don't forget about all the ignored stuff in the final code though.

I continue to transcribe the definition of \lstinline{cd} interactively in Agda.
\begin{hscode}\SaveRestoreHook
\column{B}{@{}>{\hspre}l<{\hspost}@{}}%
\column{10}{@{}>{\hspre}l<{\hspost}@{}}%
\column{15}{@{}>{\hspre}l<{\hspost}@{}}%
\column{26}{@{}>{\hspre}l<{\hspost}@{}}%
\column{31}{@{}>{\hspre}l<{\hspost}@{}}%
\column{42}{@{}>{\hspre}l<{\hspost}@{}}%
\column{52}{@{}>{\hspre}l<{\hspost}@{}}%
\column{E}{@{}>{\hspre}l<{\hspost}@{}}%
\>[B]{}\Varid{cd}\;\mathbin{:}\;\highlight{suppressed}{\mathrm 1\;\unskip\le\ignorenext\;\Var k}\;\Varid{→}\;\highlight{suppressed}{\Var k\;\unskip<\ignorenext\;\Var n}\;\Varid{→}\;\Conid{B}\;\unskip^{\Var n}_{\Var k}\;\Var a\;\Varid{→}\;\Conid{B}\;\unskip^{\Var n}_{\mathrm 1\Varid{+}\Var k}\;(\Conid{Vec}\;(\mathrm 1\;\Varid{+}\;\Var k)\;\Var a){}\<[E]%
\\
\>[B]{}\Varid{cd}\;(\Con{bin}\;{}\<[15]%
\>[15]{}(\Con{tip_s}\;\Var y)\;{}\<[31]%
\>[31]{}(\Con{tip_z}\;\Var z){}\<[42]%
\>[42]{})\;\mathrel{=}\;\Con{tip_s}\;{}\<[52]%
\>[52]{}(\Var y\;\Varid{∷}\;\Var z\;\Varid{∷}\;\Varid{[]}){}\<[E]%
\\
\>[B]{}\Varid{cd}\;(\Con{bin}\;{}\<[15]%
\>[15]{}(\Con{tip_s}\;\Var y)\;{}\<[26]%
\>[26]{}\Var u\;\unskip@\ignorenext\;{}\<[31]%
\>[31]{}(\Con{bin}\;\anonymous \;\anonymous ){}\<[42]%
\>[42]{})\;\mathrel{=}\;\Con{tip_s}\;{}\<[52]%
\>[52]{}(\Var y\;\Varid{∷}\;\Varid{unTip}_{\Conid{B}}\;(\Varid{cd}\;\Var u)){}\<[E]%
\\
\>[B]{}\Varid{cd}\;(\Con{bin}\;{}\<[10]%
\>[10]{}\Var t\;\unskip@\ignorenext\;{}\<[15]%
\>[15]{}(\Con{bin}\;\anonymous \;\anonymous )\;{}\<[31]%
\>[31]{}(\Con{tip_z}\;\Var z){}\<[42]%
\>[42]{})\;\mathrel{=}\;\Con{bin}\;{}\<[52]%
\>[52]{}(\Varid{cd}\;\Var t)\;(\Varid{map}_{\Conid{B}}\;(\anonymous \kern.5pt\Varid{∷}\;(\Var z\;\Varid{∷}\;\Varid{[]}))\;\Var t){}\<[E]%
\\
\>[B]{}\Varid{cd}\;(\Con{bin}\;{}\<[10]%
\>[10]{}\Var t\;\unskip@\ignorenext\;{}\<[15]%
\>[15]{}(\Con{bin}\;\anonymous \;\anonymous )\;{}\<[26]%
\>[26]{}\Var u\;\unskip@\ignorenext\;{}\<[31]%
\>[31]{}(\Con{bin}\;\anonymous \;\anonymous ){}\<[42]%
\>[42]{})\;\mathrel{=}\;\Con{bin}\;{}\<[52]%
\>[52]{}(\Varid{cd}\;\Var t)\;(\Varid{zip}_{\Conid{B}}\kern-1.75pt\Conid{With}\;\anonymous \kern.5pt\Varid{∷}\anonymous \;\Var t\;(\Varid{cd}\;\Var u)){}\<[E]%
\ColumnHook
\end{hscode}\resethooks
In the first case, \ensuremath{\Con{bin}\;(\Con{tip_s}\;\Var y)\;(\Con{tip_z}\;\Var z)}, Agda conveniently figures out for me whether a \lstinline{Tip} in the pattern should be a \ensuremath{\Con{tip_s}} or \ensuremath{\Con{tip_z}}.
I expect Agda to fill in the right constructor for the goal type \ensuremath{\Conid{B}\;\unskip^{\mathrm 2}_{\mathrm 2}\;(\Conid{Vec}\;\mathrm 2\;\Var a)} too, but Agda complains that it cannot decide between \ensuremath{\Con{tip_s}} and \ensuremath{\Con{bin}}.
Indeed, \ensuremath{\Conid{B}\;\unskip^{\mathrm 2}_{\mathrm 2}\;(\Conid{Vec}\;\mathrm 2\;\Var a)} matches the result type of both \ensuremath{\Con{tip_s}} and \ensuremath{\Con{bin}}.
But \ensuremath{\Con{bin}} is actually impossible because its left subtree would have type \ensuremath{\Conid{B}\;\unskip^{\mathrm 1\Varid{+}\Var k}_{\mathrm 2\Varid{+}\Var k}\;\Var a}, and this type is uninhabited.
So the indices of~\ensuremath{\Conid{B}} still determine the constructor, though not as directly as in the case of \ensuremath{\Conid{Vec}}.

I go through the cases \ensuremath{\Con{bin}\;(\Con{tip_s}\;\Var y)\;\Var u} and \ensuremath{\Con{bin}\;\Var t\;(\Con{tip_z}\;\Var z)} without difficulties, after supplying the definitions of \ensuremath{\Varid{unTip}_{\Conid{B}}\;\mathbin{:}\;\Conid{B}\;\unskip^{\Var n}_{\Var n}\;\Var a\;\Varid{→}\;\Var a} and \ensuremath{\Varid{map}_{\Conid{B}}\;\mathbin{:}\;(\Var a\;\Varid{→}\;\Var b)\;\Varid{→}\;\Conid{B}\;\unskip^{\Var n}_{\Var k}\;\Var a\;\Varid{→}\;\Conid{B}\;\unskip^{\Var n}_{\Var k}\;\Var b}.
In the final \ensuremath{\Con{bin}\;\Var t\;\Var u} case, the result should be a \ensuremath{\Con{bin}}, and the left subtree \ensuremath{\Varid{cd}\;\Var t} is accepted by Agda.
Slightly anxiously, I start constructing the right subtree.
Agda tells me that \ensuremath{\Var t\;\mathbin{:}\;\Conid{B}\;\unskip^{\mathrm 1\Varid{+}\Var n}_{\mathrm 2\Varid{+}\Var k}\;\Var a} and \ensuremath{\Var u\;\mathbin{:}\;\Conid{B}\;\unskip^{\mathrm 1\Varid{+}\Var n}_{\mathrm 1\Varid{+}\Var k}\;\Var a}.
When I ask what type \ensuremath{\Varid{cd}\;\Var u} has, Agda responds with \ensuremath{\Conid{B}\;\unskip^{\mathrm 1\Varid{+}\Var n}_{\mathrm 2\Varid{+}\Var k}\;(\Conid{Vec}\;(\mathrm 2\;\Varid{+}\;\Var k)\;\Var a)}.
That's the same shape as~\ensuremath{\Var t}, so \ensuremath{\Var t}~and \ensuremath{\Varid{cd}\;\Var u} can be safely zipped together using a shape-preserving \ensuremath{\Varid{zip}_{\Conid{B}}\kern-1.75pt\Conid{With}\;\mathbin{:}\;(\Var a\;\Varid{→}\;\Var b\;\Varid{→}\;\Var c)\;\Varid{→}\;\Conid{B}\;\unskip^{\Var n}_{\Var k}\;\Var a\;\Varid{→}\;\Conid{B}\;\unskip^{\Var n}_{\Var k}\;\Var b\;\Varid{→}\;\Conid{B}\;\unskip^{\Var n}_{\Var k}\;\Var c}.

I've gone through the whole definition!
So, as I guessed, the binomial shape constraint holds throughout \ensuremath{\Varid{cd}}.
What's nice is that I didn't need to do much.
The type checker took care of most of the proof, using the indices to keep track of the tree shapes throughout the transcription.

\subsection{Properties in Types}
\label{sec:BT}

So much for the shape.
But how do I know that the contents are correct~--- that \lstinline{cd} is correctly rearranging the values?

The input to \lstinline{cd} is a tree of values associated with $k$-sublists (for some $1 \le k < n$), and these values are rearranged in relation to \ensuremath{(\mathrm 1\;\Varid{+}\;\Var k)}-sublists.
These values can have any type~\lstinline{a}, which is \emph{parametrically} quantified in the type of \lstinline{cd}.
The parametric quantification ensures that \lstinline{cd} cannot do different things to different~\lstinline{a}'s (because there's no way to do a case analysis on~\lstinline{a}).
So if I can specify what \lstinline{cd} should do for a specific choice of~\lstinline{a}, then \lstinline{cd} will have to do the same thing for all~\lstinline{a}'s.
Well, I suppose a natural choice is \ensuremath{\Var k}-sublists themselves.
I'll define Richard's trees~(\cref{fig:map_g_cd}) but just put \ensuremath{\Var k}-sublists in them, and then specify that \lstinline{cd} should transform level~\ensuremath{\Var k} to level $k{}^{1\kern-.2em}/_{\!2}$.

In Haskell, I suppose Richard might have defined the tree of \lstinline{k}-sublists of a list \lstinline{xs} as\csp\lstinline{choose k xs}, where the function \lstinline{choose} is defined by
\begin{lstlisting}
choose :: Int -> L a -> B (L a)
choose    0  _                     = Tip []
choose (k+1) xs | length xs == k+1 = Tip xs
choose (k+1) (xs ++ [x])           = Bin (choose (k+1) xs)
                                         (mapB (++[x]) (choose k xs))
\end{lstlisting}
The pattern-matching structure of \lstinline{choose} is the same as how \ensuremath{\Conid{B}\;\unskip^{\Var n}_{\Var k}} analyses its indices~(\cref{sec:shape}), except that here I'm working with a list rather than just its length.
The function generalises Richard's \lstinline{dc} that computes the immediate sublists~(\cref{sec:algorithms}):\csp\lstinline{dc xs}\csp can be redefined as\csp\lstinline{flatten (choose (length xs - 1) xs)}, where
\begin{lstlisting}
flatten :: B a -> L a
flatten (Tip x)   = [x]
flatten (Bin t u) = flatten t ++ flatten u
\end{lstlisting}
Then Richard could have specified \lstinline{cd} by
\begin{equation}
\text{\lstinline{cd (choose k xs)}} \equals \text{\lstinline{mapB (flatten . choose k) (choose (k+1) xs)}}
\tag{$\ast$}
\label{eq:cd-spec}
\end{equation}
Informally: given all the \lstinline{k}-sublists of an \lstinline{n}-list \lstinline{xs}, \lstinline{cd} should rearrange and duplicate them into the appropriate positions for the \lstinline{(k+1)}-sublists, where in the position for a particular \lstinline{(k+1)}-sublist, the result should be the list of all its immediate sublists as computed by\csp\lstinline{flatten . choose k}.

I could finish what Richard might have done in his paper by deriving the definition of \lstinline{cd} from the specification~(\cref{eq:cd-spec}).
Maybe I could switch to \ensuremath{\Conid{B}\;\unskip^{\Var n}_{\Var k}} throughout to make the shapes clear.
But there's going to be a large amount of tedious equational reasoning\ldots
I'm not thrilled by the prospect.

\pause

My editor is still running Agda and showing the shape-indexed version of \ensuremath{\Varid{cd}}, with \ensuremath{\Conid{B}\;\unskip^{\Var n}_{\Var k}} in its type~(\cref{sec:shape}).
The whole point of programming with \emph{inductive families}~\citep{Dybjer-inductive-families} such as \ensuremath{\Conid{B}\;\unskip^{\Var n}_{\Var k}} is to `say more and prove less': encode more properties in the indices, so that those properties are automatically taken care of as programs construct and deconstruct indexed data, requiring fewer manual proofs.
Instead of just the shapes, maybe it's possible to extend \ensuremath{\Conid{B}\;\unskip^{\Var n}_{\Var k}} and encode the \emph{entire} specification~(\cref{eq:cd-spec}) of \lstinline{cd} in its type?

What the specification~(\cref{eq:cd-spec}) says is basically that \lstinline{cd} should transform a tree produced by \lstinline{choose} into another one also produced by \lstinline{choose} and then processed using a \lstinline{mapB}.
I suppose I could force a tree to be the result of \lstinline{mapB h (choose k xs)} for some~\lstinline{h} (which is the common form of the input and output trees of \lstinline{cd}) by adding \lstinline{h} and \lstinline{xs} as indices to \ensuremath{\Conid{B}\;\unskip^{\Var n}_{\Var k}}, and imposing equality constraints on the tree elements:
\begin{hscode}\SaveRestoreHook
\column{B}{@{}>{\hspre}l<{\hspost}@{}}%
\column{3}{@{}>{\hspre}l<{\hspost}@{}}%
\column{5}{@{}>{\hspre}l<{\hspost}@{}}%
\column{11}{@{}>{\hspre}l<{\hspost}@{}}%
\column{36}{@{}>{\hspre}l<{\hspost}@{}}%
\column{61}{@{}>{\hspre}l<{\hspost}@{}}%
\column{72}{@{}>{\hspre}l<{\hspost}@{}}%
\column{78}{@{}>{\hspre}l<{\hspost}@{}}%
\column{86}{@{}>{\hspre}l<{\hspost}@{}}%
\column{E}{@{}>{\hspre}l<{\hspost}@{}}%
\>[3]{}\Keyword{data}\;\Conid{B}^\prime\;\mathbin{:}\;(\Var n\;\Var k\;\mathbin{:}\;\Conid{ℕ})\;(\Var b\;\mathbin{:}\;\Conid{Set})\;\Varid{→}\;(\Conid{Vec}\;\Var k\;\Var a\;\Varid{→}\;\Var b)\;\Varid{→}\;\Conid{Vec}\;\Var n\;\Var a\;\Varid{→}\;\Conid{Set}\;\Keyword{where}{}\<[E]%
\\
\>[3]{}\hsindent{2}{}\<[5]%
\>[5]{}\Con{tip_z}\;{}\<[11]%
\>[11]{}\mathbin{:}\;(\Var y\;\mathbin{:}\;\Var b)\;(\Var e\;\mathbin{:}\;\Var y\;\unskip\equiv\ignorenext\;\Var h\;\Varid{[]}{}\<[36]%
\>[36]{})\;{}\<[61]%
\>[61]{}\Varid{→}\;\Conid{B}^\prime\;\Var n\;{}\<[78]%
\>[78]{}\Con{zero}\;{}\<[86]%
\>[86]{}\Var b\;\Var h\;\Var xs{}\<[E]%
\\
\>[3]{}\hsindent{2}{}\<[5]%
\>[5]{}\Con{tip_s}\;{}\<[11]%
\>[11]{}\mathbin{:}\;(\Var y\;\mathbin{:}\;\Var b)\;(\Var e\;\mathbin{:}\;\Var y\;\unskip\equiv\ignorenext\;\Var h\;\Var xs{}\<[36]%
\>[36]{})\;{}\<[61]%
\>[61]{}\Varid{→}\;\Conid{B}^\prime\;(\Con{suc}\;{}\<[72]%
\>[72]{}\Var k)\;({}\<[78]%
\>[78]{}\Con{suc}\;\Var k)\;{}\<[86]%
\>[86]{}\Var b\;\Var h\;\Var xs{}\<[E]%
\\
\>[3]{}\hsindent{2}{}\<[5]%
\>[5]{}\Con{bin}\;{}\<[11]%
\>[11]{}\mathbin{:}\;\Conid{B}^\prime\;\Var n\;(\Con{suc}\;\Var k)\;\Var b\;\Var h\;\Var xs\;\Varid{→}\;\Conid{B}^\prime\;\Var n\;\Var k\;\Var b\;(\Var h\;\unskip\mathrel\cdot\ignorenext\;(\Var x\;\Varid{∷}\anonymous ))\;\Var xs\;{}\<[61]%
\>[61]{}\Varid{→}\;\Conid{B}^\prime\;(\Con{suc}\;{}\<[72]%
\>[72]{}\Var n)\;({}\<[78]%
\>[78]{}\Con{suc}\;\Var k)\;{}\<[86]%
\>[86]{}\Var b\;\Var h\;(\Var x\;\Varid{∷}\;\Var xs){}\<[E]%
\ColumnHook
\end{hscode}\resethooks
It's a rather complex extension of \ensuremath{\Conid{B}\;\unskip^{\Var n}_{\Var k}}, but I think I'll be fine if I stick to the same programming methodology: perform `pattern matching' on the indices, and specify what should be in the tree in each case.
In the first case, \lstinline{choose} returns \lstinline{Tip []}, so the tree should be a \ensuremath{\Con{tip_z}}, and its element~\ensuremath{\Var y} should be accompanied by a proof~\ensuremath{\Var e} that \ensuremath{\Var y}~equals \ensuremath{\Var h\;\Varid{[]}}, so that \ensuremath{\Con{tip_z}\;\Var y\;\Var e} `is' \lstinline{mapB h (Tip [])}.
The second case is similar.
In the third case, the first thing I do is switch from Richard's snoc pattern\csp\lstinline{xs ++ [x]}\csp to a cons index \ensuremath{\Var x\;\Varid{∷}\;\Var xs} --- this just `reverses' the list and shouldn't break anything, as long as the snoc in\csp\lstinline{mapB (++[x])}\csp is also switched to a cons to match.
The first inductive call of \lstinline{choose} easily translates into the type of the left subtree.
The right subtree should be the result of\csp\lstinline{map h (map (x:) (choose k xs))}.
Luckily, the two maps can be fused into a single\csp\lstinline{map (h . (x:))}.
So the type of the second subtree uses the index \ensuremath{\Var h\;\unskip\mathrel\cdot\ignorenext\;(\Var x\;\Varid{∷}\anonymous )} instead of~\ensuremath{\Var h}.

Even though I sort of derived~\ensuremath{\Conid{B}^\prime} from a specification and know it should work, I still feel an urge to see with my own eyes that \ensuremath{\Conid{B}^\prime}~does work as intended.
So I write a tree with `holes' (missing parts of a program) at the element positions, and let Agda tell me what types the holes should have:
\begin{hscode}\SaveRestoreHook
\column{B}{@{}>{\hspre}l<{\hspost}@{}}%
\column{15}{@{}>{\hspre}l<{\hspost}@{}}%
\column{21}{@{}>{\hspre}l<{\hspost}@{}}%
\column{28}{@{}>{\hspre}l<{\hspost}@{}}%
\column{E}{@{}>{\hspre}l<{\hspost}@{}}%
\>[B]{}\Varid{test}\Conid{B}^\prime\;\mathbin{:}\;\{\mskip1.5mu \Var b\;\mathbin{:}\;\Conid{Set}\mskip1.5mu\}\;\{\mskip1.5mu \Var h\;\mathbin{:}\;\Conid{Vec}\;\mathrm 2\;\Conid{Char}\;\Varid{→}\;\Var b\mskip1.5mu\}\;\Varid{→}\;\Conid{B}^\prime{}^{\Varid{4}}_{\mathrm 2}\;\Var b\;\Var h\;\text{\ttfamily \char34 abcd\char34}{}\<[E]%
\\
\>[B]{}\Varid{test}\Conid{B}^\prime\;\mathrel{=}\;\Con{bin}\;{}\<[15]%
\>[15]{}(\Con{bin}\;{}\<[21]%
\>[21]{}(\Con{tip_s}\;\highlight{goal}{\textbf\{\,\Var b\,\textbf\}_{\kern1pt\textbf{0}}}\;\highlight{goal}{\textbf\{\,\mathrm{?_0}\;\unskip\equiv\ignorenext\;\Var h\;\text{\ttfamily \char34 cd\char34}\,\textbf\}_{\kern1pt\textbf{1}}})\;{}\<[E]%
\\
\>[21]{}(\Con{bin}\;(\Con{tip_s}\;\highlight{goal}{\textbf\{\,\Var b\,\textbf\}_{\kern1pt\textbf{2}}}\;\highlight{goal}{\textbf\{\,\mathrm{?_2}\;\unskip\equiv\ignorenext\;\Var h\;\text{\ttfamily \char34 bd\char34}\,\textbf\}_{\kern1pt\textbf{3}}})\;(\Con{tip_z}\;\highlight{goal}{\textbf\{\,\Var b\,\textbf\}_{\kern1pt\textbf{4}}}\;\highlight{goal}{\textbf\{\,\mathrm{?_4}\;\unskip\equiv\ignorenext\;\Var h\;\text{\ttfamily \char34 bc\char34}\,\textbf\}_{\kern1pt\textbf{5}}})))\;{}\<[E]%
\\
\>[15]{}(\Con{bin}\;{}\<[21]%
\>[21]{}(\Con{bin}\;(\Con{tip_s}\;\highlight{goal}{\textbf\{\,\Var b\,\textbf\}_{\kern1pt\textbf{6}}}\;\highlight{goal}{\textbf\{\,\mathrm{?_6}\;\unskip\equiv\ignorenext\;\Var h\;\text{\ttfamily \char34 ad\char34}\,\textbf\}_{\kern1pt\textbf{7}}})\;(\Con{tip_z}\;\highlight{goal}{\textbf\{\,\Var b\,\textbf\}_{\kern1pt\textbf{8}}}\;\highlight{goal}{\textbf\{\,\mathrm{?_8}\;\unskip\equiv\ignorenext\;\Var h\;\text{\ttfamily \char34 ac\char34}\,\textbf\}_{\kern1pt\textbf{9}}}))\;{}\<[E]%
\\
\>[21]{}(\Con{tip_z}\;{}\<[28]%
\>[28]{}\highlight{goal}{\textbf\{\,\Var b\,\textbf\}_{\kern1pt\textbf{10}}}\;\highlight{goal}{\textbf\{\,\mathrm{?_{10}}\;\unskip\equiv\ignorenext\;\Var h\;\text{\ttfamily \char34 ab\char34}\,\textbf\}_{\kern1pt\textbf{11}}})){}\<[E]%
\ColumnHook
\end{hscode}\resethooks
The goal types of the even-numbered holes are all~\ensuremath{\Var b}, and the odd-numbered holes require proofs that the even-numbered holes are equal to \ensuremath{\Var h\;\Var zs} for all the \ensuremath{\mathrm 2}-sublists \ensuremath{\Var zs} of \ensuremath{\text{\ttfamily \char34 abcd\char34}}.
It works!

\pause

\ensuremath{\Conid{B}^\prime}~doesn't look bad, but I can't help raising an eyebrow.
With yet more effort, I suppose I could refine the type of \ensuremath{\Varid{cd}} to use~\ensuremath{\Conid{B}^\prime} and encode the full specification~(\cref{eq:cd-spec}).
But the refined \ensuremath{\Varid{cd}} would need to manipulate the equality proofs in those trees, and maybe eventually I'd still be doing essentially the same tedious equational reasoning that I wanted to avoid.

Another problem is that the upgraded \ensuremath{\Varid{cd}} would only work on trees of sublists, whereas the original \lstinline{cd} in Haskell works on trees of \emph{any} type of values.
Indeed, the specification~(\cref{eq:cd-spec}) talks about the behaviour of \lstinline{cd} on trees of sublists only.
By encoding the specification in the type, I'd actually restrict \ensuremath{\Varid{cd}} to trees of sublists.
That doesn't sound too useful.

Still, I can't take my eyes off the definition of~\ensuremath{\Conid{B}^\prime}.
The way it absorbs the definition of \lstinline{choose} looks right.
If only the elements weren't restricted to pairs of the form \ensuremath{\Var y\;\mathbin{:}\;\Var b} and \ensuremath{\Var e\;\mathbin{:}\;\Var y\;\unskip\equiv\ignorenext\;\Var h\;\Var zs}\ldots

\medskip
A lightbulb lights up above my head.
\begin{hscode}\SaveRestoreHook
\column{B}{@{}>{\hspre}l<{\hspost}@{}}%
\column{3}{@{}>{\hspre}l<{\hspost}@{}}%
\column{9}{@{}>{\hspre}l<{\hspost}@{}}%
\column{55}{@{}>{\hspre}l<{\hspost}@{}}%
\column{66}{@{}>{\hspre}l<{\hspost}@{}}%
\column{72}{@{}>{\hspre}l<{\hspost}@{}}%
\column{80}{@{}>{\hspre}l<{\hspost}@{}}%
\column{E}{@{}>{\hspre}l<{\hspost}@{}}%
\>[B]{}\Keyword{data}\;\Conid{BT}\;\mathbin{:}\;(\Var n\;\Var k\;\mathbin{:}\;\Conid{ℕ})\;\Varid{→}\;(\Conid{Vec}\;\Var k\;\Var a\;\Varid{→}\;\Conid{Set})\;\Varid{→}\;\Conid{Vec}\;\Var n\;\Var a\;\Varid{→}\;\Conid{Set}\;\Keyword{where}{}\<[E]%
\\
\>[B]{}\hsindent{3}{}\<[3]%
\>[3]{}\Con{tip_z}\;{}\<[9]%
\>[9]{}\mathbin{:}\;\Var p\;\Varid{[]}\;{}\<[55]%
\>[55]{}\Varid{→}\;\Conid{BT}\;\Var n\;{}\<[72]%
\>[72]{}\Con{zero}\;{}\<[80]%
\>[80]{}\Var p\;\Var xs{}\<[E]%
\\
\>[B]{}\hsindent{3}{}\<[3]%
\>[3]{}\Con{tip_s}\;{}\<[9]%
\>[9]{}\mathbin{:}\;\Var p\;\Var xs\;{}\<[55]%
\>[55]{}\Varid{→}\;\Conid{BT}\;(\Con{suc}\;{}\<[66]%
\>[66]{}\Var k)\;({}\<[72]%
\>[72]{}\Con{suc}\;\Var k)\;{}\<[80]%
\>[80]{}\Var p\;\Var xs{}\<[E]%
\\
\>[B]{}\hsindent{3}{}\<[3]%
\>[3]{}\Con{bin}\;{}\<[9]%
\>[9]{}\mathbin{:}\;\Conid{BT}\;\Var n\;(\Con{suc}\;\Var k)\;\Var p\;\Var xs\;\Varid{→}\;\Conid{BT}\;\Var n\;\Var k\;(\Var p\;\unskip\mathrel\cdot\ignorenext\;(\Var x\;\Varid{∷}\anonymous ))\;\Var xs\;{}\<[55]%
\>[55]{}\Varid{→}\;\Conid{BT}\;(\Con{suc}\;{}\<[66]%
\>[66]{}\Var n)\;({}\<[72]%
\>[72]{}\Con{suc}\;\Var k)\;{}\<[80]%
\>[80]{}\Var p\;(\Var x\;\Varid{∷}\;\Var xs){}\<[E]%
\ColumnHook
\end{hscode}\resethooks
Just generalise the element type!
More specifically, generalise that to a \emph{type family} \ensuremath{\Var p\;\mathbin{:}\;\Conid{Vec}\;\Var k\;\Var a\;\Varid{→}\;\Conid{Set}} indexed by \ensuremath{\Var k}-sublists. Then \ensuremath{\Conid{B}\;\unskip^{\Var n}_{\Var k}\;\Var a} becomes a special case by specialising~\ensuremath{\Var p} to \ensuremath{\Varid{const}\;\Var a} (and supplying any \ensuremath{\Var n}-list as the last index), and similarly \ensuremath{\Conid{B}^\prime{}^{\Var n}_{\Var k}\;\Var b\;\Var h\;\Var xs} by specialising \ensuremath{\Var p} to \ensuremath{\Varid{λ}\;\Var zs\;\Varid{→}\;\Conid{Σ}[\kern-2pt\;\Var y\;\unskip\in\ignorenext\;\Var b\;\kern-2pt]\;\Var y\;\unskip\equiv\ignorenext\;\Var h\;\Var zs}!

I wasn't expecting this generalisation.
After taking a moment to calm down, I look more closely at this new, unifying data type.
The index~\ensuremath{\Var p} in \ensuremath{\Conid{BT}} replaces~\ensuremath{\Var h} in~\ensuremath{\Conid{B}^\prime}, and is similarly applied to all the sublists of a particular length.
What's different is that \ensuremath{\Var p}~is applied to a sublist to get the whole element type in a tip.
So, in general, all the elements in a tree of type \ensuremath{\Conid{BT}^{\Var n}_{\Var k}\;\Var p\;\Var xs} have \emph{distinct} types, which are \ensuremath{\Var p\;\Var ys} for all the \ensuremath{\Var k}-sublists \ensuremath{\Var ys} of the \ensuremath{\Var n}-list \ensuremath{\Var xs}.
To see an example:
\begin{hscode}\SaveRestoreHook
\column{B}{@{}>{\hspre}l<{\hspost}@{}}%
\column{15}{@{}>{\hspre}l<{\hspost}@{}}%
\column{E}{@{}>{\hspre}l<{\hspost}@{}}%
\>[B]{}\Varid{testBT}\;\mathbin{:}\;\{\mskip1.5mu \Var p\;\mathbin{:}\;\Conid{Vec}\;\mathrm 2\;\Conid{Char}\;\Varid{→}\;\Conid{Set}\mskip1.5mu\}\;\Varid{→}\;\Conid{BT}^{\Varid{4}}_{\mathrm 2}\;\Var p\;\text{\ttfamily \char34 abcd\char34}{}\<[E]%
\\
\>[B]{}\Varid{testBT}\;\mathrel{=}\;\Con{bin}\;{}\<[15]%
\>[15]{}(\Con{bin}\;(\Con{tip_s}\;\highlight{goal}{\textbf\{\,\Var p\;\text{\ttfamily \char34 cd\char34}\,\textbf\}_{\kern1pt\textbf{0}}})\;(\Con{bin}\;(\Con{tip_s}\;\highlight{goal}{\textbf\{\,\Var p\;\text{\ttfamily \char34 bd\char34}\,\textbf\}_{\kern1pt\textbf{1}}})\;(\Con{tip_z}\;\highlight{goal}{\textbf\{\,\Var p\;\text{\ttfamily \char34 bc\char34}\,\textbf\}_{\kern1pt\textbf{2}}})))\;{}\<[E]%
\\
\>[15]{}(\Con{bin}\;(\Con{bin}\;(\Con{tip_s}\;\highlight{goal}{\textbf\{\,\Var p\;\text{\ttfamily \char34 ad\char34}\,\textbf\}_{\kern1pt\textbf{3}}})\;(\Con{tip_z}\;\highlight{goal}{\textbf\{\,\Var p\;\text{\ttfamily \char34 ac\char34}\,\textbf\}_{\kern1pt\textbf{4}}}))\;(\Con{tip_z}\;\highlight{goal}{\textbf\{\,\Var p\;\text{\ttfamily \char34 ab\char34}\,\textbf\}_{\kern1pt\textbf{5}}})){}\<[E]%
\ColumnHook
\end{hscode}\resethooks
It's simpler to think of a tree of type \ensuremath{\Conid{BT}^{\Var n}_{\Var k}\;\Var p\;\Var xs} as a table with all the \ensuremath{\Var k}-sublists of~\ensuremath{\Var xs} as the keys. For each key~\ensuremath{\Var ys}, there's an entry of type \ensuremath{\Var p\;\Var ys}.
(So the `T' in \ensuremath{\Conid{BT}} stands for both `tree' and `table'.)

This \ensuremath{\Conid{BT}} definition is really intriguing\ldots
Most likely, there is a way to derive \ensuremath{\Conid{BT}} from \lstinline{choose}, and that'll work for a whole class of functions.
The type of \ensuremath{\Conid{BT}^{\Var n}_{\Var k}\;\mathbin{:}\;(\Var p\;\mathbin{:}\;\Conid{Vec}\;\Var k\;\Var a\;\Varid{→}\;\Conid{Set})\;\Varid{→}\;\Conid{Vec}\;\Var n\;\Var a\;\Varid{→}\;\Conid{Set}} looks just like a continuation-passing version of some \ensuremath{\Conid{Vec}\;\Var n\;\Var a\;\unskip\leadsto\ignorenext\;\Conid{Vec}\;\Var k\;\Var a}, which would be the type of a version of \lstinline{choose} that nondeterministically returns a \ensuremath{\Var k}-sublist of an \ensuremath{\Var n}-list.
And the index~\ensuremath{\Var p} works like a continuation too.
Take the (expanded) type of \ensuremath{\Con{bin}} for example:
\begin{hscode}\SaveRestoreHook
\column{B}{@{}>{\hspre}l<{\hspost}@{}}%
\column{6}{@{}>{\hspre}l<{\hspost}@{}}%
\column{10}{@{}>{\hspre}l<{\hspost}@{}}%
\column{21}{@{}>{\hspre}l<{\hspost}@{}}%
\column{25}{@{}>{\hspre}l<{\hspost}@{}}%
\column{32}{@{}>{\hspre}l<{\hspost}@{}}%
\column{47}{@{}>{\hspre}l<{\hspost}@{}}%
\column{51}{@{}>{\hspre}l<{\hspost}@{}}%
\column{E}{@{}>{\hspre}l<{\hspost}@{}}%
\>[B]{}\Con{bin}\;{}\<[6]%
\>[6]{}\mathbin{:}\;{}\<[10]%
\>[10]{}\Conid{BT}\;\Var n\;(\Con{suc}\;{}\<[21]%
\>[21]{}\Var k)\;{}\<[25]%
\>[25]{}(\Varid{λ}\;\Var ys\;{}\<[32]%
\>[32]{}\Varid{→}\;\Var p\;\Var ys)\;{}\<[47]%
\>[47]{}\Var xs\;{}\<[E]%
\\
\>[6]{}\kern-.345em\mathrlap{\to}\;{}\<[10]%
\>[10]{}\Conid{BT}\;\Var n\;{}\<[21]%
\>[21]{}\Var k\;{}\<[25]%
\>[25]{}(\Varid{λ}\;\Var zs\;{}\<[32]%
\>[32]{}\Varid{→}\;\Var p\;(\Var x\;\Varid{∷}\;\Var zs))\;{}\<[47]%
\>[47]{}\Var xs\;{}\<[51]%
\>[51]{}\Varid{→}\;\Conid{BT}\;(\Con{suc}\;\Var n)\;(\Con{suc}\;\Var k)\;\Var p\;(\Var x\;\Varid{∷}\;\Var xs){}\<[E]%
\ColumnHook
\end{hscode}\resethooks
I can read it as `to compute a \ensuremath{(\mathrm 1\;\Varid{+}\;\Var k)}-sublist of \ensuremath{\Var x\;\Varid{∷}\;\Var xs} and pass it to continuation~\ensuremath{\Var p}, either compute a \ensuremath{(\mathrm 1\;\Varid{+}\;\Var k)}-sublist \ensuremath{\Var ys} of \ensuremath{\Var xs} and pass \ensuremath{\Var ys} to \ensuremath{\Var p} directly, or compute a \ensuremath{\Var k}-sublist \ensuremath{\Var zs} of \ensuremath{\Var xs} and pass \ensuremath{\Var x\;\Varid{∷}\;\Var zs} to~\ensuremath{\Var p}'.
All the results from \ensuremath{\Var p} are then collected in a tree as the indices of the element types.
A simpler and familiar example (which can be found in the Agda standard library) is
\begin{hscode}\SaveRestoreHook
\column{B}{@{}>{\hspre}l<{\hspost}@{}}%
\column{3}{@{}>{\hspre}l<{\hspost}@{}}%
\column{8}{@{}>{\hspre}l<{\hspost}@{}}%
\column{28}{@{}>{\hspre}l<{\hspost}@{}}%
\column{E}{@{}>{\hspre}l<{\hspost}@{}}%
\>[B]{}\Keyword{data}\;\Conid{All}\;\mathbin{:}\;(\Var a\;\Varid{→}\;\Conid{Set})\;\Varid{→}\;\Conid{Vec}\;\Var n\;\Var a\;\Varid{→}\;\Conid{Set}\;\Keyword{where}{}\<[E]%
\\
\>[B]{}\hsindent{3}{}\<[3]%
\>[3]{}\Varid{[]}\;{}\<[8]%
\>[8]{}\mathbin{:}\;{}\<[28]%
\>[28]{}\Conid{All}\;\Var p\;\Varid{[]}{}\<[E]%
\\
\>[B]{}\hsindent{3}{}\<[3]%
\>[3]{}\anonymous \kern.5pt\Varid{∷}\anonymous \;{}\<[8]%
\>[8]{}\mathbin{:}\;\Var p\;\Var x\;\Varid{→}\;\Conid{All}\;\Var p\;\Var xs\;\Varid{→}\;{}\<[28]%
\>[28]{}\Conid{All}\;\Var p\;(\Var x\;\Varid{∷}\;\Var xs){}\<[E]%
\ColumnHook
\end{hscode}\resethooks
This should be derivable from the function that nondeterministically returns an element of a list.
I'm onto something general --- maybe it's interesting enough for
a research
paper!

\subsection{Specifications as Types}
\label{sec:spec}

That paper will have to wait though.
I've still got a problem to solve: how do I use \ensuremath{\Conid{BT}} to specify \lstinline{cd}?

What's special about \ensuremath{\Conid{BT}} is that the element types are indexed by sublists, so I know from the type of an element which sublist it is associated with.
That is, I can now directly say `values associated with sublists' and rearrange these values, rather than indirectly specify the rearrangement in terms of sublists and then extend to other types of values through parametricity.
\ensuremath{\Conid{BT}^{\Var n}_{\Var k}\;\Var p\;\Var xs} is the type of a tree of \ensuremath{\Var p}-typed values associated with the \ensuremath{\Var k}-sublists of \ensuremath{\Var xs}, and that's precisely the intended meaning of \lstinline{cd}'s input.
What about the output?
It should tabulate the \ensuremath{(\mathrm 1\;\Varid{+}\;\Var k)}-sublists of \ensuremath{\Var xs}, so the type should be \ensuremath{\Conid{BT}^{\Var n}_{\mathrm 1\Varid{+}\Var k}\;\Var q\;\Var xs} for some \ensuremath{\Var q\;\mathbin{:}\;\Conid{Vec}\;(\mathrm 1\;\Varid{+}\;\Var k)\;\Var a\;\Varid{→}\;\Conid{Set}}.
For each sublist \ensuremath{\Var ys\;\mathbin{:}\;\Conid{Vec}\;(\mathrm 1\;\Varid{+}\;\Var k)\;\Var a}, I~want a list of \ensuremath{\Var p}-typed values associated with the immediate sublists of \ensuremath{\Var ys}, which are \ensuremath{\Var k}-sublists.
Or, instead of a list, I can just use a tree of type \ensuremath{\Conid{BT}^{\mathrm 1\Varid{+}\Var k}_{\Var k}\;\Var p\;\Var ys}.
Therefore the whole type is
\begin{hscode}\SaveRestoreHook
\column{B}{@{}>{\hspre}l<{\hspost}@{}}%
\column{E}{@{}>{\hspre}l<{\hspost}@{}}%
\>[B]{}\Varid{retabulate}\;\mathbin{:}\;\highlight{suppressed}{\Var k\;\unskip<\ignorenext\;\Var n}\;\Varid{→}\;\Conid{BT}^{\Var n}_{\Var k}\;\Var p\;\Var xs\;\Varid{→}\;\Conid{BT}^{\Var n}_{\mathrm 1\Varid{+}\Var k}\;(\Conid{BT}^{\mathrm 1\Varid{+}\Var k}_{\Var k}\;\Var p)\;\Var xs{}\<[E]%
\ColumnHook
\end{hscode}\resethooks
which is parametric in~\ensuremath{\Var p} (so, like the Haskell version of \lstinline{cd}, the elements can have any types).
I think it's time to rename \lstinline{cd} to something more meaningful, and decide to use `\ensuremath{\Varid{retabulate}}' because I'm moving values in a table into appropriate positions in a new (nested) table with a new tabulation scheme.
And a side condition \ensuremath{\Var k\;\unskip<\ignorenext\;\Var n} is needed to guarantee that the output shape \ensuremath{\Conid{BT}^{\Var n}_{\mathrm 1\Varid{+}\Var k}} is valid.

The type of \ensuremath{\Varid{retabulate}} looks like a sensible refinement of the type of \lstinline{cd}, except that I'm letting \ensuremath{\Varid{retabulate}} return a tree of trees, rather than a tree of lists. Could that change be too drastic?
Hm\ldots actually, no --- the shape of \ensuremath{\Conid{BT}^{\mathrm 1\Varid{+}\Var k}_{\Var k}} is always a (nonempty) list!
If \ensuremath{\Var k}~is \ensuremath{\Con{zero}}, a \ensuremath{\Conid{BT}^{\mathrm 1}_{\mathrm 0}}-tree has to be a \ensuremath{\Con{tip_z}}.
Otherwise, a \ensuremath{\Conid{BT}^{\mathrm 2\Varid{+}\Var k}_{\mathrm 1\Varid{+}\Var k}}-tree has to take the form \ensuremath{\Con{bin}\;(\Con{tip_s}\;\Var y)\;\Var t}.
This expression is in fact a cons-like operation:
\begin{hscode}\SaveRestoreHook
\column{B}{@{}>{\hspre}l<{\hspost}@{}}%
\column{E}{@{}>{\hspre}l<{\hspost}@{}}%
\>[B]{}\anonymous \kern.5pt\Varid{∷}_{\Conid{BT}}\anonymous \;\mathbin{:}\;\Var p\;\Var xs\;\Varid{→}\;\Conid{BT}\;\unskip^{\mathrm 1\Varid{+}\Var k}_{\Var k}\;(\Var p\;\unskip\mathrel\cdot\ignorenext\;(\Var x\;\Varid{∷}\anonymous ))\;\Var xs\;\Varid{→}\;\Conid{BT}\;\unskip^{\mathrm 2\Varid{+}\Var k}_{\mathrm 1\Varid{+}\Var k}\;\Var p\;(\Var x\;\Varid{∷}\;\Var xs){}\<[E]%
\\
\>[B]{}\Var y\;\Varid{∷}_{\Conid{BT}}\;\Var t\;\mathrel{=}\;\Con{bin}\;(\Con{tip_s}\;\Var y)\;\Var t{}\<[E]%
\ColumnHook
\end{hscode}\resethooks
To construct a table indexed by all the immediate sublists of \ensuremath{\Var x\;\Varid{∷}\;\Var xs}, I need an entry for \ensuremath{\Var xs}, the immediate sublist without~\ensuremath{\Var x}, and a table of entries for all the other immediate sublists, which are \ensuremath{\Var x\;\Varid{∷}\;\Var ws} for all the immediate sublists \ensuremath{\Var ws} of \ensuremath{\Var xs}.

I go on to define \ensuremath{\Varid{retabulate}}.
Its type is much more informative than that of \lstinline{cd}.
Rather than transcribing \lstinline{cd}, can I use the type of \ensuremath{\Varid{retabulate}} to guide me through the implementation?
I type the left-hand side into the editor, leave the right-hand side as a hole, and perform some case splitting on the input tree:
\begin{hscode}\SaveRestoreHook
\column{B}{@{}>{\hspre}l<{\hspost}@{}}%
\column{22}{@{}>{\hspre}l<{\hspost}@{}}%
\column{32}{@{}>{\hspre}l<{\hspost}@{}}%
\column{35}{@{}>{\hspre}l<{\hspost}@{}}%
\column{40}{@{}>{\hspre}l<{\hspost}@{}}%
\column{51}{@{}>{\hspre}l<{\hspost}@{}}%
\column{54}{@{}>{\hspre}l<{\hspost}@{}}%
\column{E}{@{}>{\hspre}l<{\hspost}@{}}%
\>[B]{}\Varid{retabulate}\;(\Con{tip_z}\;\anonymous )\;{}\<[54]%
\>[54]{}\mathrel{=}\;\highlight{goal}{\textbf\{\,~~\,\textbf\}_{\kern1pt\textbf{0}}}{}\<[E]%
\\
\>[B]{}\Varid{retabulate}\;(\Con{tip_s}\;\anonymous )\;{}\<[54]%
\>[54]{}\mathrel{=}\;\highlight{goal}{\textbf\{\,~~\,\textbf\}_{\kern1pt\textbf{1}}}{}\<[E]%
\\
\>[B]{}\Varid{retabulate}\;(\Con{bin}\;{}\<[22]%
\>[22]{}(\Con{tip_s}\;\anonymous )\;{}\<[35]%
\>[35]{}\anonymous {}\<[51]%
\>[51]{})\;{}\<[54]%
\>[54]{}\mathrel{=}\;\highlight{goal}{\textbf\{\,~~\,\textbf\}_{\kern1pt\textbf{2}}}{}\<[E]%
\\
\>[B]{}\Varid{retabulate}\;(\Con{bin}\;{}\<[22]%
\>[22]{}(\Con{bin}\;\anonymous \;\anonymous {}\<[32]%
\>[32]{})\;{}\<[40]%
\>[40]{}(\Con{tip_z}\;\anonymous ){}\<[51]%
\>[51]{})\;{}\<[54]%
\>[54]{}\mathrel{=}\;\highlight{goal}{\textbf\{\,~~\,\textbf\}_{\kern1pt\textbf{3}}}{}\<[E]%
\\
\>[B]{}\Varid{retabulate}\;(\Con{bin}\;{}\<[22]%
\>[22]{}(\Con{bin}\;\anonymous \;\anonymous {}\<[32]%
\>[32]{})\;{}\<[40]%
\>[40]{}(\Con{tip_s}\;\anonymous ){}\<[51]%
\>[51]{})\;{}\<[54]%
\>[54]{}\mathrel{=}\;\highlight{goal}{\textbf\{\,~~\,\textbf\}_{\kern1pt\textbf{4}}}{}\<[E]%
\\
\>[B]{}\Varid{retabulate}\;(\Con{bin}\;\Var t\;\unskip@\ignorenext\;{}\<[22]%
\>[22]{}(\Con{bin}\;\anonymous \;\anonymous {}\<[32]%
\>[32]{})\;{}\<[35]%
\>[35]{}\Var u\;\unskip@\ignorenext\;{}\<[40]%
\>[40]{}(\Con{bin}\;\anonymous \;\anonymous ){}\<[51]%
\>[51]{})\;{}\<[54]%
\>[54]{}\mathrel{=}\;\highlight{goal}{\textbf\{\,\Conid{BT}^{\mathrm 2\Varid{+}\Var n}_{\mathrm 3\Varid{+}\Var k}\;(\Conid{BT}^{\mathrm 3\Varid{+}\Var k}_{\mathrm 2\Varid{+}\Var k}\;\Var p\;(\Var x\;\Varid{∷}\;{\Var x}_{1}\;\Varid{∷}\;\Var xs))\,\textbf\}_{\kern1pt\textbf{5}}}{}\<[E]%
\ColumnHook
\end{hscode}\resethooks
The cases for \ensuremath{\Con{tip_s}\;\anonymous } and \ensuremath{\Con{bin}\;\anonymous \;(\Con{tip_s}\;\anonymous )} can be eliminated immediately since the side condition \ensuremath{\Var k\;\unskip<\ignorenext\;\Var n} is violated.
I go straight to the last and the most difficult case, \ensuremath{\Con{bin}\;\Var t\;\Var u}, where \ensuremath{\Var t}~and~\ensuremath{\Var u} are both constructed by \ensuremath{\Con{bin}}.
Their types are
\begin{hscode}\SaveRestoreHook
\column{B}{@{}>{\hspre}l<{\hspost}@{}}%
\column{4}{@{}>{\hspre}l<{\hspost}@{}}%
\column{E}{@{}>{\hspre}l<{\hspost}@{}}%
\>[B]{}\Var t\;{}\<[4]%
\>[4]{}\mathbin{:}\;\Conid{BT}\;\unskip^{\mathrm 1\Varid{+}\Var n}_{\mathrm 2\Varid{+}\Var k}\;\Var p\;({\Var x}_{1}\;\Varid{∷}\;\Var xs){}\<[E]%
\\
\>[B]{}\Var u\;{}\<[4]%
\>[4]{}\mathbin{:}\;\Conid{BT}\;\unskip^{\mathrm 1\Varid{+}\Var n}_{\mathrm 1\Varid{+}\Var k}\;(\Var p\;\unskip\mathrel\cdot\ignorenext\;(\Var x\;\Varid{∷}\anonymous ))\;({\Var x}_{1}\;\Varid{∷}\;\Var xs){}\<[E]%
\ColumnHook
\end{hscode}\resethooks
Neither of them have the right shape to be used immediately, so in \ensuremath{\highlight{goal}{\textbf\{\,~~\,\textbf\}_{\kern1pt\textbf{5}}}} I try starting with a \ensuremath{\Con{bin}}:
\begin{hscode}\SaveRestoreHook
\column{B}{@{}>{\hspre}l<{\hspost}@{}}%
\column{53}{@{}>{\hspre}l<{\hspost}@{}}%
\column{E}{@{}>{\hspre}l<{\hspost}@{}}%
\>[B]{}\Varid{retabulate}\;(\Con{bin}\;\Var t\;\unskip@\ignorenext\;(\Con{bin}\;\anonymous \;\anonymous )\;\Var u\;\unskip@\ignorenext\;(\Con{bin}\;\anonymous \;\anonymous ))\;\mathrel{=}\;\Con{bin}\;{}\<[53]%
\>[53]{}\highlight{goal}{\textbf\{\,\Conid{BT}\;\unskip^{\mathrm 1\Varid{+}\Var n}_{\mathrm 3\Varid{+}\Var k}\;(\Conid{BT}\;\unskip^{\mathrm 3\Varid{+}\Var k}_{\mathrm 2\Varid{+}\Var k}\;\Var p)\;({\Var x}_{1}\;\Varid{∷}\;\Var xs)\,\textbf\}_{\kern1pt\textbf{6}}}\;{}\<[E]%
\\
\>[53]{}\highlight{goal}{\textbf\{\,\Conid{BT}\;\unskip^{\mathrm 1\Varid{+}\Var n}_{\mathrm 2\Varid{+}\Var k}\;(\Conid{BT}\;\unskip^{\mathrm 3\Varid{+}\Var k}_{\mathrm 2\Varid{+}\Var k}\;\Var p\;\unskip\mathrel\cdot\ignorenext\;(\Var x\;\Varid{∷}\anonymous ))\;({\Var x}_{1}\;\Varid{∷}\;\Var xs)\,\textbf\}_{\kern1pt\textbf{7}}}{}\<[E]%
\ColumnHook
\end{hscode}\resethooks
\ensuremath{\highlight{goal}{\textbf\{\,~~\,\textbf\}_{\kern1pt\textbf{6}}}} is directly fulfilled by the induction hypothesis \ensuremath{\Varid{retabulate}\;\Var t}!
That's a good sign.

What can I put in \ensuremath{\highlight{goal}{\textbf\{\,~~\,\textbf\}_{\kern1pt\textbf{7}}}}?
Prompted by the success with \ensuremath{\highlight{goal}{\textbf\{\,~~\,\textbf\}_{\kern1pt\textbf{6}}}}, I try the induction hypothesis for~\ensuremath{\Var u}:
\begin{hscode}\SaveRestoreHook
\column{B}{@{}>{\hspre}l<{\hspost}@{}}%
\column{E}{@{}>{\hspre}l<{\hspost}@{}}%
\>[B]{}\Varid{retabulate}\;\Var u\;\mathbin{:}\;\Conid{BT}\;\unskip^{\mathrm 1\Varid{+}\Var n}_{\mathrm 2\Varid{+}\Var k}\;(\Conid{BT}\;\unskip^{\mathrm 2\Varid{+}\Var k}_{\mathrm 1\Varid{+}\Var k}\;(\Var p\;\unskip\mathrel\cdot\ignorenext\;(\Var x\;\Varid{∷}\anonymous )))\;({\Var x}_{1}\;\Varid{∷}\;\Var xs){}\<[E]%
\ColumnHook
\end{hscode}\resethooks
Hm.
Its type has the same outer shape \ensuremath{\Conid{BT}\;\unskip^{\mathrm 1\Varid{+}\Var n}_{\mathrm 2\Varid{+}\Var k}\;\anonymous \;({\Var x}_{1}\;\Varid{∷}\;\Var xs)} that I want in the goal type, but the element types don't match\ldots

To fill out \ensuremath{\highlight{goal}{\textbf\{\,~~\,\textbf\}_{\kern1pt\textbf{7}}}} I really have to pay more attention to what the types say.
I stare at the types, and, slowly, they start to make sense.
\begin{itemize}[leftmargin=*]
\item The outer shape \ensuremath{\Conid{BT}\;\unskip^{\mathrm 1\Varid{+}\Var n}_{\mathrm 2\Varid{+}\Var k}\;\anonymous \;({\Var x}_{1}\;\Varid{∷}\;\Var xs)} tells me that I'm dealing with tables indexed by the \ensuremath{(\mathrm 2\;\Varid{+}\;\Var k)}-sublists \ensuremath{\Var zs} of \ensuremath{{\Var x}_{1}\;\Varid{∷}\;\Var xs}.
\item For each \ensuremath{\Var zs}, I need to construct a table of \ensuremath{\Var p}-typed entries indexed by all the immediate sublists of \ensuremath{\Var x\;\Varid{∷}\;\Var zs}, because that's what the element types \ensuremath{\Conid{BT}\;\unskip^{\mathrm 3\Varid{+}\Var k}_{\mathrm 2\Varid{+}\Var k}\;\Var p\;\unskip\mathrel\cdot\ignorenext\;(\Var x\;\Varid{∷}\anonymous )} in the goal type mean.
\item What do I have in \ensuremath{\Varid{retabulate}\;\Var u}?
I have a table of entries indexed by \ensuremath{\Var x\;\Varid{∷}\;\Var ws} for all the immediate sublists \ensuremath{\Var ws} of \ensuremath{\Var zs} in \ensuremath{\Varid{retabulate}\;\Var u} --- that's what the element types \ensuremath{\Conid{BT}\;\unskip^{\mathrm 2\Varid{+}\Var k}_{\mathrm 1\Varid{+}\Var k}\;(\Var p\;\unskip\mathrel\cdot\ignorenext\;(\Var x\;\Varid{∷}\anonymous ))} mean.
\item What's the relationship between these sublists \ensuremath{\Var x\;\Varid{∷}\;\Var ws} and the table I need to construct, which is indexed by the immediate sublists of \ensuremath{\Var x\;\Varid{∷}\;\Var zs}?
Oh, right, the sublists \ensuremath{\Var x\;\Varid{∷}\;\Var ws} are the immediate sublists of \ensuremath{\Var x\;\Varid{∷}\;\Var zs} with the first element~\ensuremath{\Var x}!
So I've already got most of the entries I need.
\item I'm still missing an entry for the immediate sublist of \ensuremath{\Var x\;\Varid{∷}\;\Var zs} without~\ensuremath{\Var x}, which is \ensuremath{\Var zs}.
Do I have that?
I search through the context.
The type of~\ensuremath{\Var t} draws my attention: it has the familiar outer shape \ensuremath{\Conid{BT}\;\unskip^{\mathrm 1\Varid{+}\Var n}_{\mathrm 2\Varid{+}\Var k}\;\anonymous \;({\Var x}_{1}\;\Varid{∷}\;\Var xs)}.
What entries are in~\ensuremath{\Var t}?
Its type tells me that there's an entry for each \ensuremath{(\mathrm 2\;\Varid{+}\;\Var k)}-sublist \ensuremath{\Var zs} of \ensuremath{{\Var x}_{1}\;\Varid{∷}\;\Var xs}.
That's precisely what I'm missing!
\end{itemize}
So, for each \ensuremath{\Var zs}, I can construct a table indexed by all the immediate sublists of \ensuremath{\Var x\;\Varid{∷}\;\Var zs} by combining an entry for \ensuremath{\Var zs} (the immediate sublist without~\ensuremath{\Var x}) from~\ensuremath{\Var t} with a table of entries for the other immediate sublists with~\ensuremath{\Var x} from \ensuremath{\Varid{retabulate}\;\Var u}.
An entry and a table of entries --- aren't they exactly the arguments of \ensuremath{\anonymous \kern.5pt\Varid{∷}_{\Conid{BT}}\anonymous }\,?
Indeed, I can fulfil \ensuremath{\highlight{goal}{\textbf\{\,~~\,\textbf\}_{\kern1pt\textbf{7}}}} by combining all the corresponding entries in~\ensuremath{\Var t} and \ensuremath{\Varid{retabulate}\;\Var u} (that share the same index \ensuremath{\Var zs}) using \ensuremath{\anonymous \kern.5pt\Varid{∷}_{\Conid{BT}}\anonymous }\,, that is, \ensuremath{\Varid{zip}_{\Conid{BT}}\kern-1pt\Conid{With}\;\anonymous \kern.5pt\Varid{∷}_{\Conid{BT}}\anonymous \;\Var t\;(\Varid{retabulate}\;\Var u)}!

The rest of the cases are much easier, and I come up with a definition of \ensuremath{\Varid{retabulate}},
\begin{hscode}\SaveRestoreHook
\column{B}{@{}>{\hspre}l<{\hspost}@{}}%
\column{18}{@{}>{\hspre}l<{\hspost}@{}}%
\column{23}{@{}>{\hspre}l<{\hspost}@{}}%
\column{30}{@{}>{\hspre}l<{\hspost}@{}}%
\column{34}{@{}>{\hspre}l<{\hspost}@{}}%
\column{39}{@{}>{\hspre}l<{\hspost}@{}}%
\column{50}{@{}>{\hspre}l<{\hspost}@{}}%
\column{53}{@{}>{\hspre}l<{\hspost}@{}}%
\column{61}{@{}>{\hspre}l<{\hspost}@{}}%
\column{E}{@{}>{\hspre}l<{\hspost}@{}}%
\>[B]{}\Varid{retabulate}\;\mathbin{:}\;\highlight{suppressed}{\Var k\;\unskip<\ignorenext\;\Var n}\;\Varid{→}\;\Conid{BT}\;\unskip^{\Var n}_{\Var k}\;\Var p\;\Var xs\;\Varid{→}\;\Conid{BT}\;\unskip^{\Var n}_{\mathrm 1\Varid{+}\Var k}\;(\Conid{BT}\;\unskip^{\mathrm 1\Varid{+}\Var k}_{\Var k}\;\Var p)\;\Var xs{}\<[E]%
\\
\>[B]{}\Varid{retabulate}\;\{\mskip1.5mu \Var xs\;\unskip=\ignorenext\;\anonymous \;\Varid{∷}\;\Varid{[]}{}\<[30]%
\>[30]{}\mskip1.5mu\}\;(\Con{tip_z}\;\Var y)\;{}\<[53]%
\>[53]{}\mathrel{=}\;\Con{tip_s}\;{}\<[61]%
\>[61]{}(\Con{tip_z}\;\Var y){}\<[E]%
\\
\>[B]{}\Varid{retabulate}\;\{\mskip1.5mu \Var xs\;\unskip=\ignorenext\;\anonymous \;\Varid{∷}\;\anonymous \;\Varid{∷}\;\anonymous {}\<[30]%
\>[30]{}\mskip1.5mu\}\;(\Con{tip_z}\;\Var y)\;{}\<[53]%
\>[53]{}\mathrel{=}\;\Con{bin}\;{}\<[61]%
\>[61]{}(\Varid{retabulate}\;(\Con{tip_z}\;\Var y))\;(\Con{tip_z}\;(\Con{tip_z}\;\Var y)){}\<[E]%
\\
\>[B]{}\Varid{retabulate}\;(\Con{bin}\;{}\<[23]%
\>[23]{}(\Con{tip_s}\;\Var y)\;{}\<[34]%
\>[34]{}\Var u{}\<[50]%
\>[50]{})\;{}\<[53]%
\>[53]{}\mathrel{=}\;\Con{tip_s}\;{}\<[61]%
\>[61]{}(\Var y\;\Varid{∷}_{\Conid{BT}}\;\Var u){}\<[E]%
\\
\>[B]{}\Varid{retabulate}\;(\Con{bin}\;{}\<[18]%
\>[18]{}\Var t\;\unskip@\ignorenext\;{}\<[23]%
\>[23]{}(\Con{bin}\;\anonymous \;\anonymous )\;{}\<[39]%
\>[39]{}(\Con{tip_z}\;\Var z){}\<[50]%
\>[50]{})\;{}\<[53]%
\>[53]{}\mathrel{=}\;\Con{bin}\;{}\<[61]%
\>[61]{}(\Varid{retabulate}\;\Var t)\;(\Varid{map}_{\Conid{BT}}\;(\anonymous \kern.5pt\Varid{∷}_{\Conid{BT}}\;(\Con{tip_z}\;\Var z))\;\Var t){}\<[E]%
\\
\>[B]{}\Varid{retabulate}\;(\Con{bin}\;{}\<[18]%
\>[18]{}\Var t\;\unskip@\ignorenext\;{}\<[23]%
\>[23]{}(\Con{bin}\;\anonymous \;\anonymous )\;{}\<[34]%
\>[34]{}\Var u\;\unskip@\ignorenext\;{}\<[39]%
\>[39]{}(\Con{bin}\;\anonymous \;\anonymous ){}\<[50]%
\>[50]{})\;{}\<[53]%
\>[53]{}\mathrel{=}\;\Con{bin}\;{}\<[61]%
\>[61]{}(\Varid{retabulate}\;\Var t)\;{}\<[E]%
\\
\>[61]{}(\Varid{zip}_{\Conid{BT}}\kern-1pt\Conid{With}\;\anonymous \kern.5pt\Varid{∷}_{\Conid{BT}}\anonymous \;\Var t\;(\Varid{retabulate}\;\Var u)){}\<[E]%
\ColumnHook
\end{hscode}\resethooks
where the map and zip functions are the expected ones:
\begin{hscode}\SaveRestoreHook
\column{B}{@{}>{\hspre}l<{\hspost}@{}}%
\column{12}{@{}>{\hspre}l<{\hspost}@{}}%
\column{17}{@{}>{\hspre}l<{\hspost}@{}}%
\column{E}{@{}>{\hspre}l<{\hspost}@{}}%
\>[B]{}\Varid{map}_{\Conid{BT}}\;{}\<[12]%
\>[12]{}\mathbin{:}\;({}\<[17]%
\>[17]{}\Varid{∀}\;\{\mskip1.5mu \Var ys\mskip1.5mu\}\;\Varid{→}\;\Var p\;\Var ys\;\Varid{→}\;\Var q\;\Var ys)\;\Varid{→}\;\Varid{∀}\;\{\mskip1.5mu \Var xs\mskip1.5mu\}\;\Varid{→}\;\Conid{BT}^{\Var n}_{\Var k}\;\Var p\;\Var xs\;\Varid{→}\;\Conid{BT}^{\Var n}_{\Var k}\;\Var q\;\Var xs{}\<[E]%
\\
\>[B]{}\Varid{zip}_{\Conid{BT}}\kern-1pt\Conid{With}\;{}\<[12]%
\>[12]{}\mathbin{:}\;({}\<[17]%
\>[17]{}\Varid{∀}\;\{\mskip1.5mu \Var ys\mskip1.5mu\}\;\Varid{→}\;\Var p\;\Var ys\;\Varid{→}\;\Var q\;\Var ys\;\Varid{→}\;\Var r\;\Var ys)\;{}\<[E]%
\\
\>[12]{}\kern-.345em\mathrlap{\to}\;{}\<[17]%
\>[17]{}\Varid{∀}\;\{\mskip1.5mu \Var xs\mskip1.5mu\}\;\Varid{→}\;\Conid{BT}^{\Var n}_{\Var k}\;\Var p\;\Var xs\;\Varid{→}\;\Conid{BT}^{\Var n}_{\Var k}\;\Var q\;\Var xs\;\Varid{→}\;\Conid{BT}^{\Var n}_{\Var k}\;\Var r\;\Var xs{}\<[E]%
\ColumnHook
\end{hscode}\resethooks


Amazingly, the more informative type of \ensuremath{\Varid{retabulate}} did help me to develop its definition.
As I filled in the holes, I didn't feel I had much of a choice --- in a good way, because that reflected the precision of the type.
Furthermore, I discovered a new presentation that varies slightly from Richard's \lstinline{cd}!
The first two cases of \lstinline{cd} are subsumed by the third case, \ensuremath{\Con{bin}\;(\Con{tip_s}\;\Var y)\;\Var u}, of \ensuremath{\Varid{retabulate}}.
The second case of \lstinline{cd} recursively traverses the given tree to convert it to a list, which is not needed in \ensuremath{\Varid{retabulate}}, because it yields a tree of trees.
Therefore the first two cases of \lstinline{cd} can be unified into one.
I forgot to include the side condition \ensuremath{\mathrm 1\;\unskip\le\ignorenext\;\Var k} in \ensuremath{\Varid{retabulate}}, but that leads to two new \ensuremath{\Con{tip_z}} cases instead of preventing me from completing the definition.
Whereas \lstinline{cd} has to start from level~$1$ of the sublist lattice~(\cref{fig:sublists-lattice}), this pair of cases of \ensuremath{\Varid{retabulate}} is capable of producing a level-$1$ table (with as many elements as \ensuremath{\Var xs}) from a level-$0$ table, which is a \ensuremath{\Con{tip_z}}.
This is due to \ensuremath{\Var xs} now being available as an index, providing the missing context for \ensuremath{\Varid{retabulate}}.


In fact, looking at the type more closely, I suspect that the extensional behaviour of \ensuremath{\Varid{retabulate}} is completely determined by the type (so the type works as a nice and tight specification): the shape of the output table is completely determined by the indices; moreover, all the input elements have distinct types in general, so each element in the output table has to be the only input element with the required type --- there is no choice to make.
Formally, the proof will most likely be based on parametricity (and might be similar to \varcitet{Voigtlander-BX-for-free}{'s}).
That'll probably be a fun exercise\ldots but I'll leave that for another day.

\subsection{Precision of Types}
\label{sec:td-and-bu-in-Agda}

Right now I'm more eager to understand why the bottom-up algorithm \lstinline{bu} equals the top-down \lstinline{td}~(\cref{sec:algorithms,sec:bu}).
Will dependent types continue to be helpful?
I should try and find out by transcribing the two algorithms into Agda too.

The first thing to do is make the type of the two algorithms as precise as the type of \ensuremath{\Varid{retabulate}}.
I~start from the combining function~\lstinline{g}.
Its type \lstinline{L s -> s}
has been making me shudder involuntarily whenever I see it: it says so little about what \lstinline{g}~should do.
The intention ---~that \lstinline{g}~should compute the solution for a list from those of its immediate sublists~--- is nowhere to be seen.

But now I have the right vocabulary to state the intention precisely in Agda.
I can use \ensuremath{\Conid{BT}} to quantify over all sublists of a particular length.
And to say `a solution for a list' (instead of just `a solution') I should switch from a type to a type family
\begin{hscode}\SaveRestoreHook
\column{B}{@{}>{\hspre}l<{\hspost}@{}}%
\column{E}{@{}>{\hspre}l<{\hspost}@{}}%
\>[B]{}\Var s\;\mathbin{:}\;\Varid{∀}\;\{\mskip1.5mu \Var k\mskip1.5mu\}\;\Varid{→}\;\Conid{Vec}\;\Var k\;\Var a\;\Varid{→}\;\Conid{Set}{}\<[E]%
\ColumnHook
\end{hscode}\resethooks
such that \ensuremath{\Var s\;\Var ys} is the type of solutions for \ensuremath{\Var ys}.
So
\begin{temp}
\begin{hscode}\SaveRestoreHook
\column{B}{@{}>{\hspre}l<{\hspost}@{}}%
\column{E}{@{}>{\hspre}l<{\hspost}@{}}%
\>[B]{}\Var g\;\mathbin{:}\;\Varid{∀}\;\{\mskip1.5mu \Var k\mskip1.5mu\}\;\Varid{→}\;\{\mskip1.5mu \Var ys\;\mathbin{:}\;\Conid{Vec}\;(\mathrm 2\;\Varid{+}\;\Var k)\;\Var a\mskip1.5mu\}\;\Varid{→}\;\Conid{BT}\;\unskip^{\mathrm 2\Varid{+}\Var k}_{\mathrm 1\Varid{+}\Var k}\;\Var s\;\Var ys\;\Varid{→}\;\Var s\;\Var ys{}\<[E]%
\ColumnHook
\end{hscode}\resethooks
\end{temp}
I look at the type with a satisfied smile --- now that's what I call a nice and informative type!
It says concisely and precisely what \ensuremath{\Var g}~should do: compute a solution for any \ensuremath{\Var ys\;\mathbin{:}\;\Conid{Vec}\;(\mathrm 2\;\Varid{+}\;\Var k)\;\Var a} from a table of solutions for all the length-\ensuremath{(\mathrm 1\;\Varid{+}\;\Var k)} (that is, immediate) sublists of \ensuremath{\Var ys}.

The smile quickly turns into a frown though.
I still don't feel comfortable with \ensuremath{\Conid{BT}\;\unskip^{\mathrm 2\Varid{+}\Var k}_{\mathrm 1\Varid{+}\Var k}}.
The indices are apparently not the most general ones --- why not \ensuremath{\Conid{BT}\;\unskip^{\mathrm 1\Varid{+}\Var k}_{\Var k}}?

I delete~({\color{red}\ding{56}}) the type of~\ensuremath{\Var g} and start pondering.
I wrote \ensuremath{\Conid{BT}\;\unskip^{\mathrm 2\Varid{+}\Var k}_{\mathrm 1\Varid{+}\Var k}} in that type because that was what Richard wanted to say.
Richard used singleton lists as the base cases instead of the empty list, so \lstinline{g}~was applied to solutions for sublists of length at least~\ensuremath{\mathrm 1}, hence the subscript \ensuremath{\mathrm 1\;\Varid{+}\;\Var k}.
But the most general type
\begin{hscode}\SaveRestoreHook
\column{B}{@{}>{\hspre}l<{\hspost}@{}}%
\column{E}{@{}>{\hspre}l<{\hspost}@{}}%
\>[B]{}\Var g\;\mathbin{:}\;\Varid{∀}\;\{\mskip1.5mu \Var k\mskip1.5mu\}\;\Varid{→}\;\{\mskip1.5mu \Var ys\;\mathbin{:}\;\Conid{Vec}\;(\mathrm 1\;\Varid{+}\;\Var k)\;\Var a\mskip1.5mu\}\;\Varid{→}\;\Conid{BT}\;\unskip^{\mathrm 1\Varid{+}\Var k}_{\Var k}\;\Var s\;\Var ys\;\Varid{→}\;\Var s\;\Var ys{}\<[E]%
\ColumnHook
\end{hscode}\resethooks
looks just fine.
In particular, when \ensuremath{\Var k}~is~\ensuremath{\mathrm 0}, the type says that \ensuremath{\Var g}~should compute a solution for a singleton list from a solution for the empty list (the only immediate sublist of a singleton list), which seems reasonable\ldots

\begin{aha}
\ldots And indeed it is reasonable!
\end{aha}

When I discovered the extra \ensuremath{\Con{tip_z}} cases of \ensuremath{\Varid{retabulate}}~(\cref{sec:spec}), I saw that it may be possible to start from level~0 of the sublist lattice~(\cref{fig:sublists-lattice}) after all.
Now it's confirmed.
Instead of starting with solutions for singleton lists, I can start with a given solution for the empty list
\begin{temp}
\begin{hscode}\SaveRestoreHook
\column{B}{@{}>{\hspre}l<{\hspost}@{}}%
\column{E}{@{}>{\hspre}l<{\hspost}@{}}%
\>[B]{}\Var e\;\mathbin{:}\;\Var s\;\Varid{[]}{}\<[E]%
\ColumnHook
\end{hscode}\resethooks
\end{temp}
at level~0, and work upwards using~\ensuremath{\Var g}.
There's no problem going from level~0 to level~1, because there's now additional context in the indices so that \ensuremath{\Var g}~knows for which singleton list a solution should be computed.
Making types precise has helped me to find a more natural and general form of recursive computation over immediate sublists!

And it's not just any recursive computation --- it's now an alternative \emph{induction principle} for lists.
The base case~\ensuremath{\Var e} is still about the empty list, and the inductive case~\ensuremath{\Var g} assumes that the induction hypothesis holds for all the immediate sublists.
(I~guess I've been instinctively drawn towards induction principles, in common with most dependently typed programmers.)
\begin{temp}
\begin{hscode}\SaveRestoreHook
\column{B}{@{}>{\hspre}l<{\hspost}@{}}%
\column{3}{@{}>{\hspre}l<{\hspost}@{}}%
\column{14}{@{}>{\hspre}c<{\hspost}@{}}%
\column{14E}{@{}l@{}}%
\column{17}{@{}>{\hspre}l<{\hspost}@{}}%
\column{21}{@{}>{\hspre}l<{\hspost}@{}}%
\column{E}{@{}>{\hspre}l<{\hspost}@{}}%
\>[B]{}\Conid{ImmediateSublistInduction}\;\mathbin{:}\;\Conid{Set₁}{}\<[E]%
\\
\>[B]{}\Conid{ImmediateSublistInduction}\;\mathrel{=}\;{}\<[E]%
\\
\>[B]{}\hsindent{3}{}\<[3]%
\>[3]{}\{\mskip1.5mu \Var a\;\mathbin{:}\;\Conid{Set}\mskip1.5mu\}\;{}\<[14]%
\>[14]{}({}\<[14E]%
\>[17]{}\Var s\;{}\<[21]%
\>[21]{}\mathbin{:}\;\Varid{∀}\;\{\mskip1.5mu \Var k\mskip1.5mu\}\;\Varid{→}\;\Conid{Vec}\;\Var k\;\Var a\;\Varid{→}\;\Conid{Set})\;{}\<[E]%
\\
\>[14]{}({}\<[14E]%
\>[17]{}\Var e\;{}\<[21]%
\>[21]{}\mathbin{:}\;\Var s\;\Varid{[]})\;{}\<[E]%
\\
\>[14]{}({}\<[14E]%
\>[17]{}\Var g\;{}\<[21]%
\>[21]{}\mathbin{:}\;\Varid{∀}\;\{\mskip1.5mu \Var k\mskip1.5mu\}\;\Varid{→}\;\{\mskip1.5mu \Var ys\;\mathbin{:}\;\Conid{Vec}\;(\mathrm 1\;\Varid{+}\;\Var k)\;\Var a\mskip1.5mu\}\;\Varid{→}\;\Conid{BT}\;\unskip^{\mathrm 1\Varid{+}\Var k}_{\Var k}\;\Var s\;\Var ys\;\Varid{→}\;\Var s\;\Var ys)\;{}\<[E]%
\\
\>[B]{}\hsindent{3}{}\<[3]%
\>[3]{}\{\mskip1.5mu \Var n\;\mathbin{:}\;\Conid{ℕ}\mskip1.5mu\}\;{}\<[14]%
\>[14]{}({}\<[14E]%
\>[17]{}\Var xs\;{}\<[21]%
\>[21]{}\mathbin{:}\;\Conid{Vec}\;\Var n\;\Var a)\;\Varid{→}\;\Var s\;\Var xs{}\<[E]%
\ColumnHook
\end{hscode}\resethooks
\end{temp}

\pause

\ensuremath{\Conid{ImmediateSublistInduction}} looks very nice, but I'm still worried: will it be possible to transcribe \lstinline{td} and \lstinline{bu} for this type easily, that is, without ugly stuff like type conversions (\ensuremath{\Varid{subst}}, \ensuremath{\Keyword{rewrite}}, etc)?

Sadly, there may never be any theory that tells me quickly whether or not a type admits pretty programs.
The only way to find out is to try writing some.
I start transcribing \lstinline{td}~(\cref{sec:algorithms}).
The only component for which I still don't have a dependently typed version is \lstinline{dc}, which I've generalised to \lstinline{choose}~(\cref{sec:spec}).
Now I can give it a precise type:
\begin{hscode}\SaveRestoreHook
\column{B}{@{}>{\hspre}l<{\hspost}@{}}%
\column{E}{@{}>{\hspre}l<{\hspost}@{}}%
\>[B]{}\Varid{choose}\;\mathbin{:}\;(\Var n\;\Var k\;\mathbin{:}\;\Conid{ℕ})\;\Varid{→}\;\Var k\;\unskip\mathrel{\le_\uparrow}\ignorenext\;\Var n\;\Varid{→}\;(\Var xs\;\mathbin{:}\;\Conid{Vec}\;\Var n\;\Var a)\;\Varid{→}\;\Conid{BT}\;\unskip^{\Var n}_{\Var k}\;\Conid{Exactly}\;\Var xs{}\<[E]%
\ColumnHook
\end{hscode}\resethooks
The key ingredient is the \ensuremath{\Conid{BT}\;\unskip^{\Var n}_{\Var k}} in the result type, which tabulates all the \ensuremath{\Var k}-sublists as the indices of the element types.
I just need to plug in this data type
\begin{hscode}\SaveRestoreHook
\column{B}{@{}>{\hspre}l<{\hspost}@{}}%
\column{3}{@{}>{\hspre}l<{\hspost}@{}}%
\column{E}{@{}>{\hspre}l<{\hspost}@{}}%
\>[B]{}\Keyword{data}\;\Conid{Exactly}\;\mathbin{:}\;\Var a\;\Varid{→}\;\Conid{Set}\;\Keyword{where}{}\<[E]%
\\
\>[B]{}\hsindent{3}{}\<[3]%
\>[3]{}\Con{exactly}\;\mathbin{:}\;(\Var x\;\mathbin{:}\;\Var a)\;\Varid{→}\;\Conid{Exactly}\;\Var x{}\<[E]%
\ColumnHook
\end{hscode}\resethooks
to say that the elements in the resulting table should be exactly the tabulated indices.
In contrast to the Haskell version, I need to make the Agda version of \ensuremath{\Varid{choose}} total by saying explicitly that the length~\ensuremath{\Var n} of the list \ensuremath{\Var xs} is at least~\ensuremath{\Var k}, so that there are enough elements to choose from.
Somewhat notoriously, there are many versions of natural number inequality.
To align with the inductive structure of \lstinline{choose}, I pick the data type \ensuremath{\Var m\;\unskip\mathrel{\le_\uparrow}\ignorenext\;\Var n} where \ensuremath{\Var m}~remains fixed throughout the definition and serves as an `origin', and an inhabitant of \ensuremath{\Var m\;\unskip\mathrel{\le_\uparrow}\ignorenext\;\Var n} is the distance (which is essentially a natural number) that \ensuremath{\Var n}~is away from the origin~\ensuremath{\Var m}:
\begin{hscode}\SaveRestoreHook
\column{B}{@{}>{\hspre}l<{\hspost}@{}}%
\column{3}{@{}>{\hspre}l<{\hspost}@{}}%
\column{9}{@{}>{\hspre}l<{\hspost}@{}}%
\column{21}{@{}>{\hspre}l<{\hspost}@{}}%
\column{E}{@{}>{\hspre}l<{\hspost}@{}}%
\>[B]{}\Keyword{data}\;\anonymous \kern.5pt{\le_\uparrow}\anonymous \;\mathbin{:}\;\Conid{ℕ}\;\Varid{→}\;\Conid{ℕ}\;\Varid{→}\;\Conid{Set}\;\Keyword{where}{}\<[E]%
\\
\>[B]{}\hsindent{3}{}\<[3]%
\>[3]{}\Con{zero}\;{}\<[9]%
\>[9]{}\mathbin{:}\;{}\<[21]%
\>[21]{}\Var m\;\unskip\mathrel{\le_\uparrow}\ignorenext\;\Var m{}\<[E]%
\\
\>[B]{}\hsindent{3}{}\<[3]%
\>[3]{}\Con{suc}\;{}\<[9]%
\>[9]{}\mathbin{:}\;\Var m\;\unskip\mathrel{\le_\uparrow}\ignorenext\;\Var n\;\Varid{→}\;{}\<[21]%
\>[21]{}\Var m\;\unskip\mathrel{\le_\uparrow}\ignorenext\;\Con{suc}\;\Var n{}\<[E]%
\ColumnHook
\end{hscode}\resethooks

Given the precise type, transcribing \lstinline{choose} is pretty much straightforward:
\begin{hscode}\SaveRestoreHook
\column{B}{@{}>{\hspre}l<{\hspost}@{}}%
\column{3}{@{}>{\hspre}l<{\hspost}@{}}%
\column{17}{@{}>{\hspre}c<{\hspost}@{}}%
\column{17E}{@{}l@{}}%
\column{20}{@{}>{\hspre}l<{\hspost}@{}}%
\column{28}{@{}>{\hspre}l<{\hspost}@{}}%
\column{31}{@{}>{\hspre}l<{\hspost}@{}}%
\column{39}{@{}>{\hspre}l<{\hspost}@{}}%
\column{49}{@{}>{\hspre}l<{\hspost}@{}}%
\column{57}{@{}>{\hspre}l<{\hspost}@{}}%
\column{70}{@{}>{\hspre}c<{\hspost}@{}}%
\column{70E}{@{}l@{}}%
\column{E}{@{}>{\hspre}l<{\hspost}@{}}%
\>[B]{}\Varid{choose}\;\mathbin{:}\;(\Var n\;\Var k\;\mathbin{:}\;\Conid{ℕ})\;\Varid{→}\;\Var k\;\unskip\mathrel{\le_\uparrow}\ignorenext\;\Var n\;\Varid{→}\;(\Var xs\;\mathbin{:}\;\Conid{Vec}\;\Var n\;\Var a)\;\Varid{→}\;\Conid{BT}\;\unskip^{\Var n}_{\Var k}\;\Conid{Exactly}\;\Var xs{}\<[E]%
\\
\>[B]{}\Varid{choose}\;\anonymous \;{}\<[20]%
\>[20]{}\Con{zero}\;{}\<[28]%
\>[28]{}\anonymous \;{}\<[39]%
\>[39]{}\anonymous \;{}\<[49]%
\>[49]{}\mathrel{=}\;\Con{tip_z}\;{}\<[57]%
\>[57]{}(\Con{exactly}\;\Varid{[]}{}\<[70]%
\>[70]{}){}\<[70E]%
\\
\>[B]{}\Varid{choose}\;(\Con{suc}\;\Var k)\;{}\<[17]%
\>[17]{}({}\<[17E]%
\>[20]{}\Con{suc}\;\Var k)\;{}\<[31]%
\>[31]{}\Con{zero}\;{}\<[39]%
\>[39]{}\Var xs\;{}\<[49]%
\>[49]{}\mathrel{=}\;\Con{tip_s}\;{}\<[57]%
\>[57]{}(\Con{exactly}\;\Var xs{}\<[70]%
\>[70]{}){}\<[70E]%
\\
\>[B]{}\Varid{choose}\;(\Con{suc}\;\Var n)\;{}\<[17]%
\>[17]{}({}\<[17E]%
\>[20]{}\Con{suc}\;\Var k)\;{}\<[28]%
\>[28]{}({}\<[31]%
\>[31]{}\Con{suc}\;\Var d)\;{}\<[39]%
\>[39]{}(\Var x\;\Varid{∷}\;\Var xs)\;{}\<[49]%
\>[49]{}\mathrel{=}\;{}\<[E]%
\\
\>[B]{}\hsindent{3}{}\<[3]%
\>[3]{}\Con{bin}\;(\Varid{choose}\;\Var n\;(\Con{suc}\;\Var k)\;\Var d\;\Var xs)\;(\Varid{map}_{\Conid{BT}}\;(\Varid{map}_{\Conid{Exactly}}\;(\Var x\;\Varid{∷}\anonymous ))\;(\Varid{choose}\;\Var n\;\Var k\;(\Varid{incr}\;\Var d)\;\Var xs)){}\<[E]%
\\
\>[B]{}\hsindent{3}{}\<[3]%
\>[3]{}\mbox{\onelinecomment  \hspace{.4em}\ensuremath{\Varid{map}_{\Conid{Exactly}}\;\mathbin{:}\;(\Var f\;\mathbin{:}\;\Var a\;\Varid{→}\;\Var b)\;\Varid{→}\;\{\mskip1.5mu \Var x\;\mathbin{:}\;\Var a\mskip1.5mu\}\;\Varid{→}\;\Conid{Exactly}\;\Var x\;\Varid{→}\;\Conid{Exactly}\;(\Var f\;\Var x)}}{}\<[E]%
\ColumnHook
\end{hscode}\resethooks
The function performs an induction on~\ensuremath{\Var k}, like the Haskell version.
The second and third cases are an inner induction on the distance that \ensuremath{\Var n}~is away from \ensuremath{\Con{suc}\;\Var k}.
In the second case the distance is \ensuremath{\Con{zero}}, meaning that \ensuremath{\Var n}~is (at the origin) \ensuremath{\Con{suc}\;\Var k}, so \ensuremath{\Var xs} has the right length and can be directly returned.
Otherwise the distance is \ensuremath{\Con{suc}\;\Var d} in the third case, where the first inductive call is on~\ensuremath{\Var d}, and the second inductive call is on~\ensuremath{\Var k} while the distance is unchanged, but the type \ensuremath{\Con{suc}\;\Var k\;\unskip\mathrel{\le_\uparrow}\ignorenext\;\Con{suc}\;\Var n} needs to be adjusted to \ensuremath{\Var k\;\unskip\mathrel{\le_\uparrow}\ignorenext\;\Var n} (by invoking \ensuremath{\Varid{incr}\;\mathbin{:}\;\Con{suc}\;\Var m\;\unskip\mathrel{\le_\uparrow}\ignorenext\;\Var n\;\Varid{→}\;\Var m\;\unskip\mathrel{\le_\uparrow}\ignorenext\;\Var n} on~\ensuremath{\Var d}).

\pause

I step back and take another look at \ensuremath{\Varid{choose}}.
There's one thing that bothers me: \ensuremath{\Conid{Exactly}\;\Var x} is a type that has a unique inhabitant, so I could've used the unit type~\ensuremath{\Varid{⊤}} as the element types instead, and I'd still give the same amount of information, which is none!
That doesn't make a lot of sense --- I thought I was computing all the \ensuremath{\Var k}-sublists and returning them in a table, but somehow those sublists didn't really matter, and I could just return a blank table of type \ensuremath{\Conid{BT}\;\unskip^{\Var n}_{\Var k}\;(\Varid{const}\;\Varid{⊤})\;\Var xs}\ldots?

\begin{aha}
Hold on, it's actually making sense\ldots
\end{aha}

It's because all the information is already in the table structure of \ensuremath{\Conid{BT}}.
Indeed, I can write
\begin{hscode}\SaveRestoreHook
\column{B}{@{}>{\hspre}l<{\hspost}@{}}%
\column{E}{@{}>{\hspre}l<{\hspost}@{}}%
\>[B]{}\Varid{map}_{\Conid{BT}}\;(\Varid{λ}\;\{\mskip1.5mu \{\mskip1.5mu \Var ys\mskip1.5mu\}\;\Con{tt}\;\Varid{→}\;\Con{exactly}\;\Var ys\mskip1.5mu\})\;\mathbin{:}\;\Conid{BT}\;\unskip^{\Var n}_{\Var k}\;(\Varid{const}\;\Varid{⊤})\;\Var xs\;\Varid{→}\;\Conid{BT}\;\unskip^{\Var n}_{\Var k}\;\Conid{Exactly}\;\Var xs{}\<[E]%
\ColumnHook
\end{hscode}\resethooks
to recover a table of \ensuremath{\Conid{Exactly}}s from just a blank table by replacing every \ensuremath{\Con{tt}} (the unique inhabitant of~\ensuremath{\Varid{⊤}}) with the index \ensuremath{\Var ys} there.
What \ensuremath{\Varid{choose}} does is not really compute the sublists --- \ensuremath{\Conid{BT}} has already `computed' them.
Instead, \ensuremath{\Varid{choose}} merely affirms that there is a table indexed by all the \ensuremath{\Var k}-sublists of an \ensuremath{\Var n}-list whenever \ensuremath{\Var k\;\unskip\mathrel{\le_\uparrow}\ignorenext\;\Var n}.
The elements in the table don't matter, and might as well be \ensuremath{\Con{tt}}.

So, instead of \ensuremath{\Varid{choose}}, I can use
\begin{temp}
\begin{hscode}\SaveRestoreHook
\column{B}{@{}>{\hspre}l<{\hspost}@{}}%
\column{18}{@{}>{\hspre}l<{\hspost}@{}}%
\column{26}{@{}>{\hspre}l<{\hspost}@{}}%
\column{29}{@{}>{\hspre}l<{\hspost}@{}}%
\column{45}{@{}>{\hspre}l<{\hspost}@{}}%
\column{E}{@{}>{\hspre}l<{\hspost}@{}}%
\>[B]{}\Varid{blank}\;\mathbin{:}\;(\Var n\;\Var k\;\mathbin{:}\;\Conid{ℕ})\;\Varid{→}\;\Var k\;\unskip\mathrel{\le_\uparrow}\ignorenext\;\Var n\;\Varid{→}\;\{\mskip1.5mu \Var xs\;\mathbin{:}\;\Conid{Vec}\;\Var n\;\Var a\mskip1.5mu\}\;\Varid{→}\;\Conid{BT}\;\unskip^{\Var n}_{\Var k}\;(\Varid{const}\;\Varid{⊤})\;\Var xs{}\<[E]%
\\
\>[B]{}\Varid{blank}\;\anonymous \;{}\<[18]%
\>[18]{}\Con{zero}\;{}\<[26]%
\>[26]{}\anonymous \;{}\<[45]%
\>[45]{}\mathrel{=}\;\Con{tip_z}\;\Con{tt}{}\<[E]%
\\
\>[B]{}\Varid{blank}\;(\Con{suc}\;\Var k)\;({}\<[18]%
\>[18]{}\Con{suc}\;\Var k)\;{}\<[29]%
\>[29]{}\Con{zero}\;{}\<[45]%
\>[45]{}\mathrel{=}\;\Con{tip_s}\;\Con{tt}{}\<[E]%
\\
\>[B]{}\Varid{blank}\;(\Con{suc}\;\Var n)\;({}\<[18]%
\>[18]{}\Con{suc}\;\Var k)\;{}\<[26]%
\>[26]{}({}\<[29]%
\>[29]{}\Con{suc}\;\Var d)\;\{\mskip1.5mu \anonymous \;\Varid{∷}\;\anonymous \mskip1.5mu\}\;{}\<[45]%
\>[45]{}\mathrel{=}\;\Con{bin}\;(\Varid{blank}\;\Var n\;(\Con{suc}\;\Var k)\;\Var d)\;(\Varid{blank}\;\Var n\;\Var k\;(\Varid{incr}\;\Var d)){}\<[E]%
\ColumnHook
\end{hscode}\resethooks
\end{temp}
to construct a blank table indexed by all the immediate sublists in the inductive case of \ensuremath{\Varid{td}}, where I'll then compute and fill in solutions for all the immediate sublists by invoking \ensuremath{\Varid{td}} inductively, and finally invoke~\ensuremath{\Var g} to combine all those solutions.
And the base case simply returns~\ensuremath{\Var e}.
\begin{temp}
\begin{hscode}\SaveRestoreHook
\column{B}{@{}>{\hspre}l<{\hspost}@{}}%
\column{18}{@{}>{\hspre}l<{\hspost}@{}}%
\column{24}{@{}>{\hspre}l<{\hspost}@{}}%
\column{E}{@{}>{\hspre}l<{\hspost}@{}}%
\>[B]{}\Varid{td}\;\mathbin{:}\;\Conid{ImmediateSublistInduction}{}\<[E]%
\\
\>[B]{}\Varid{td}\;\Var s\;\Var e\;\Var g\;\{\mskip1.5mu \Con{zero}{}\<[18]%
\>[18]{}\mskip1.5mu\}\;\Varid{[]}\;{}\<[24]%
\>[24]{}\mathrel{=}\;\Var e{}\<[E]%
\\
\>[B]{}\Varid{td}\;\Var s\;\Var e\;\Var g\;\{\mskip1.5mu \Con{suc}\;\Var n{}\<[18]%
\>[18]{}\mskip1.5mu\}\;\Var xs\;{}\<[24]%
\>[24]{}\mathrel{=}\;\Var g\;(\Varid{map}_{\Conid{BT}}\;(\Varid{λ}\;\{\mskip1.5mu \{\mskip1.5mu \Var ys\mskip1.5mu\}\;\Con{tt}\;\Varid{→}\;\Varid{td}\;\Var s\;\Var e\;\Var g\;\{\mskip1.5mu \Var n\mskip1.5mu\}\;\Var ys\mskip1.5mu\})\;(\Varid{blank}\;\unskip^{\mathrm 1\Varid{+}\Var n}_{\Var n}\;(\Con{suc}\;\Con{zero}))){}\<[E]%
\ColumnHook
\end{hscode}\resethooks
\end{temp}

I look aghast at the monster I've created.
Sure, the definition type-checks, but oh my\ldots it's terribly ugly.
The problems are cosmetic though, and should be easy to fix.
\begin{inlineenum}
\item The induction is on~\ensuremath{\Var n}, which shouldn't have been an implicit argument.
\item In the base case, I have to match \ensuremath{\Var xs} with~\ensuremath{\Varid{[]}} to make it type-correct to return~\ensuremath{\Var e}, but that could've been avoided if I set \ensuremath{\Var e\;\mathbin{:}\;\{\mskip1.5mu \Var xs\;\mathbin{:}\;\Conid{Vec}\;\mathrm 0\;\Var a\mskip1.5mu\}\;\Varid{→}\;\Var s\;\Var xs}.
\item In the inductive case, \ensuremath{\Var xs} doesn't need to be explicit because it's passed around only implicitly in the indices on the right-hand side.
\item I can get rid of the~\ensuremath{\Varid{λ}} around the inductive call to \ensuremath{\Varid{td}} if I make \ensuremath{\Var ys} implicit and add a dummy \ensuremath{\Varid{⊤}}~argument.
\end{inlineenum}
In fact, if I add a dummy \ensuremath{\Varid{⊤}}~argument to \ensuremath{\Var e}~and \ensuremath{\Varid{blank}} as well, I can make the definition point-free like in Richard's paper --- a temptation I cannot resist.
So I revise the induction principle,
\begin{hscode}\SaveRestoreHook
\column{B}{@{}>{\hspre}l<{\hspost}@{}}%
\column{3}{@{}>{\hspre}l<{\hspost}@{}}%
\column{14}{@{}>{\hspre}c<{\hspost}@{}}%
\column{14E}{@{}l@{}}%
\column{17}{@{}>{\hspre}l<{\hspost}@{}}%
\column{20}{@{}>{\hspre}l<{\hspost}@{}}%
\column{E}{@{}>{\hspre}l<{\hspost}@{}}%
\>[B]{}\Conid{ImmediateSublistInduction}\;\mathbin{:}\;\Conid{Set₁}{}\<[E]%
\\
\>[B]{}\Conid{ImmediateSublistInduction}\;\mathrel{=}\;{}\<[E]%
\\
\>[B]{}\hsindent{3}{}\<[3]%
\>[3]{}\{\mskip1.5mu \Var a\;\mathbin{:}\;\Conid{Set}\mskip1.5mu\}\;{}\<[14]%
\>[14]{}({}\<[14E]%
\>[17]{}\Var s\;{}\<[20]%
\>[20]{}\mathbin{:}\;\Varid{∀}\;\{\mskip1.5mu \Var k\mskip1.5mu\}\;\Varid{→}\;\Conid{Vec}\;\Var k\;\Var a\;\Varid{→}\;\Conid{Set})\;{}\<[E]%
\\
\>[14]{}({}\<[14E]%
\>[17]{}\Var e\;{}\<[20]%
\>[20]{}\mathbin{:}\;\{\mskip1.5mu \Var ys\;\mathbin{:}\;\Conid{Vec}\;\mathrm 0\;\Var a\mskip1.5mu\}\;\Varid{→}\;\Varid{⊤}\;\Varid{→}\;\Var s\;\Var ys)\;{}\<[E]%
\\
\>[14]{}({}\<[14E]%
\>[17]{}\Var g\;{}\<[20]%
\>[20]{}\mathbin{:}\;\Varid{∀}\;\{\mskip1.5mu \Var k\mskip1.5mu\}\;\Varid{→}\;\{\mskip1.5mu \Var ys\;\mathbin{:}\;\Conid{Vec}\;(\mathrm 1\;\Varid{+}\;\Var k)\;\Var a\mskip1.5mu\}\;\Varid{→}\;\Conid{BT}\;\unskip^{\mathrm 1\Varid{+}\Var k}_{\Var k}\;\Var s\;\Var ys\;\Varid{→}\;\Var s\;\Var ys)\;{}\<[E]%
\\
\>[14]{}({}\<[14E]%
\>[17]{}\Var n\;{}\<[20]%
\>[20]{}\mathbin{:}\;\Conid{ℕ})\;\{\mskip1.5mu \Var xs\;\mathbin{:}\;\Conid{Vec}\;\Var n\;\Var a\mskip1.5mu\}\;\Varid{→}\;\Varid{⊤}\;\Varid{→}\;\Var s\;\Var xs{}\<[E]%
\ColumnHook
\end{hscode}\resethooks
add the dummy \ensuremath{\Varid{⊤}}~argument to \ensuremath{\Varid{blank}} (also ignoring the inequality argument from now on),
\begin{hscode}\SaveRestoreHook
\column{B}{@{}>{\hspre}l<{\hspost}@{}}%
\column{E}{@{}>{\hspre}l<{\hspost}@{}}%
\>[B]{}\Varid{blank}\;\mathbin{:}\;(\Var n\;\Var k\;\mathbin{:}\;\Conid{ℕ})\;\Varid{→}\;\highlight{suppressed}{\Var k\;\unskip\mathrel{\le_\uparrow}\ignorenext\;\Var n}\;\Varid{→}\;\{\mskip1.5mu \Var xs\;\mathbin{:}\;\Conid{Vec}\;\Var n\;\Var a\mskip1.5mu\}\;\Varid{→}\;\Varid{⊤}\;\Varid{→}\;\Conid{BT}\;\unskip^{\Var n}_{\Var k}\;(\Varid{const}\;\Varid{⊤})\;\Var xs{}\<[E]%
\ColumnHook
\end{hscode}\resethooks
and get my point-free \ensuremath{\Varid{td}}:
\begin{hscode}\SaveRestoreHook
\column{B}{@{}>{\hspre}l<{\hspost}@{}}%
\column{13}{@{}>{\hspre}l<{\hspost}@{}}%
\column{21}{@{}>{\hspre}l<{\hspost}@{}}%
\column{E}{@{}>{\hspre}l<{\hspost}@{}}%
\>[B]{}\Varid{td}\;\mathbin{:}\;\Conid{ImmediateSublistInduction}{}\<[E]%
\\
\>[B]{}\Varid{td}\;\Var s\;\Var e\;\Var g\;{}\<[13]%
\>[13]{}\Con{zero}\;{}\<[21]%
\>[21]{}\mathrel{=}\;\Var e{}\<[E]%
\\
\>[B]{}\Varid{td}\;\Var s\;\Var e\;\Var g\;({}\<[13]%
\>[13]{}\Con{suc}\;\Var n)\;{}\<[21]%
\>[21]{}\mathrel{=}\;\Var g\;\unskip\mathrel\cdot\ignorenext\;\Varid{map}_{\Conid{BT}}\;(\Varid{td}\;\Var s\;\Var e\;\Var g\;\Var n)\;\unskip\mathrel\cdot\ignorenext\;\Varid{blank}\;\unskip^{\mathrm 1\Varid{+}\Var n}_{\Var n}{}\<[E]%
\ColumnHook
\end{hscode}\resethooks
The revised \ensuremath{\Conid{ImmediateSublistInduction}} may not be too user-friendly, but that can be amended later (when there's actually a user).

\pause

And it'd be wonderful if the revised \ensuremath{\Conid{ImmediateSublistInduction}} worked for \ensuremath{\Varid{bu}} too!
I proceed to transcribe \lstinline{bu}~(\cref{sec:bu}):
\begin{hscode}\SaveRestoreHook
\column{B}{@{}>{\hspre}l<{\hspost}@{}}%
\column{3}{@{}>{\hspre}l<{\hspost}@{}}%
\column{10}{@{}>{\hspre}l<{\hspost}@{}}%
\column{20}{@{}>{\hspre}l<{\hspost}@{}}%
\column{28}{@{}>{\hspre}l<{\hspost}@{}}%
\column{E}{@{}>{\hspre}l<{\hspost}@{}}%
\>[B]{}\Varid{bu}\;\mathbin{:}\;\Conid{ImmediateSublistInduction}{}\<[E]%
\\
\>[B]{}\Varid{bu}\;\Var s\;\Var e\;\Var g\;\Var n\;\mathrel{=}\;\Varid{unTip}\;\unskip\mathrel\cdot\ignorenext\;\Varid{loop}\;\mathrm 0\;\highlight{goal}{\textbf\{\,\mathrm 0\;\unskip\mathrel{_\downarrow\kern-.8pt{\le}}\ignorenext\;\Var n\,\textbf\}_{\kern1pt\textbf{0}}}\;\unskip\mathrel\cdot\ignorenext\;\Varid{map}_{\Conid{BT}}\;\Var e\;\unskip\mathrel\cdot\ignorenext\;\Varid{blank}\;\unskip^{\Var n}_{\mathrm 0}{}\<[E]%
\\
\>[B]{}\hsindent{3}{}\<[3]%
\>[3]{}\Keyword{where}\;{}\<[10]%
\>[10]{}\Varid{loop}\;\mathbin{:}\;(\Var k\;\mathbin{:}\;\Conid{ℕ})\;\Varid{→}\;\Var k\;\unskip\mathrel{_\downarrow\kern-.8pt{\le}}\ignorenext\;\Var n\;\Varid{→}\;\Conid{BT}\;\unskip^{\Var n}_{\Var k}\;\Var s\;\Var xs\;\Varid{→}\;\Conid{BT}\;\unskip^{\Var n}_{\Var n}\;\Var s\;\Var xs{}\<[E]%
\\
\>[10]{}\Varid{loop}\;\Var n\;{}\<[20]%
\>[20]{}\Con{zero}\;{}\<[28]%
\>[28]{}\mathrel{=}\;\Varid{id}{}\<[E]%
\\
\>[10]{}\Varid{loop}\;\Var k\;({}\<[20]%
\>[20]{}\Con{suc}\;\Var d)\;{}\<[28]%
\>[28]{}\mathrel{=}\;\Varid{loop}\;(\mathrm 1\;\Varid{+}\;\Var k)\;\Var d\;\unskip\mathrel\cdot\ignorenext\;\Varid{map}_{\Conid{BT}}\;\Var g\;\unskip\mathrel\cdot\ignorenext\;\Varid{retabulate}{}\<[E]%
\ColumnHook
\end{hscode}\resethooks
I construct the initial table by reusing \ensuremath{\Varid{blank}} to create a blank level-\ensuremath{\mathrm 0} table and using \ensuremath{\Varid{map}_{\Conid{BT}}\;\Var e} to fill in the initial solution for the empty list.
Then the \ensuremath{\Varid{loop}} function increases the level to~\ensuremath{\Var n} by repeatedly retabulating a level-\ensuremath{\Var k} table as a level-\ensuremath{(\mathrm 1\;\Varid{+}\;\Var k)} table and filling in solutions for all the length-\ensuremath{(\mathrm 1\;\Varid{+}\;\Var k)} sublists using \ensuremath{\Varid{map}_{\Conid{BT}}\;\Var g}.
Finally, a solution for the whole input list is extracted from the level-\ensuremath{\Var n} table using
\begin{hscode}\SaveRestoreHook
\column{B}{@{}>{\hspre}l<{\hspost}@{}}%
\column{19}{@{}>{\hspre}l<{\hspost}@{}}%
\column{26}{@{}>{\hspre}l<{\hspost}@{}}%
\column{E}{@{}>{\hspre}l<{\hspost}@{}}%
\>[B]{}\Varid{unTip}\;\mathbin{:}\;\Conid{BT}\;\unskip^{\Var n}_{\Var n}\;\Var p\;\Var xs\;\Varid{→}\;\Var p\;\Var xs{}\<[E]%
\\
\>[B]{}\Varid{unTip}\;{}\<[19]%
\>[19]{}(\Con{tip_s}\;{}\<[26]%
\>[26]{}\Var p)\;\mathrel{=}\;\Var p{}\<[E]%
\\
\>[B]{}\Varid{unTip}\;\{\mskip1.5mu \Var xs\;\unskip=\ignorenext\;\Varid{[]}\mskip1.5mu\}\;{}\<[19]%
\>[19]{}(\Con{tip_z}\;{}\<[26]%
\>[26]{}\Var p)\;\mathrel{=}\;\Var p{}\<[E]%
\ColumnHook
\end{hscode}\resethooks
The \ensuremath{\Con{bin}} case is impossible and ignored.
(Hm, \ensuremath{\Varid{retabulate}} and \ensuremath{\Varid{unTip}} smell comonadic, and maybe the indices constitute a grading~\citep{grading}\ldots but I can't get distracted now.)

The argument/counter~\ensuremath{\Var k} of \ensuremath{\Varid{loop}} should satisfy the invariant \ensuremath{\Var k\;\unskip\mathrel{_\downarrow\kern-.8pt{\le}}\ignorenext\;\Var n}.
Again, this version of natural number inequality is chosen to align with the inductive structure of \ensuremath{\Varid{loop}}.
The data type \ensuremath{\Var m\;\unskip\mathrel{_\downarrow\kern-.8pt{\le}}\ignorenext\;\Var n} is dual to \ensuremath{\anonymous \kern.5pt{\le_\uparrow}\anonymous } in the sense that this time it is~\ensuremath{\Var n} that is fixed throughout the definition, and \ensuremath{\Var m}~moves away from~\ensuremath{\Var n}:
\begin{hscode}\SaveRestoreHook
\column{B}{@{}>{\hspre}l<{\hspost}@{}}%
\column{3}{@{}>{\hspre}l<{\hspost}@{}}%
\column{9}{@{}>{\hspre}l<{\hspost}@{}}%
\column{25}{@{}>{\hspre}l<{\hspost}@{}}%
\column{28}{@{}>{\hspre}l<{\hspost}@{}}%
\column{E}{@{}>{\hspre}l<{\hspost}@{}}%
\>[B]{}\Keyword{data}\;\anonymous \kern.5pt{_\downarrow\kern-.8pt{\le}}\anonymous \;\mathbin{:}\;\Conid{ℕ}\;\Varid{→}\;\Conid{ℕ}\;\Varid{→}\;\Conid{Set}\;\Keyword{where}{}\<[E]%
\\
\>[B]{}\hsindent{3}{}\<[3]%
\>[3]{}\Con{zero}\;{}\<[9]%
\>[9]{}\mathbin{:}\;{}\<[25]%
\>[25]{}\Var n\;{}\<[28]%
\>[28]{}{_\downarrow\kern-.8pt{\le}}\;\Var n{}\<[E]%
\\
\>[B]{}\hsindent{3}{}\<[3]%
\>[3]{}\Con{suc}\;{}\<[9]%
\>[9]{}\mathbin{:}\;\Con{suc}\;\Var m\;\unskip\mathrel{_\downarrow\kern-.8pt{\le}}\ignorenext\;\Var n\;\Varid{→}\;{}\<[25]%
\>[25]{}\Var m\;{}\<[28]%
\>[28]{}{_\downarrow\kern-.8pt{\le}}\;\Var n{}\<[E]%
\ColumnHook
\end{hscode}\resethooks
Then \ensuremath{\Varid{loop}} simply performs induction on the distance \ensuremath{\Var k\;\unskip\mathrel{_\downarrow\kern-.8pt{\le}}\ignorenext\;\Var n}; the counter~\ensuremath{\Var k} goes up as the distance decreases in inductive calls, and eventually reaches~\ensuremath{\Var n} when the distance becomes \ensuremath{\Con{zero}}.
The remaining \ensuremath{\highlight{goal}{\textbf\{\,\mathrm 0\;\unskip\mathrel{_\downarrow\kern-.8pt{\le}}\ignorenext\;\Var n\,\textbf\}_{\kern1pt\textbf{0}}}} is actually nontrivial, but the Agda standard library covers that.


\subsection{Equality from Types}
\label{sec:equality-from-types}

Okay, I've made the type of both \ensuremath{\Varid{td}} and \ensuremath{\Varid{bu}} precise.
How does this help me prove \ensuremath{\Varid{td}} equals \ensuremath{\Varid{bu}}?
The definitions still look rather different except for their type\ldots

\begin{aha}
\ldots And the type is an induction principle.
\end{aha}

\vspace{-\medskipamount}

\begin{aha}
Is it possible to have extensionally different implementations of an induction principle?
\end{aha}

Let me think about the induction principle of natural numbers.
\begin{hscode}\SaveRestoreHook
\column{B}{@{}>{\hspre}l<{\hspost}@{}}%
\column{16}{@{}>{\hspre}l<{\hspost}@{}}%
\column{17}{@{}>{\hspre}l<{\hspost}@{}}%
\column{25}{@{}>{\hspre}l<{\hspost}@{}}%
\column{E}{@{}>{\hspre}l<{\hspost}@{}}%
\>[B]{}\Conid{ℕ-Induction}\;\mathbin{:}\;\Conid{Set₁}{}\<[E]%
\\
\>[B]{}\Conid{ℕ-Induction}\;\mathrel{=}\;{}\<[16]%
\>[16]{}(\Var p\;\mathbin{:}\;\Conid{ℕ}\;\Varid{→}\;\Conid{Set})\;(\Var{pz}\;\mathbin{:}\;\Var p\;\Con{zero})\;(\Var{ps}\;\mathbin{:}\;\Varid{∀}\;\{\mskip1.5mu \Var m\mskip1.5mu\}\;\Varid{→}\;\Var p\;\Var m\;\Varid{→}\;\Var p\;(\Con{suc}\;\Var m))\;{}\<[E]%
\\
\>[16]{}(\Var n\;\mathbin{:}\;\Conid{ℕ})\;\Varid{→}\;\Var p\;\Var n{}\<[E]%
\\[\blanklineskip]%
\>[B]{}\Varid{ind}\;\mathbin{:}\;\Conid{ℕ-Induction}{}\<[E]%
\\
\>[B]{}\Varid{ind}\;\Var p\;\Var{pz}\;\Var{ps}\;{}\<[17]%
\>[17]{}\Con{zero}\;{}\<[25]%
\>[25]{}\mathrel{=}\;\Var{pz}{}\<[E]%
\\
\>[B]{}\Varid{ind}\;\Var p\;\Var{pz}\;\Var{ps}\;({}\<[17]%
\>[17]{}\Con{suc}\;\Var n)\;{}\<[25]%
\>[25]{}\mathrel{=}\;\Var{ps}\;(\Varid{ind}\;\Var p\;\Var{pz}\;\Var{ps}\;\Var n){}\<[E]%
\ColumnHook
\end{hscode}\resethooks
The motive~\ensuremath{\Var p} is parametrically quantified, so a proof of \ensuremath{\Var p\;\Var n} has to be \ensuremath{\Var n}~applications of \ensuremath{\Var{ps}} to \ensuremath{\Var{pz}}.
There are intensionally different ways to construct that (\ensuremath{\Varid{ind}} versus a tail-recursive implementation, for example), but extensionally they're all the same.

Of course, parametricity is needed to prove that formally.
I look up \varcitet{Bernardy-proofs-for-free}{'s} translation, which (after a bit of simplification) gives the following statement:
\begin{hscode}\SaveRestoreHook
\column{B}{@{}>{\hspre}l<{\hspost}@{}}%
\column{3}{@{}>{\hspre}l<{\hspost}@{}}%
\column{8}{@{}>{\hspre}l<{\hspost}@{}}%
\column{36}{@{}>{\hspre}l<{\hspost}@{}}%
\column{41}{@{}>{\hspre}l<{\hspost}@{}}%
\column{E}{@{}>{\hspre}l<{\hspost}@{}}%
\>[B]{}\Conid{ℕ-Induction-unary-parametricity}\;\mathbin{:}\;\Conid{ℕ-Induction}\;\Varid{→}\;\Conid{Set₁}{}\<[E]%
\\
\>[B]{}\Conid{ℕ-Induction-unary-parametricity}\;\Var f\;\mathrel{=}\;{}\<[E]%
\\
\>[B]{}\hsindent{3}{}\<[3]%
\>[3]{}\{\mskip1.5mu \Var p\;{}\<[8]%
\>[8]{}\mathbin{:}\;\Conid{ℕ}\;\Varid{→}\;\Conid{Set}\mskip1.5mu\}\;{}\<[36]%
\>[36]{}(\Var q\;{}\<[41]%
\>[41]{}\mathbin{:}\;\Varid{∀}\;\{\mskip1.5mu \Var m\mskip1.5mu\}\;\Varid{→}\;\Var p\;\Var m\;\Varid{→}\;\Conid{Set})\;{}\<[E]%
\\
\>[B]{}\hsindent{3}{}\<[3]%
\>[3]{}\{\mskip1.5mu \Var{pz}\;{}\<[8]%
\>[8]{}\mathbin{:}\;\Var p\;\Con{zero}\mskip1.5mu\}\;{}\<[36]%
\>[36]{}(\Var{qz}\;{}\<[41]%
\>[41]{}\mathbin{:}\;\Var q\;\Var{pz})\;{}\<[E]%
\\
\>[B]{}\hsindent{3}{}\<[3]%
\>[3]{}\{\mskip1.5mu \Var{ps}\;{}\<[8]%
\>[8]{}\mathbin{:}\;\Varid{∀}\;\{\mskip1.5mu \Var m\mskip1.5mu\}\;\Varid{→}\;\Var p\;\Var m\;\Varid{→}\;\Var p\;(\Con{suc}\;\Var m)\mskip1.5mu\}\;{}\<[36]%
\>[36]{}(\Var{qs}\;{}\<[41]%
\>[41]{}\mathbin{:}\;\Varid{∀}\;\{\mskip1.5mu \Var m\mskip1.5mu\}\;\{\mskip1.5mu \Var x\;\mathbin{:}\;\Var p\;\Var m\mskip1.5mu\}\;\Varid{→}\;\Var q\;\Var x\;\Varid{→}\;\Var q\;(\Var{ps}\;\Var x))\;{}\<[E]%
\\
\>[B]{}\hsindent{3}{}\<[3]%
\>[3]{}\{\mskip1.5mu \Var n\;{}\<[8]%
\>[8]{}\mathbin{:}\;\Conid{ℕ}\mskip1.5mu\}\;\Varid{→}\;\Var q\;(\Var f\;\Var p\;\Var{pz}\;\Var{ps}\;\Var n){}\<[E]%
\ColumnHook
\end{hscode}\resethooks
Unary parametricity can be thought of as adding an invariant~(\ensuremath{\Var q}) to a parametrically quantified type or type family; this invariant is assumed to hold for any first-order input~(\ensuremath{\Var{qz}}) and be preserved by any higher-order input~(\ensuremath{\Var{qs}}), and is guaranteed to hold for the output.
Now choose the invariant that any proof of \ensuremath{\Var p\;\Var m} is equal to the one produced by \ensuremath{\Varid{ind}\;\Var p\;\Var{pz}\;\Var{ps}\;\Var m}, and that's it:
\begin{hscode}\SaveRestoreHook
\column{B}{@{}>{\hspre}l<{\hspost}@{}}%
\column{3}{@{}>{\hspre}l<{\hspost}@{}}%
\column{6}{@{}>{\hspre}l<{\hspost}@{}}%
\column{E}{@{}>{\hspre}l<{\hspost}@{}}%
\>[B]{}\Conid{ℕ-Induction-uniqueness-from-parametricity}\;\mathbin{:}\;{}\<[E]%
\\
\>[B]{}\hsindent{6}{}\<[6]%
\>[6]{}(\Var f\;\mathbin{:}\;\Conid{ℕ-Induction})\;\Varid{→}\;\Conid{ℕ-Induction-unary-parametricity}\;\Var f\;{}\<[E]%
\\
\>[B]{}\hsindent{3}{}\<[3]%
\>[3]{}\Varid{→}\;{}\<[6]%
\>[6]{}(\Var p\;\mathbin{:}\;\Conid{ℕ}\;\Varid{→}\;\Conid{Set})\;(\Var{pz}\;\mathbin{:}\;\Var p\;\Con{zero})\;(\Var{ps}\;\mathbin{:}\;\Varid{∀}\;\{\mskip1.5mu \Var m\mskip1.5mu\}\;\Varid{→}\;\Var p\;\Var m\;\Varid{→}\;\Var p\;(\Con{suc}\;\Var m))\;(\Var n\;\mathbin{:}\;\Conid{ℕ})\;{}\<[E]%
\\
\>[B]{}\hsindent{3}{}\<[3]%
\>[3]{}\Varid{→}\;{}\<[6]%
\>[6]{}\Var f\;\Var p\;\Var{pz}\;\Var{ps}\;\Var n\;\unskip\equiv\ignorenext\;\Varid{ind}\;\Var p\;\Var{pz}\;\Var{ps}\;\Var n{}\<[E]%
\\
\>[B]{}\Conid{ℕ-Induction-uniqueness-from-parametricity}\;\Var f\;\Var{param}\;\Var p\;\Var{pz}\;\Var{ps}\;\Var n\;\mathrel{=}\;{}\<[E]%
\\
\>[B]{}\hsindent{3}{}\<[3]%
\>[3]{}\Var{param}\;(\Varid{λ}\;\{\mskip1.5mu \Var m\mskip1.5mu\}\;\Var x\;\Varid{→}\;\Var x\;\unskip\equiv\ignorenext\;\Varid{ind}\;\Var p\;\Var{pz}\;\Var{ps}\;\Var m)\;\Con{refl}\;(\Varid{cong}\;\Var{ps}){}\<[E]%
\\
\>[B]{}\hsindent{3}{}\<[3]%
\>[3]{}\mbox{\onelinecomment  \hspace{.3em}\ensuremath{\Con{refl}\;\mathbin{:}\;\{\mskip1.5mu \Var x\;\mathbin{:}\;\Var a\mskip1.5mu\}\;\Varid{→}\;\Var x\;\unskip\equiv\ignorenext\;\Var x};\hspace{.7em}\ensuremath{\Varid{cong}\;\mathbin{:}\;(\Var f\;\mathbin{:}\;\Var a\;\Varid{→}\;\Var b)\;\Varid{→}\;\{\mskip1.5mu \Var x\;\Var y\;\mathbin{:}\;\Var a\mskip1.5mu\}\;\Varid{→}\;\Var x\;\unskip\equiv\ignorenext\;\Var y\;\Varid{→}\;\Var f\;\Var x\;\unskip\equiv\ignorenext\;\Var f\;\Var y}}{}\<[E]%
\ColumnHook
\end{hscode}\resethooks
The same argument works for \ensuremath{\Conid{ImmediateSublistInduction}} --- any function of the type satisfying unary parametricity is pointwise equal to \ensuremath{\Varid{td}}.
I finish the Agda proofs for both induction principles in a dreamlike state.

\begin{aha}
Yeah, I have a proof that \ensuremath{\Varid{td}} equals \ensuremath{\Varid{bu}}.
\end{aha}

Well, strictly speaking I don't have one yet.
(Vanilla) Agda doesn't have internal parametricity~\citep{Van-Muylder-internal-parametricity}, so I'd need to prove the parametricity of \ensuremath{\Varid{bu}}, painfully.
But there shouldn't be any surprise.

Somehow I feel empty though.
I was expecting a more traditional proof based on equational reasoning.
This kind of proof may require more work, but allows me to compare what \ensuremath{\Varid{td}} and \ensuremath{\Varid{bu}} do \emph{intensionally}.
That's an aspect overlooked from the parametricity perspective.
Despite having a proof now, I think I'm going to have to delve into the definitions of \ensuremath{\Varid{td}} and \ensuremath{\Varid{bu}} anyway, to get a clearer picture of their relationship.

\section{Diagrams}


Another hint Richard left was `naturality', a category-theoretic notion which he used a lot in his paper.
In functional programming, naturality usually stems from parametric polymorphism: all parametric functions, such as \lstinline{cd} and \lstinline{unTip}, satisfy naturality.
I've got some parametric functions too, such as \ensuremath{\Varid{retabulate}} and \ensuremath{\Varid{unTip}}.
Their dependent function types with all the indices are more advanced than Richard's types though, and the simply typed form of naturality Richard used no longer makes sense.
But one nice thing about category theory is its adaptability --- all I need to figure out is which category I'm in, and then I'll be able to work out what naturality means for my functions systematically within the world of category theory.

And, if the key is naturality, now I have an additional tool that Richard didn't: string diagrams.
I've seen how dramatically string diagrams simplify proofs using naturality, so it's probably worthwhile to take a look at the two algorithms from a string-diagrammatic perspective.

But before I get to string diagrams, I need to work through some basic category theory\ldots

\subsection{From Categories to String Diagrams}
\label{sec:basic-category-theory}

Functional programmers are familiar with types and functions, and know when functions can be composed sequentially --- when adjacent functions meet at the same type.
And it's possible to compose an empty sequence of functions, in which case the result is the identity function.
\emph{Categories} are settings in which the same intuition about sequential composition works.
Instead of types and functions, the categorical programmer can switch to work with some \emph{objects} and \emph{morphisms} specified by a category, where each morphism is labelled with a source object and a target object (like the source and target types of a function), and morphisms can be composed sequentially when adjacent morphisms meet at the same object.
And, like identity functions, there are identity morphisms too.
Working in a new setting that's identified as a category
is a blessing for the functional programmer: it means that the programmer can still rely on some of their intuitions about types and functions to navigate in the new setting.
More importantly, some notions that prove to be useful in functional programming (such as naturality) can be defined generically on categories and systematically transported to other settings.

A clue about the kind of category I'm in is that I'm tempted to say `\ensuremath{\Varid{retabulate}} transforms a tree of \ensuremath{\Var p}'s to a tree of trees of \ensuremath{\Var p}'s'.
When a simply typed functional programmer says `a tree of something', that `something' is a type, that is, an object in the familiar category of types and functions.
But here \ensuremath{\Var p}~is not a type.
It's a type family.
So I've landed in a different kind of category where the objects are type families.

There are quite a few versions of `categories of families'.
I go through the types of the components used in the algorithms~(\cref{sec:td-and-bu-in-Agda}) to find a common form, and it seems that the simplest version suffices: given an index type \ensuremath{\Var a\;\mathbin{:}\;\Conid{Set}}, a category of families \ensuremath{\text{\textbf{\textsf{Fam}}}\;\Var a} has objects of type
\begin{hscode}\SaveRestoreHook
\column{B}{@{}>{\hspre}l<{\hspost}@{}}%
\column{E}{@{}>{\hspre}l<{\hspost}@{}}%
\>[B]{}\Conid{Fam}\;\mathbin{:}\;\Conid{Set}\;\Varid{→}\;\Conid{Set₁}{}\<[E]%
\\
\>[B]{}\Conid{Fam}\;\Var a\;\mathrel{=}\;\Var a\;\Varid{→}\;\Conid{Set}{}\<[E]%
\ColumnHook
\end{hscode}\resethooks
and morphisms of type
\begin{hscode}\SaveRestoreHook
\column{B}{@{}>{\hspre}l<{\hspost}@{}}%
\column{E}{@{}>{\hspre}l<{\hspost}@{}}%
\>[B]{}\anonymous {\rightrightarrows}\anonymous \;\mathbin{:}\;\Conid{Fam}\;\Var a\;\Varid{→}\;\Conid{Fam}\;\Var a\;\Varid{→}\;\Conid{Set}{}\<[E]%
\\
\>[B]{}\Var p\;\unskip\rightrightarrows\ignorenext\;\Var q\;\mathrel{=}\;\Varid{∀}\;\{\mskip1.5mu \Var x\mskip1.5mu\}\;\Varid{→}\;\Var p\;\Var x\;\Varid{→}\;\Var q\;\Var x{}\<[E]%
\ColumnHook
\end{hscode}\resethooks
That is, a morphism from~\ensuremath{\Var p} to~\ensuremath{\Var q} is a family of functions between corresponding types (with the same index) in \ensuremath{\Var p}~and~\ensuremath{\Var q}.
Everything in the definition is parametrised by the index type~\ensuremath{\Var a}, so actually I'm working in not just one but many related categories of families, with different index types.
These categories are still inherently types and functions, so it's no surprise that their sequential composition works in the way familiar to the functional programmer.

With the definition of \ensuremath{\text{\textbf{\textsf{Fam}}}}, now I can rewrite the parametric function types of \ensuremath{\Varid{retabulate}} and \ensuremath{\Varid{unTip}} to look more like the ones in Haskell:
\begin{hscode}\SaveRestoreHook
\column{B}{@{}>{\hspre}l<{\hspost}@{}}%
\column{13}{@{}>{\hspre}l<{\hspost}@{}}%
\column{16}{@{}>{\hspre}l<{\hspost}@{}}%
\column{30}{@{}>{\hspre}l<{\hspost}@{}}%
\column{E}{@{}>{\hspre}l<{\hspost}@{}}%
\>[B]{}\Varid{retabulate}\;{}\<[13]%
\>[13]{}\mathbin{:}\;{}\<[16]%
\>[16]{}\Conid{BT}\;\unskip^{\Var n}_{\Var k}\;\Var p\;{}\<[30]%
\>[30]{}{\rightrightarrows}\;\Conid{BT}\;\unskip^{\Var n}_{\mathrm 1\Varid{+}\Var k}\;(\Conid{BT}\;\unskip^{\mathrm 1\Varid{+}\Var k}_{\Var k}\;\Var p){}\<[E]%
\\
\>[B]{}\Varid{unTip}\;{}\<[13]%
\>[13]{}\mathbin{:}\;{}\<[16]%
\>[16]{}\Conid{BT}\;\unskip^{\Var n}_{\Var n}\;\Var p\;{}\<[30]%
\>[30]{}{\rightrightarrows}\;\Var p{}\<[E]%
\ColumnHook
\end{hscode}\resethooks
I can fit \ensuremath{\Varid{blank}\;\unskip^{\Var n}_{\Var k}} into \ensuremath{\text{\textbf{\textsf{Fam}}}\;(\Conid{Vec}\;\Var n\;\Var a)} by lifting its \ensuremath{\Varid{⊤}}~argument to a type family \ensuremath{\Varid{const}\;\Varid{⊤}} (that is, an object of the category):
\begin{hscode}\SaveRestoreHook
\column{B}{@{}>{\hspre}l<{\hspost}@{}}%
\column{E}{@{}>{\hspre}l<{\hspost}@{}}%
\>[B]{}\Varid{blank}\;\unskip^{\Var n}_{\Var k}\;\mathbin{:}\;\Varid{const}\;\Varid{⊤}\;\unskip\rightrightarrows\ignorenext\;\Conid{BT}\;\unskip^{\Var n}_{\Var k}\;(\Varid{const}\;\Varid{⊤}){}\<[E]%
\ColumnHook
\end{hscode}\resethooks
The base and inductive cases of \ensuremath{\Conid{ImmediateSublistInduction}} fit into these \ensuremath{\text{\textbf{\textsf{Fam}}}} categories too: given \ensuremath{\Var a\;\mathbin{:}\;\Conid{Set}} and \ensuremath{\Var s\;\mathbin{:}\;\Varid{∀}\;\{\mskip1.5mu \Var k\mskip1.5mu\}\;\Varid{→}\;\Conid{Fam}\;(\Conid{Vec}\;\Var k\;\Var a)}, I can write
\begin{hscode}\SaveRestoreHook
\column{B}{@{}>{\hspre}l<{\hspost}@{}}%
\column{4}{@{}>{\hspre}l<{\hspost}@{}}%
\column{20}{@{}>{\hspre}l<{\hspost}@{}}%
\column{E}{@{}>{\hspre}l<{\hspost}@{}}%
\>[B]{}\Var g\;{}\<[4]%
\>[4]{}\mathbin{:}\;\Conid{BT}\;\unskip^{\mathrm 1\Varid{+}\Var k}_{\Var k}\;\Var s\;{}\<[20]%
\>[20]{}{\rightrightarrows}\;\Var s{}\<[E]%
\\
\>[B]{}\Var e\;{}\<[4]%
\>[4]{}\mathbin{:}\;\Varid{const}\;\Varid{⊤}\;{}\<[20]%
\>[20]{}{\rightrightarrows}\;\Var s\;\{\mskip1.5mu \mathrm 0\mskip1.5mu\}{}\<[E]%
\ColumnHook
\end{hscode}\resethooks
(It's important to say explicitly that the target of~\ensuremath{\Var e} is \ensuremath{\Var s\;\{\mskip1.5mu \mathrm 0\mskip1.5mu\}\;\mathbin{:}\;\Conid{Fam}\;(\Conid{Vec}\;\mathrm 0\;\Var a)} to make it clear that \ensuremath{\Var e}~gives a solution for the empty list.)

\pause

I haven't done much really.
It's just a bit of abstraction that hides part of the indices, and might even be described as cosmetic.
What's important is that, by fitting my programs into the \ensuremath{\text{\textbf{\textsf{Fam}}}} categories, I can start talking about them in categorical language.
In particular, I want to talk about naturality.
That means I should look for \emph{functors} and \emph{natural transformations} in my programs.

A parametric data type such as \ensuremath{\Conid{BT}^{\Var n}_{\Var k}} is categorically the object part of a \emph{functor}, which maps objects in a category to objects in a possibly different category.
In the case of \ensuremath{\Conid{BT}^{\Var n}_{\Var k}}, the functor goes from \ensuremath{\text{\textbf{\textsf{Fam}}}\;(\Conid{Vec}\;\Var k\;\Var a)} to \ensuremath{\text{\textbf{\textsf{Fam}}}\;(\Conid{Vec}\;\Var n\;\Var a)} --- indeed I can rewrite the type of \ensuremath{\Conid{BT}^{\Var n}_{\Var k}} as
\begin{hscode}\SaveRestoreHook
\column{B}{@{}>{\hspre}l<{\hspost}@{}}%
\column{E}{@{}>{\hspre}l<{\hspost}@{}}%
\>[B]{}\Conid{BT}^{\Var n}_{\Var k}\;\mathbin{:}\;\Conid{Fam}\;(\Conid{Vec}\;\Var k\;\Var a)\;\Varid{→}\;\Conid{Fam}\;(\Conid{Vec}\;\Var n\;\Var a){}\<[E]%
\ColumnHook
\end{hscode}\resethooks
The \ensuremath{\Conid{BT}}-typed trees are made up of the constructors of \ensuremath{\Conid{BT}} and hold elements of types~\ensuremath{\Var p}.
A categorical insight is that the constructors constitute an \emph{independent} layer of data added by the functor outside the elements.
The independence of this functor layer is described formally by the definitions of functors and natural transformations.

One aspect of this independence is that the functor layer can stay the same and impervious to whatever is happening at the inner layer.
Categorically, `whatever is happening' means an arbitrary morphism.
In the case of \ensuremath{\Conid{BT}}, the inner layer (the elements) may be changed by some arbitrary morphism of type \ensuremath{\Var p\;\unskip\rightrightarrows\ignorenext\;\Var q}, and that can always be lifted to a morphism of type \ensuremath{\Conid{BT}^{\Var n}_{\Var k}\;\Var p\;\unskip\rightrightarrows\ignorenext\;\Conid{BT}^{\Var n}_{\Var k}\;\Var q} that doesn't change the functor layer (the tree constructors).
This lifting is the morphism part of a functor, and is the `map' function that comes with any (normal) parametric data type.
I've already had a map function for \ensuremath{\Conid{BT}}, and indeed its type can be rewritten as
\begin{hscode}\SaveRestoreHook
\column{B}{@{}>{\hspre}l<{\hspost}@{}}%
\column{E}{@{}>{\hspre}l<{\hspost}@{}}%
\>[B]{}\Varid{map}_{\Conid{BT}}\;\mathbin{:}\;(\Var p\;\unskip\rightrightarrows\ignorenext\;\Var q)\;\Varid{→}\;(\Conid{BT}^{\Var n}_{\Var k}\;\Var p\;\unskip\rightrightarrows\ignorenext\;\Conid{BT}^{\Var n}_{\Var k}\;\Var q){}\<[E]%
\ColumnHook
\end{hscode}\resethooks
In a sense, a lifted morphism such as \ensuremath{\Varid{map}_{\Conid{BT}}\;\Var f} is essentially just~\ensuremath{\Var f} since \ensuremath{\Varid{map}_{\Conid{BT}}\;\Var f} does nothing to the functor layer.
So when \ensuremath{\Var f}~is a composition, that composition shows up at the level of lifted morphisms too.
Formally, this is stated as a \emph{functoriality} equation:
\begin{hscode}\SaveRestoreHook
\column{B}{@{}>{\hspre}l<{\hspost}@{}}%
\column{E}{@{}>{\hspre}l<{\hspost}@{}}%
\>[B]{}\Varid{map}_{\Conid{BT}}\;(\Var f^\prime\;\Varid{·}\;\Var f)\;\mathrel{=}\;\Varid{map}_{\Conid{BT}}\;\Var f^\prime\;\Varid{·}\;\Varid{map}_{\Conid{BT}}\;\Var f{}\<[E]%
\ColumnHook
\end{hscode}\resethooks
(Also \ensuremath{\Varid{map}_{\Conid{BT}}\;\Varid{id}\;\unskip=\ignorenext\;\Varid{id}} in the degenerate case of composing no morphisms.)

The functor layer may also be changed by \emph{natural transformations} independently of whatever is happening at the inner layer.
In \ensuremath{\text{\textbf{\textsf{Fam}}}} categories, a natural transformation has type \ensuremath{\Varid{∀}\;\{\mskip1.5mu \Var p\mskip1.5mu\}\;\Varid{→}\;\Var F\;\Var p\;\unskip\rightrightarrows\ignorenext\;\Var G\;\Var p} for some functors \ensuremath{\Var F}~and~\ensuremath{\Var G}, and transforms an \ensuremath{\Var F}-layer to a \ensuremath{\Var G}-layer without changing the inner layer~\ensuremath{\Var p}, whatever \ensuremath{\Var p}~is.
For example, \ensuremath{\Varid{retabulate}} transforms the functor layer \ensuremath{\Conid{BT}^{\Var n}_{\Var k}} to a \emph{composition} of functors \ensuremath{\Conid{BT}^{\Var n}_{\mathrm 1\Varid{+}\Var k}\;\Varid{·}\;\Conid{BT}^{\mathrm 1\Varid{+}\Var k}_{\Var k}} (which can be regarded as two functor layers) without changing~\ensuremath{\Var p}.
Indeed, \ensuremath{\Varid{retabulate}} transforms only the tree constructors and doesn't change the elements to something else.
Moreover, this transformation of the functor layer does not interfere with whatever is happening at the inner layer.
Again `whatever is happening' amounts to a quantification over all morphisms: for any \ensuremath{\Var f\;\mathbin{:}\;\Var p\;\unskip\rightrightarrows\ignorenext\;\Var q} happening at the inner layer, if \ensuremath{\Varid{retabulate}} is happening at the functor layer too, it doesn't make a difference whether \ensuremath{\Var f}~or \ensuremath{\Varid{retabulate}} happens first, because they happen at independent layers.
Formally, this is stated as a \emph{naturality} equation (where \ensuremath{\Var f}~needs to be lifted appropriately):
\begin{hscode}\SaveRestoreHook
\column{B}{@{}>{\hspre}l<{\hspost}@{}}%
\column{E}{@{}>{\hspre}l<{\hspost}@{}}%
\>[B]{}\Varid{retabulate}\;\Varid{·}\;\Varid{map}_{\Conid{BT}}\;\Var f\;\mathrel{=}\;\Varid{map}_{\Conid{BT}}\;(\Varid{map}_{\Conid{BT}}\;\Var f)\;\Varid{·}\;\Varid{retabulate}{}\<[E]%
\ColumnHook
\end{hscode}\resethooks



\pause

With functor composition, in general there can be many functor layers in an object (like the target of \ensuremath{\Varid{retabulate}}), and all these layers can be transformed independently by natural transformations.
The best way of managing this structure is to use \emph{string diagrams}.
In string diagrams, functors are drawn as wires, and natural transformations are drawn as dots with input functors/wires attached below and output functors/wires above.
(I learned string diagrams mainly from \citet{Coecke-PQP}, so my string diagrams have inputs below and outputs on top.)
The natural transformations I've got are \ensuremath{\Varid{retabulate}} and \ensuremath{\Varid{unTip}}, and I can draw their types as
\[ \input{pics/retabulate-unTip.tikz} \]
String diagrams focus on the functor layers and represent them explicitly as a bunch of wires --- functor composition is represented as juxtaposition of wires, and the identity functor is omitted (it is drawn as no wires).
As a string diagram, \ensuremath{\Varid{retabulate}} has one input wire labelled \ensuremath{\Conid{BT}^{\Var n}_{\Var k}} and two output wires \ensuremath{\Conid{BT}^{\Var n}_{\mathrm 1\Varid{+}\Var k}} and \ensuremath{\Conid{BT}^{\mathrm 1\Varid{+}\Var k}_{\Var k}}, since it transforms a \ensuremath{\Conid{BT}^{\Var n}_{\Var k}} layer to two layers \ensuremath{\Conid{BT}^{\Var n}_{\mathrm 1\Varid{+}\Var k}\;\Varid{·}\;\Conid{BT}^{\mathrm 1\Varid{+}\Var k}_{\Var k}}.
The diagram of \ensuremath{\Varid{unTip}} goes from one wire to none, since \ensuremath{\Varid{unTip}} transforms \ensuremath{\Conid{BT}^{\Var n}_{\Var n}} to the identity functor.
Indeed, what \ensuremath{\Varid{unTip}} does is get rid of the \ensuremath{\Con{tip_z}} or \ensuremath{\Con{tip_s}} constructor.

\begin{wrapfigure}{r}{.075\textwidth}
\ctikzfig{pics/vertical-right}
\end{wrapfigure}
Whereas functor composition is arranged horizontally, sequential composition of natural transformations goes vertically.
Given transformations \ensuremath{\alpha\;\mathbin{:}\;\Varid{∀}\;\{\mskip1.5mu \Var p\mskip1.5mu\}\;\Varid{→}\;\Var F\;\Var p\;\unskip\rightrightarrows\ignorenext\;\Var G\;\Var p} and \ensuremath{\beta\;\mathbin{:}\;\Varid{∀}\;\{\mskip1.5mu \Var p\mskip1.5mu\}\;\Varid{→}\;\Var G\;\Var p\;\unskip\rightrightarrows\ignorenext\;\Var H\;\Var p}, their sequential composition \ensuremath{\beta\;\unskip\mathrel\cdot\ignorenext\;\alpha\;\mathbin{:}\;\Varid{∀}\;\{\mskip1.5mu \Var p\mskip1.5mu\}\;\Varid{→}\;\Var F\;\Var p\;\unskip\rightrightarrows\ignorenext\;\Var H\;\Var p} is drawn in a string diagram as \ensuremath{\alpha}~and~\ensuremath{\beta} juxtaposed vertically and sharing the middle wire with label~\ensuremath{\Var G} (obscuring a section of the wire).

The power of string diagrams becomes evident when things happen in both the horizontal and vertical dimensions.
For example, suppose there are two layers \ensuremath{\Var F}~and~\ensuremath{\Var F^\prime}, where the outer layer~\ensuremath{\Var F} should be transformed by~\ensuremath{\alpha} and the inner layer~\ensuremath{\Var F^\prime} by \ensuremath{\alpha^\prime\;\mathbin{:}\;\Varid{∀}\;\{\mskip1.5mu \Var p\mskip1.5mu\}\;\Varid{→}\;\Var F^\prime\;\Var p\;\unskip\rightrightarrows\ignorenext\;\Var G^\prime\;\Var p}.
There are two ways of doing this: either \ensuremath{\Varid{map}\;\unskip_{\Var G}\;\alpha^\prime\;\unskip\mathrel\cdot\ignorenext\;\alpha}, where the outer layer~\ensuremath{\Var F} is transformed to~\ensuremath{\Var G} first, or \ensuremath{\alpha\;\unskip\mathrel\cdot\ignorenext\;\Varid{map}\;\unskip_{\Var F}\;\alpha^\prime}, where the inner layer~\ensuremath{\Var F^\prime} is transformed to~\ensuremath{\Var G^\prime} first.
The two ways are equal by naturality of~\ensuremath{\alpha}, but the equality can be seen more directly with string diagrams:
\[ \input{pics/horizontal-definitions.tikz} \]
The \ensuremath{\Varid{map}} means skipping over the outer/left functor and transforming the inner/right functor; so in the diagrams, \ensuremath{\alpha^\prime}~is applied to the inner/right wire.
(I've added dashed lines to emphasise that both diagrams are constructed as the sequential composition of two transformations.)
By placing layers of functors in a separate dimension, it's much easier to see which layers are being transformed, and determine whether two sequentially composed transformations are in fact applied independently, so that their order of application can be swapped.
This is abstracted as a diagrammatic reasoning principle: dots in a diagram can be moved upwards or downwards, possibly changing their vertical positions relative to other dots (while stretching or shrinking the wires, which can be thought of as elastic strings), and the (extensional) meaning of the diagram will remain the same.

\pause

\begin{figure}
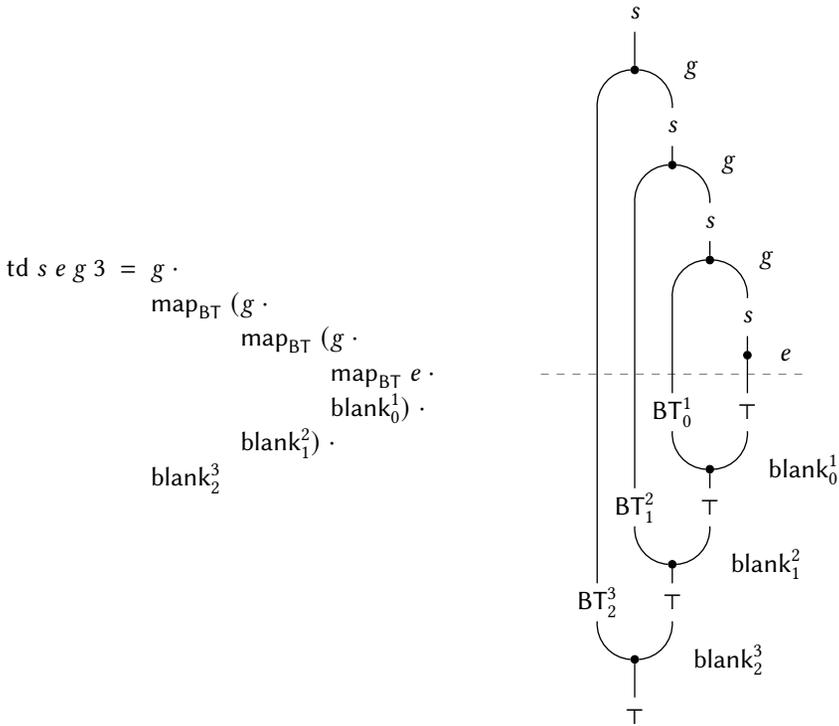

\begin{center}
\begin{varwidth}{\textwidth}
\setlength{\mathindent}{0em}
\begin{hscode}\SaveRestoreHook
\column{B}{@{}>{\hspre}l<{\hspost}@{}}%
\column{15}{@{}>{\hspre}l<{\hspost}@{}}%
\column{22}{@{}>{\hspre}c<{\hspost}@{}}%
\column{22E}{@{}l@{}}%
\column{25}{@{}>{\hspre}l<{\hspost}@{}}%
\column{32}{@{}>{\hspre}c<{\hspost}@{}}%
\column{32E}{@{}l@{}}%
\column{35}{@{}>{\hspre}l<{\hspost}@{}}%
\column{E}{@{}>{\hspre}l<{\hspost}@{}}%
\>[B]{}\Varid{td}\;\Var s\;\Var e\;\Var g\;\mathrm 3\;\mathrel{=}\;{}\<[15]%
\>[15]{}\Var g\;\unskip\mathrel\cdot\ignorenext\;{}\<[E]%
\\
\>[15]{}\Varid{map}_{\Conid{BT}}\;{}\<[22]%
\>[22]{}({}\<[22E]%
\>[25]{}\Var g\;\unskip\mathrel\cdot\ignorenext{}\<[E]%
\\
\>[25]{}\Varid{map}_{\Conid{BT}}\;{}\<[32]%
\>[32]{}({}\<[32E]%
\>[35]{}\Var g\;\unskip\mathrel\cdot\ignorenext{}\<[E]%
\\
\>[35]{}\Varid{map}_{\Conid{BT}}\;\Var e\;\unskip\mathrel\cdot\ignorenext{}\<[E]%
\\
\>[35]{}\Varid{blank}\;\unskip^{\mathrm 1}_{\mathrm 0})\;\unskip\mathrel\cdot\ignorenext{}\<[E]%
\\
\>[25]{}\Varid{blank}\;\unskip^{\mathrm 2}_{\mathrm 1})\;\unskip\mathrel\cdot\ignorenext\;{}\<[E]%
\\
\>[15]{}\Varid{blank}\;\unskip^{\mathrm 3}_{\mathrm 2}{}\<[E]%
\ColumnHook
\end{hscode}\resethooks
\end{varwidth}%
\hspace{.1\textwidth}%
\begin{varwidth}{\textwidth}
\ctikzfig{pics/td}
\end{varwidth}
\end{center}
\caption{A special case of the top-down algorithm as a string diagram.}
\label{fig:td-diagram}
\end{figure}

I want to draw \ensuremath{\Varid{td}} and \ensuremath{\Varid{bu}} as string diagrams.
However, some of their components, namely \ensuremath{\Varid{blank}}, \ensuremath{\Var e}, and~\ensuremath{\Var g}, are not natural transformations.
Technically, only natural transformations can go into string diagrams.
But I'm still tempted to draw those components intuitively as
\[ \input{pics/blank-g-e.tikz} \]
After a bit of thought, I come up with some technical justification.
Any morphism \ensuremath{\Var f\;\mathbin{:}\;\Var p\;\unskip\rightrightarrows\ignorenext\;\Var q} can be lifted to have the type \ensuremath{\Varid{∀}\;\{\mskip1.5mu \Var r\mskip1.5mu\}\;\Varid{→}\;(\Varid{const}\;\Var p)\;\Var r\;\unskip\rightrightarrows\ignorenext\;(\Varid{const}\;\Var q)\;\Var r}, and become a natural transformation from \ensuremath{\Varid{const}\;\Var p} to \ensuremath{\Varid{const}\;\Var q}.
It's fine to leave the lifting implicit and just write \ensuremath{\Var p}~and~\ensuremath{\Var q} for wire labels, since it's usually clear that \ensuremath{\Var p}~and~\ensuremath{\Var q} are not functors and need to be lifted.
For example, \ensuremath{\Var s}~is a type family, which is an object in a \ensuremath{\text{\textbf{\textsf{Fam}}}} category, and needs to be lifted to \ensuremath{\Varid{const}\;\Var s} to be a functor.
For~\ensuremath{\Varid{⊤}} there's one more step: \ensuremath{\Varid{⊤}}~abbreviates the type family \ensuremath{\Varid{const}\;\Varid{⊤}}, which is an object in a \ensuremath{\text{\textbf{\textsf{Fam}}}} category, and needs to be further lifted to \ensuremath{\Varid{const}\;(\Varid{const}\;\Varid{⊤})} to be a functor.
It's kind of technical, but in the end these diagrams are okay.



\subsection{Diagrammatic Reasoning}
\label{sec:diagrammatic-reasoning}

All the abstract nonsense took me some time.
But I still don't know whether string diagrams will actually help me to understand the two algorithms (\cref{sec:td-and-bu-in-Agda}).
It's time to find out.

I'm not confident enough to work with the full recursive definitions straight away, so I take the special case \ensuremath{\Varid{td}\;\Var s\;\Var e\;\Var g\;\mathrm 3} of the top-down algorithm and unfold it into a deeply nested expression \ensuremath{\Var g\;\unskip\mathrel\cdot\ignorenext\;\Varid{map}_{\Conid{BT}}\;(\ldots)\;\unskip\mathrel\cdot\ignorenext\;\Varid{blank}\;\unskip^{\mathrm 3}_{\mathrm 2}} (as on the left of \cref{fig:td-diagram}).
The rightmost component \ensuremath{\Varid{blank}\;\unskip^{\mathrm 3}_{\mathrm 2}} has type \ensuremath{\Varid{const}\;\Varid{⊤}\;\unskip\rightrightarrows\ignorenext\;\Conid{BT}\;\unskip^{\mathrm 3}_{\mathrm 2}\;(\Varid{const}\;\Varid{⊤})}, which ---~after eliding the \ensuremath{\Varid{const}}s, as I've just decided to do~--- is drawn as the bottom Y-junction in the string diagram (as on the right of \cref{fig:td-diagram}), transforming the input \ensuremath{\Varid{⊤}}-wire at the bottom into two wires \ensuremath{\Conid{BT}\;\unskip^{\mathrm 3}_{\mathrm 2}} and~\ensuremath{\Varid{⊤}}.
Then the \ensuremath{\Varid{map}_{\Conid{BT}}} means that the left wire \ensuremath{\Conid{BT}\;\unskip^{\mathrm 3}_{\mathrm 2}} is skipped over and unchanged.
The right \ensuremath{\Varid{⊤}}-wire continues to be transformed by \ensuremath{\Varid{blank}\;\unskip^{\mathrm 2}_{\mathrm 1}} and so on, and eventually becomes an \ensuremath{\Var s}-wire.
Finally, I apply~\ensuremath{\Var g} to the \ensuremath{\Conid{BT}\;\unskip^{\mathrm 3}_{\mathrm 2}}-wire I skipped over and the \ensuremath{\Var s}-wire to produce the output \ensuremath{\Var s}-wire at the top.


I step back and compare the expression and the diagram~(\cref{fig:td-diagram}).
All the \ensuremath{\Varid{map}_{\Conid{BT}}}s are gone in the diagram, because I can directly apply a transformation to the intended layers/wires, rather than count awkwardly how many outer layers I have to skip, using \ensuremath{\Varid{map}_{\Conid{BT}}} one layer at a time.
Functoriality is also transparent in the diagram, so it's slightly easier to see that \ensuremath{\Varid{td}} has two phases (which I have separated by a dashed line): the first phase constructs deeply nested blank tables, and the second phase fills and demolishes the tables inside out.

Functoriality is already somewhat transparent in the traditional expression though, thanks to the infix notation of function composition.
So I suppose I don't absolutely need the string diagram to see that \ensuremath{\Varid{td}} has two phases, although the required rewriting~(\cref{fig:functoriality-rewriting}) is not as perspicuous as just seeing the two phases in the diagram.
Moreover, there's nothing I can meaningfully move in the diagram --- all the transformations here are lifted after all.

\begin{figure}
\begin{center}
\begin{hscode}\SaveRestoreHook
\column{B}{@{}>{\hspre}l<{\hspost}@{}}%
\column{3}{@{}>{\hspre}l<{\hspost}@{}}%
\column{5}{@{}>{\hspre}l<{\hspost}@{}}%
\column{E}{@{}>{\hspre}l<{\hspost}@{}}%
\>[3]{}\Varid{td}\;\Var s\;\Var e\;\Var g\;\mathrm 3{}\<[E]%
\\
\>[B]{}\mathrel{=}\;{}\<[5]%
\>[5]{}\mbox{\commentbegin \;definition  \commentend}{}\<[E]%
\\
\>[B]{}\hsindent{3}{}\<[3]%
\>[3]{}\Var g\;\unskip\mathrel\cdot\ignorenext\;\Varid{map}_{\Conid{BT}}\;(\Var g\;\unskip\mathrel\cdot\ignorenext\;\Varid{map}_{\Conid{BT}}\;(\Var g\;\unskip\mathrel\cdot\ignorenext\;\Varid{map}_{\Conid{BT}}\;\Var e\;\unskip\mathrel\cdot\ignorenext\;\Varid{blank}\;\unskip^{\mathrm 1}_{\mathrm 0})\;\unskip\mathrel\cdot\ignorenext\;\Varid{blank}\;\unskip^{\mathrm 2}_{\mathrm 1})\;\unskip\mathrel\cdot\ignorenext\;\Varid{blank}\;\unskip^{\mathrm 3}_{\mathrm 2}{}\<[E]%
\\
\>[B]{}\mathrel{=}\;{}\<[5]%
\>[5]{}\mbox{\commentbegin \;functoriality  \commentend}{}\<[E]%
\\
\>[B]{}\hsindent{3}{}\<[3]%
\>[3]{}\Var g\;\unskip\mathrel\cdot\ignorenext\;\Varid{map}_{\Conid{BT}}\;(\Var g\;\unskip\mathrel\cdot\ignorenext\;\Varid{map}_{\Conid{BT}}\;(\Var g\;\unskip\mathrel\cdot\ignorenext\;\Varid{map}_{\Conid{BT}}\;\Var e)\;\unskip\mathrel\cdot\ignorenext\;\Varid{map}_{\Conid{BT}}\;\Varid{blank}\;\unskip^{\mathrm 1}_{\mathrm 0}\;\unskip\mathrel\cdot\ignorenext\;\Varid{blank}\;\unskip^{\mathrm 2}_{\mathrm 1})\;\unskip\mathrel\cdot\ignorenext\;\Varid{blank}\;\unskip^{\mathrm 3}_{\mathrm 2}{}\<[E]%
\\
\>[B]{}\mathrel{=}\;{}\<[5]%
\>[5]{}\mbox{\commentbegin \;functoriality  \commentend}{}\<[E]%
\\
\>[B]{}\hsindent{3}{}\<[3]%
\>[3]{}\Var g\;\unskip\mathrel\cdot\ignorenext\;\Varid{map}_{\Conid{BT}}\;(\Var g\;\unskip\mathrel\cdot\ignorenext\;\Varid{map}_{\Conid{BT}}\;(\Var g\;\unskip\mathrel\cdot\ignorenext\;\Varid{map}_{\Conid{BT}}\;\Var e))\;\unskip\mathrel\cdot\ignorenext\;{}\<[E]%
\\
\>[B]{}\hsindent{3}{}\<[3]%
\>[3]{}\Varid{map}_{\Conid{BT}}\;(\Varid{map}_{\Conid{BT}}\;\Varid{blank}\;\unskip^{\mathrm 1}_{\mathrm 0}\;\unskip\mathrel\cdot\ignorenext\;\Varid{blank}\;\unskip^{\mathrm 2}_{\mathrm 1})\;\unskip\mathrel\cdot\ignorenext\;\Varid{blank}\;\unskip^{\mathrm 3}_{\mathrm 2}{}\<[E]%
\ColumnHook
\end{hscode}\resethooks
\end{center}
\caption{Rewriting \ensuremath{\Varid{td}\;\Var s\;\Var e\;\Var g\;\mathrm 3} into two phases using functoriality.}
\label{fig:functoriality-rewriting}
\end{figure}

Hm. Maybe I'll have more luck with the bottom-up algorithm, which has `real' natural transformations?
I go on to expand \ensuremath{\Varid{bu}\;\Var s\;\Var e\;\Var g\;\mathrm 3}.
The loop in the expression unfolds into a sequence of functions, alternating between \ensuremath{\Varid{retabulate}} and \ensuremath{\Varid{map}_{\Conid{BT}}\;\Var g}.

`A sequence\ldots'
I mutter.
I shouldn't have expected anything else from unfolding a loop. But the sequential structure is so different from the deeply nested structure of \ensuremath{\Varid{td}}.

\begin{figure}
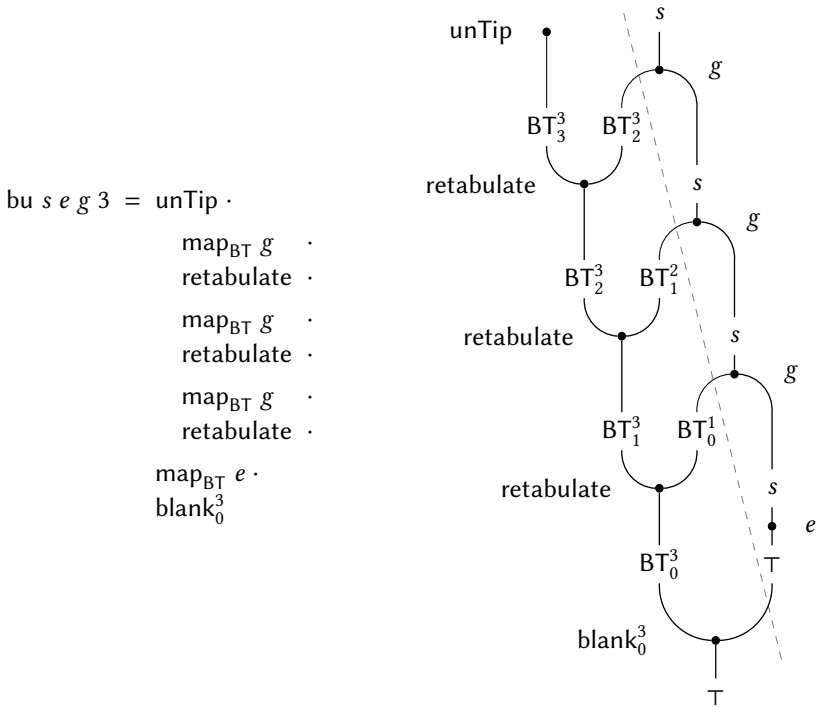

\begin{center}
\begin{varwidth}{\textwidth}
\setlength{\mathindent}{0em}
\begin{hscode}\SaveRestoreHook
\column{B}{@{}>{\hspre}l<{\hspost}@{}}%
\column{15}{@{}>{\hspre}l<{\hspost}@{}}%
\column{17}{@{}>{\hspre}l<{\hspost}@{}}%
\column{29}{@{}>{\hspre}l<{\hspost}@{}}%
\column{E}{@{}>{\hspre}l<{\hspost}@{}}%
\>[B]{}\Varid{bu}\;\Var s\;\Var e\;\Var g\;\mathrm 3\;\mathrel{=}\;{}\<[15]%
\>[15]{}\Varid{unTip}\;\unskip\mathrel\cdot\ignorenext\;{}\<[E]%
\\[\blanklineskip]%
\>[15]{}\hsindent{2}{}\<[17]%
\>[17]{}\Varid{map}_{\Conid{BT}}\;\Var g\;{}\<[29]%
\>[29]{}\unskip\mathrel\cdot\ignorenext\;{}\<[E]%
\\
\>[15]{}\hsindent{2}{}\<[17]%
\>[17]{}\Varid{retabulate}\;{}\<[29]%
\>[29]{}\unskip\mathrel\cdot\ignorenext\;{}\<[E]%
\\[\blanklineskip]%
\>[15]{}\hsindent{2}{}\<[17]%
\>[17]{}\Varid{map}_{\Conid{BT}}\;\Var g\;{}\<[29]%
\>[29]{}\unskip\mathrel\cdot\ignorenext\;{}\<[E]%
\\
\>[15]{}\hsindent{2}{}\<[17]%
\>[17]{}\Varid{retabulate}\;{}\<[29]%
\>[29]{}\unskip\mathrel\cdot\ignorenext\;{}\<[E]%
\\[\blanklineskip]%
\>[15]{}\hsindent{2}{}\<[17]%
\>[17]{}\Varid{map}_{\Conid{BT}}\;\Var g\;{}\<[29]%
\>[29]{}\unskip\mathrel\cdot\ignorenext\;{}\<[E]%
\\
\>[15]{}\hsindent{2}{}\<[17]%
\>[17]{}\Varid{retabulate}\;{}\<[29]%
\>[29]{}\unskip\mathrel\cdot\ignorenext\;{}\<[E]%
\\[\blanklineskip]%
\>[15]{}\Varid{map}_{\Conid{BT}}\;\Var e\;\unskip\mathrel\cdot\ignorenext\;{}\<[E]%
\\
\>[15]{}\Varid{blank}\;\unskip^{\mathrm 3}_{\mathrm 0}{}\<[E]%
\ColumnHook
\end{hscode}\resethooks
\end{varwidth}%
\hspace{.1\textwidth}%
\begin{varwidth}{\textwidth}
\ctikzfig{pics/bu}
\end{varwidth}
\end{center}
\caption{A special case of the bottom-up algorithm as a string diagram.}
\label{fig:bu-diagram}
\end{figure}

And then, something unexpected yet familiar appears in the transcribed diagram~(\cref{fig:bu-diagram}).

\begin{aha}
There are also two phases for table construction and demolition, and the \ensuremath{\Var g}s and \ensuremath{\Var e} in the demolition phase are \emph{exactly the same} as in \ensuremath{\Varid{td}}!
\end{aha}

The string diagram is truly helpful this time.
Now I see, as Richard hinted, that I could rewrite the traditional expression using the naturality of \ensuremath{\Varid{unTip}} and \ensuremath{\Varid{retabulate}} to push \ensuremath{\Var g}~and~\ensuremath{\Var e} to the left of the sequence and separate the two phases~(\cref{fig:naturality-rewriting}).
But in the string diagram, all those rewritings amount to nothing more than gradually pulling the two phases apart, eventually making the dashed line horizontal.
In fact I don't even bother to pull, because on this diagram I can already see simultaneously both the sequence (the dots appearing one by one vertically) and the result of rewriting the sequence using naturality.

\begin{figure}
\begin{center}
\setlength{\mathindent}{0em}
\begin{hscode}\SaveRestoreHook
\column{B}{@{}>{\hspre}l<{\hspost}@{}}%
\column{4}{@{}>{\hspre}l<{\hspost}@{}}%
\column{E}{@{}>{\hspre}l<{\hspost}@{}}%
\>[4]{}\Varid{bu}\;\Var s\;\Var e\;\Var g\;\mathrm 3{}\<[E]%
\\
\>[B]{}\mathrel{=}\;{}\<[4]%
\>[4]{}\mbox{\commentbegin \;definition  \commentend}{}\<[E]%
\\
\>[4]{}\Varid{unTip}\;\unskip\mathrel\cdot\ignorenext\;\Varid{map}_{\Conid{BT}}\;\Var g\;\unskip\mathrel\cdot\ignorenext\;\Varid{retabulate}\;\unskip\mathrel\cdot\ignorenext\;\Varid{map}_{\Conid{BT}}\;\Var g\;\unskip\mathrel\cdot\ignorenext\;\Varid{retabulate}\;\unskip\mathrel\cdot\ignorenext\;\Varid{map}_{\Conid{BT}}\;\Var g\;\unskip\mathrel\cdot\ignorenext\;\Varid{retabulate}\;\unskip\mathrel\cdot\ignorenext\;\Varid{map}_{\Conid{BT}}\;\Var e\;\unskip\mathrel\cdot\ignorenext\;\Varid{blank}\;\unskip^{\mathrm 3}_{\mathrm 0}{}\<[E]%
\\
\>[B]{}\mathrel{=}\;{}\<[4]%
\>[4]{}\mbox{\commentbegin \;naturality of \ensuremath{\Varid{unTip}}  \commentend}{}\<[E]%
\\
\>[4]{}\Var g\;\unskip\mathrel\cdot\ignorenext\;{}\<[E]%
\\
\>[4]{}\Varid{unTip}\;\unskip\mathrel\cdot\ignorenext\;\Varid{retabulate}\;\unskip\mathrel\cdot\ignorenext\;\Varid{map}_{\Conid{BT}}\;\Var g\;\unskip\mathrel\cdot\ignorenext\;\Varid{retabulate}\;\unskip\mathrel\cdot\ignorenext\;\Varid{map}_{\Conid{BT}}\;\Var g\;\unskip\mathrel\cdot\ignorenext\;\Varid{retabulate}\;\unskip\mathrel\cdot\ignorenext\;\Varid{map}_{\Conid{BT}}\;\Var e\;\unskip\mathrel\cdot\ignorenext\;\Varid{blank}\;\unskip^{\mathrm 3}_{\mathrm 0}{}\<[E]%
\\
\>[B]{}\mathrel{=}\;{}\<[4]%
\>[4]{}\mbox{\commentbegin \;naturality of \ensuremath{\Varid{retabulate}}  \commentend}{}\<[E]%
\\
\>[4]{}\Var g\;\unskip\mathrel\cdot\ignorenext\;{}\<[E]%
\\
\>[4]{}\Varid{unTip}\;\unskip\mathrel\cdot\ignorenext\;\Varid{map}_{\Conid{BT}}\;(\Varid{map}_{\Conid{BT}}\;\Var g)\;\unskip\mathrel\cdot\ignorenext\;\Varid{retabulate}\;\unskip\mathrel\cdot\ignorenext\;\Varid{retabulate}\;\unskip\mathrel\cdot\ignorenext\;\Varid{map}_{\Conid{BT}}\;\Var g\;\unskip\mathrel\cdot\ignorenext\;\Varid{retabulate}\;\unskip\mathrel\cdot\ignorenext\;\Varid{map}_{\Conid{BT}}\;\Var e\;\unskip\mathrel\cdot\ignorenext\;\Varid{blank}\;\unskip^{\mathrm 3}_{\mathrm 0}{}\<[E]%
\\
\>[B]{}\mathrel{=}\;{}\<[4]%
\>[4]{}\mbox{\commentbegin \;naturality of \ensuremath{\Varid{unTip}} again  \commentend}{}\<[E]%
\\
\>[4]{}\Var g\;\unskip\mathrel\cdot\ignorenext\;\Varid{map}_{\Conid{BT}}\;\Var g\;\unskip\mathrel\cdot\ignorenext\;{}\<[E]%
\\
\>[4]{}\Varid{unTip}\;\unskip\mathrel\cdot\ignorenext\;\Varid{retabulate}\;\unskip\mathrel\cdot\ignorenext\;\Varid{retabulate}\;\unskip\mathrel\cdot\ignorenext\;\Varid{map}_{\Conid{BT}}\;\Var g\;\unskip\mathrel\cdot\ignorenext\;\Varid{retabulate}\;\unskip\mathrel\cdot\ignorenext\;\Varid{map}_{\Conid{BT}}\;\Var e\;\unskip\mathrel\cdot\ignorenext\;\Varid{blank}\;\unskip^{\mathrm 3}_{\mathrm 0}{}\<[E]%
\\
\>[B]{}\mathrel{=}\;{}\<[4]%
\>[4]{}\mbox{\commentbegin \;similarly  \commentend}{}\<[E]%
\\
\>[4]{}\Var g\;\unskip\mathrel\cdot\ignorenext\;\Varid{map}_{\Conid{BT}}\;\Var g\;\unskip\mathrel\cdot\ignorenext\;\Varid{map}_{\Conid{BT}}\;(\Varid{map}_{\Conid{BT}}\;\Var g)\;\unskip\mathrel\cdot\ignorenext\;\Varid{map}_{\Conid{BT}}\;(\Varid{map}_{\Conid{BT}}\;(\Varid{map}_{\Conid{BT}}\;\Var e))\;\unskip\mathrel\cdot\ignorenext\;{}\<[E]%
\\
\>[4]{}\Varid{unTip}\;\unskip\mathrel\cdot\ignorenext\;\Varid{retabulate}\;\unskip\mathrel\cdot\ignorenext\;\Varid{retabulate}\;\unskip\mathrel\cdot\ignorenext\;\Varid{retabulate}\;\unskip\mathrel\cdot\ignorenext\;\Varid{blank}\;\unskip^{\mathrm 3}_{\mathrm 0}{}\<[E]%
\\
\>[B]{}\mathrel{=}\;{}\<[4]%
\>[4]{}\mbox{\commentbegin \;functoriality  \commentend}{}\<[E]%
\\
\>[4]{}\Var g\;\unskip\mathrel\cdot\ignorenext\;\Varid{map}_{\Conid{BT}}\;(\Var g\;\unskip\mathrel\cdot\ignorenext\;\Varid{map}_{\Conid{BT}}\;(\Var g\;\unskip\mathrel\cdot\ignorenext\;\Varid{map}_{\Conid{BT}}\;\Var e))\;\unskip\mathrel\cdot\ignorenext\;{}\<[E]%
\\
\>[4]{}\Varid{unTip}\;\unskip\mathrel\cdot\ignorenext\;\Varid{retabulate}\;\unskip\mathrel\cdot\ignorenext\;\Varid{retabulate}\;\unskip\mathrel\cdot\ignorenext\;\Varid{retabulate}\;\unskip\mathrel\cdot\ignorenext\;\Varid{blank}\;\unskip^{\mathrm 3}_{\mathrm 0}{}\<[E]%
\ColumnHook
\end{hscode}\resethooks
\end{center}
\caption{Rewriting \ensuremath{\Varid{bu}\;\Var s\;\Var e\;\Var g\;\mathrm 3} into two phases using naturality.}
\label{fig:naturality-rewriting}
\end{figure}


\pause

So, modulo naturality, the two algorithms have the same table demolition phase but different table construction phases.
If I can prove that their table construction phases are equal, then I'll have another proof that the two algorithms are equal, in addition to the parametricity-based proof~(\cref{sec:equality-from-types}).
For \ensuremath{\Varid{td}}, the construction phase is a right-leaning tree on the diagram, whereas for \ensuremath{\Varid{bu}} it's a left-leaning tree.
Maybe what I need is an equation about \ensuremath{\Varid{blank}} and \ensuremath{\Varid{retabulate}} that can help me to rotate a tree\ldots?
\[ \text{\lstinline{cd (choose k xs)}} \equals \text{\lstinline{mapB (flatten . choose k) (choose (k+1) xs)}} \]

\begin{figure}
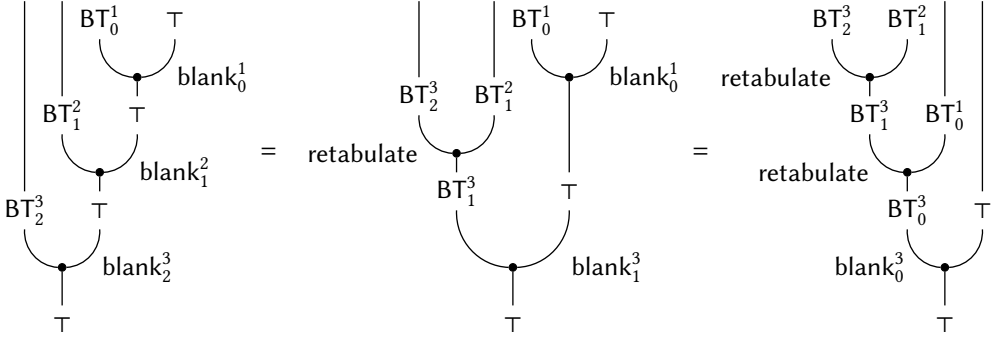

\ctikzfig{pics/rotations}
\caption{Rewriting the table construction phase of \ensuremath{\Varid{td}\;\Var s\;\Var e\;\Var g\;\mathrm 3} to that of \ensuremath{\Varid{bu}\;\Var s\;\Var e\;\Var g\;\mathrm 3} using the \ensuremath{\Varid{rotation}} equation~(\cref{eq:rotation}).}
\label{fig:rotations}
\end{figure}

The equation~(\cref{eq:cd-spec}) flashes through my mind.
Of course it has to be this equation --- I used it as a specification for \lstinline{cd}, the Haskell progenitor of \ensuremath{\Varid{retabulate}}.
How else would I introduce \ensuremath{\Varid{retabulate}} into the picture?
But first let me update this for my dependently typed string diagrams:
\begin{equation}
\input{pics/rotation.tikz}
\tag{$\ast\ast$}
\label{eq:rotation}
\end{equation}
That's a tree rotation all right!
So I should do an induction that uses this equation to rotate the right-leaning tree in \ensuremath{\Varid{td}} and obtain the left-leaning tree in \ensuremath{\Varid{bu}}~(\cref{fig:rotations}).
And then I'll need to prove the equation, meaning that I'll need to go through the definitions of \ensuremath{\Varid{retabulate}} and \ensuremath{\Varid{blank}}\ldots
Oh hell, that's a lot of work.

\begin{aha}
But wait a minute~--- do I really need to go through all this?
\end{aha}

The three functors at the top of the diagrams~(\cref{eq:rotation}) catch my attention.
In Agda, they expand to the type \ensuremath{\Conid{BT}\;\unskip^{\Var n}_{\mathrm 1\Varid{+}\Var k}\;(\Conid{BT}\;\unskip^{\mathrm 1\Varid{+}\Var k}_{\Var k}\;(\Varid{const}\;\Varid{⊤}))\;\Var xs}.
An inhabitant of this type is a table of \emph{blank} tables, so there is no choice of table entries;
and moreover the structures of all the tables are completely determined by the indices\ldots the type has a unique inhabitant!
So the equation is actually trivial to prove: the two sides are forced by the type to construct the same table. And I don't need to look into the definitions of \ensuremath{\Varid{retabulate}} and \ensuremath{\Varid{blank}} at all!

Relieved, I start to work on the proof.
The precise notion I need here is (mere) propositions~\citep[Chapter~3]{UFP-HoTT}:
\begin{hscode}\SaveRestoreHook
\column{B}{@{}>{\hspre}l<{\hspost}@{}}%
\column{E}{@{}>{\hspre}l<{\hspost}@{}}%
\>[B]{}\Varid{isProp}\;\mathbin{:}\;\Conid{Set}\;\Varid{→}\;\Conid{Set}{}\<[E]%
\\
\>[B]{}\Varid{isProp}\;\Var a\;\mathrel{=}\;\{\mskip1.5mu \Var x\;\Var y\;\mathbin{:}\;\Var a\mskip1.5mu\}\;\Varid{→}\;\Var x\;\unskip\equiv\ignorenext\;\Var y{}\<[E]%
\ColumnHook
\end{hscode}\resethooks
The type \ensuremath{\Conid{BT}\;\unskip^{\Var n}_{\Var k}\;\Var p\;\Var xs} is propositional if the element types~\ensuremath{\Var p} are propositional --- this is easy to prove by a straightforward double induction:
\begin{hscode}\SaveRestoreHook
\column{B}{@{}>{\hspre}l<{\hspost}@{}}%
\column{E}{@{}>{\hspre}l<{\hspost}@{}}%
\>[B]{}\Conid{BT-isProp}\;\mathbin{:}\;(\Varid{∀}\;\{\mskip1.5mu \Var ys\mskip1.5mu\}\;\Varid{→}\;\Varid{isProp}\;(\Var p\;\Var ys))\;\Varid{→}\;\Varid{isProp}\;(\Conid{BT}\;\unskip^{\Var n}_{\Var k}\;\Var p\;\Var xs){}\<[E]%
\ColumnHook
\end{hscode}\resethooks
And then the equation~(\cref{eq:rotation}) can be proved trivially by invoking \ensuremath{\Conid{BT-isProp}} twice:
\begin{hscode}\SaveRestoreHook
\column{B}{@{}>{\hspre}l<{\hspost}@{}}%
\column{E}{@{}>{\hspre}l<{\hspost}@{}}%
\>[B]{}\Varid{rotation}\;\mathbin{:}\;\Varid{retabulate}\;(\Varid{blank}\;\unskip^{\Var n}_{\Var k}\;\Con{tt})\;\unskip\equiv\ignorenext\;\Varid{map}_{\Conid{BT}}\;\Varid{blank}\;\unskip^{\mathrm 1\Varid{+}\Var k}_{\Var k}\;(\Varid{blank}\;\unskip^{\Var n}_{\mathrm 1\Varid{+}\Var k}\;\Con{tt}){}\<[E]%
\\
\>[B]{}\Varid{rotation}\;\mathrel{=}\;\Conid{BT-isProp}\;(\Conid{BT-isProp}\;\Con{refl}){}\<[E]%
\ColumnHook
\end{hscode}\resethooks
The side conditions of \ensuremath{\Varid{retabulate}} and \ensuremath{\Varid{blank}} (omitted above) are all universally quantified.
Usually they make proofs more complex, but not in this case because the proof doesn't look into any of the function definitions.
As long as the type is blank nested tables, the two sides of an equation can be arbitrarily complicated, and I can still prove them equal just by using \ensuremath{\Conid{BT-isProp}}.

\begin{aha}
Wait, blank nested tables --- aren't those what the construction phases of both algorithms produce?
\end{aha}

I face-palm.
It was a waste of time proving the \ensuremath{\Varid{rotation}} equation.
The construction phases of both algorithms produce blank nested tables of the same type --- \ensuremath{\Conid{BT}\;\unskip^{\mathrm 3}_{\mathrm 2}\;(\Conid{BT}\;\unskip^{\mathrm 2}_{\mathrm 1}\;(\Conid{BT}\;\unskip^{\mathrm 1}_{\mathrm 0}\;(\Varid{const}\;\Varid{⊤})))\;\Var xs} in the concrete examples I tried~(\cref{fig:td-diagram,fig:bu-diagram}).
So I can directly prove them equal using \ensuremath{\Conid{BT-isProp}} three times.
There's no need to do any rotation.


Oh well, \ensuremath{\Varid{rotation}} is still interesting because it helps to explain how the two algorithms are related intensionally: they produce the same layers of tables but in opposite orders, and \ensuremath{\Varid{rotation}} helps to show how one order can be rewritten into the other~(\cref{fig:rotations}).
It's just that a \ensuremath{\Varid{rotation}}-based proof would be quite tedious, and I don't want to go through with that.
A proof based on \ensuremath{\Conid{BT-isProp}} should be much simpler.
Conceptually I've figured it all out: both algorithms have two phases modulo naturality; their table demolition phases are exactly the same, and their table construction phases are equal due to the \ensuremath{\Conid{BT-isProp}} reasoning.
But the general proof is still going to take some work.
If I want to stick to string diagrams, I'll need to transcribe the algorithms into inductively defined diagrams.
Moreover, the \ensuremath{\Conid{BT-isProp}} reasoning is formally an induction (on the length of the input list), which needs to be worked out.
And actually, compared with a diagrammatic but informal proof, I prefer a full Agda formalisation.
That means I'll need to spell out a lot of detail, including functoriality and naturality rewriting~(\cref{fig:functoriality-rewriting,fig:naturality-rewriting}).
Whining, I finish the entire proof in Agda.
But as usual, in the end there's a dopamine hit from seeing everything checked.


\pause

Still, I can't help feeling that I’ve neglected a fundamental aspect of the problem: why the bottom-up algorithm is more efficient.
After making all the effort adopting dependent types and string diagrams, do these state-of-the-art languages help me say something about efficiency too?

String diagrams make it easier for me to see that the table construction phases of both algorithms produce the same layers of tables but in opposite orders.
Only the order used by the bottom-up algorithm allows table construction and demolition to be interleaved, and consequently the algorithm needs no more than two layers of tables at any time.
That's the crucial difference between the two algorithms.
Now I need to figure out what the difference means algorithmically.

\begin{aha}
More specifically, why is it good to keep \emph{two} layers of tables and not more?
\end{aha}

When there are multiple layers of tables of type \ensuremath{\Conid{BT}\;\unskip^{\Var n}_{\Var k}} with \ensuremath{\Var k\;\unskip<\ignorenext\;\Var n}, meaning that the input list is split into proper sublists multiple times, all the final sublists will appear (as indices in the element types) in the entire nested table multiple times --- that is, overlapping subproblems will appear.
Therefore, when I use~\ensuremath{\Var g} to fill in a nested table, I'm invoking~\ensuremath{\Var g} to compute duplicate solutions for overlapping subproblems, which is what I want to avoid.
More precisely, `using~\ensuremath{\Var g} to fill in a nested table' means applying~\ensuremath{\Var g} under at least two layers, for example \ensuremath{\Varid{map}_{\Conid{BT}}\;(\Varid{map}_{\Conid{BT}}\;\Var g)\;\mathbin{:}\;\Conid{BT}\;\unskip^{\mathrm 3}_{\mathrm 2}\;(\Conid{BT}\;\unskip^{\mathrm 2}_{\mathrm 1}\;(\Conid{BT}\;\unskip^{\mathrm 1}_{\mathrm 0}\;\Var s))\;\unskip\rightrightarrows\ignorenext\;\Conid{BT}\;\unskip^{\mathrm 3}_{\mathrm 2}\;(\Conid{BT}\;\unskip^{\mathrm 2}_{\mathrm 1}\;\Var s)}, where the result is at least two layers of tables, so there need to be at least \emph{three} layers of tables (to which \ensuremath{\Varid{map}_{\Conid{BT}}\;(\Varid{map}_{\Conid{BT}}\;\Var g)} is applied) for duplicate solutions of overlapping subproblems to be recomputed.
The bottom-up algorithm never gets to three layers of tables, and therefore avoids recomputing solutions for overlapping subproblems.

That reasoning doesn't sound too bad, although it's clear that there's much more to be done.
The whole argument is still too informal and lacks detail.
It's easy to poke holes in the reasoning --- for example, if the input list has duplicate elements, then the bottom-up algorithm won't be able to entirely avoid recomputing solutions of overlapping subproblems.
To fix this, the algorithm will need a redesign.
And of course it's tempting to explore more problem-decomposing strategies beyond immediate sublists.
Eventually I may arrive at something general about dynamic programming, which was what Richard wanted to work out in his paper.

All those are for another day, however.
I've had enough fun today.
Mostly, what I did was transcribe Richard's programs into new languages, but that helped me to reason in new ways, using more convenient tools to make sense of the programs.

I wish Richard was still around so that I could show all these to him.
He would've liked the new languages and the new ways of reasoning.



\bigpause 

\section*{Afterword}

\begingroup

This work is presented as a kind of `Socratic monologue', recording the thought processes of a functional programmer as they solve a programming mystery.
We were inspired by the science fiction novel \textit{Project Hail Mary} by Andy Weir, where the narrative masterfully weaves together intuitive presentations of scientific knowledge and the protagonist's application of that knowledge to solve the problems he is facing.
We envisaged to do something similar in this paper, although it ends up being not as leisurely and entertaining as Weir's novel, because we need to cover more technical detail, and there is very little action in our story apart from sitting in front of a computer and racking one's brains.
However, compared to the traditional rational reconstruction of a finished piece of work, we believe that this format helps both the writer and the reader to focus on currently available clues and how to make progress based on those clues by recreating the experience of solving a mystery.
In fact, our telling largely follows our actual development (tracing what \lstinline{cd} does in \cref{sec:bu}, generalising \ensuremath{\Conid{B}}~and~\ensuremath{\Conid{B}^\prime} to \ensuremath{\Conid{BT}} in \cref{sec:BT}, revising \ensuremath{\Conid{ImmediateSublistInduction}} in \cref{sec:td-and-bu-in-Agda}, realising that the \ensuremath{\Conid{BT-isProp}} argument works more generally after proving \ensuremath{\Varid{rotation}} in \cref{sec:diagrammatic-reasoning}, etc) --- that is, this paper is `based on a true story'.

The format also works well with various decisions regarding what to include in the paper, and what to omit.
We put emphasis on intuitive explanations, and give formal definitions, theorems, and proofs only when necessary: we usually rely on intuitive reasoning to tackle a problem at first, and do not hurry to write things down formally.
The
ancillary
Agda code provides the omitted formal detail though.
We have striven to keep the paper fairly self-contained: the reader should be able to get a sense of the main ideas just from the intuitive explanations.
But we do not intend this paper to be a tutorial on dependently typed programming in Agda or on category theory --- the paper is best thought of as a companion to such tutorials or textbooks, giving a larger but not overly complicated example, and applying the abstract tools to that example.
To make the paper more accessible, we have also resisted the temptation to generalise or to answer every question:
for example, we do not generalise \ensuremath{\Conid{ImmediateSublistInduction}} for dynamic programming more broadly (as \citet{Bird-zippy-tabulations} attempted to do); we leave the question of whether the type of \ensuremath{\Varid{retabulate}} uniquely determines the extensional behaviour of its inhabitants as a conjecture~(\cref{sec:spec}); and we avoid digressions into topics such as how data types like \ensuremath{\Conid{BT}} can be derived systematically~(\cref{sec:BT}) and whether \ensuremath{\Conid{BT}} is a graded comonad~(\cref{sec:td-and-bu-in-Agda}).


This work gives a dependently typed treatment of the sublists problem.
The problem has been studied in other settings.
It was one of the examples used by \citet{Bird&Hinze-nexus} when studying a technique of function memoisation using trees of shared nodes, which they called \emph{nexuses}.
\citet{Bird-zippy-tabulations} went on to study top-down and bottom-up algorithms, where the sublists problem was the final example.
To cover all the examples in the paper, \citeauthor{Bird-zippy-tabulations}'s generic bottom-up algorithm also employed a form of nexus, but it is not needed for the sublists problem and thus omitted in our work.
\citet{Mu-sublists} derived \lstinline{cd} from the specification~(\cref{eq:cd-spec}) and proved the equality between \lstinline{td} and \lstinline{bu} using traditional equational reasoning.
Neither \citet{Bird&Hinze-nexus} nor \citet{Bird-zippy-tabulations} discussed applications of the sublists problem, but \citet{Mu-sublists} observed that reduction to it is a standard technique in the algorithms community.
None of these papers used dependent types (except that \citet[Section~4.3]{Mu-sublists} mentioned the shape-indexed \ensuremath{\Conid{B}} and \ensuremath{\Varid{cd}} in \cref{sec:shape}).

The general message we want to deliver is that we can discover, explain, and prove things by writing them down in appropriate languages.
More specifically, dependent types, category theory, and string diagrams are some of those languages, and they should be in the toolbox of the mathematically inclined functional programmer.
In the case of dependent types, they can be expressive enough to replace traditional (equational) specifications and proofs.
For example, in place of \varcitet{Mu-sublists}{'s} derivation, \ensuremath{\Varid{retabulate}} can be constructed by assigning it a type having sufficient information (\cref{sec:spec}), and
the dependently typed \ensuremath{\Varid{td}} and \ensuremath{\Varid{bu}} can be proved equal simply by showing that they have the same, uniquely inhabited type~(\cref{sec:equality-from-types}).
This approach to program correctness and program equality is still under-explored, and has potential to reduce proof burdens drastically.
As for category theory, even though we use it only in a lightweight manner, it still offers a somewhat useful abstraction for managing more complex (in our case, indexed) definitions as if they were simply typed~(\cref{sec:basic-category-theory}).
More importantly, the categorical abstraction enables the use of string diagrams to simplify proofs about functoriality and naturality.
These properties are only the simplest ones that string diagrams can handle --- for other kinds of properties~\citep{Coecke-PQP,Hinze-string-diagrams} the proof simplification can be even more dramatic, although many of those properties are highly abstract.
Our comparison between diagrammatic and traditional equational reasoning (\cref{fig:td-diagram,fig:functoriality-rewriting}, and \cref{fig:bu-diagram,fig:naturality-rewriting}) should be a good, albeit modest, demonstration of the power of string diagrams in a more practical, algorithmic scenario.

\endgroup

\begin{acks}
We would like to thank Liang-Ting Chen for offering helpful suggestions about the development; Julie Summers, Royal Literary Fund Fellow at Kellogg College, Oxford, for commenting on an early draft; and Gene Tsai\todo{check again how he wants to be named} and Zhixuan Yang for proofreading a draft.
\end{acks}

\bibliographystyle{ACM-Reference-Format}
\bibliography{bib}

\end{document}

%% file: string.tikzstyles
\tikzstyle{Natural transformation}=[fill=black, draw=black, shape=circle, inner sep=1pt]

\tikzstyle{Wire}=[-, line width=.5pt]
\tikzstyle{dashed line}=[-, dashed, tikzit draw=blue, draw={rgb,255: red,128; green,128; blue,128}]

%% file: pics/retabulate-unTip.tikz
\begin{tikzpicture}
	\begin{pgfonlayer}{nodelayer}
		\node [style=Natural transformation] (0) at (0, 0) {};
		\node [style=none] (1) at (-1, 1) {};
		\node [style=none] (2) at (1, 1) {};
		\node [style=none] (3) at (0, -1) {};
		\node [style=none] (5) at (-1, 1.5) {$\mathsf{BT}^\mathit{n}_{1+\mathit{k}}$};
		\node [style=none] (6) at (1.25, 1.5) {$\mathsf{BT}^{1+\mathit{k}}_\mathit{k}$};
		\node [style=none] (7) at (0.125, -1.5) {$\mathsf{BT}^\mathit{n}_\mathit{k}$};
		\node [style=Natural transformation] (11) at (6, 0) {};
		\node [style=none] (12) at (6, -1) {};
		\node [style=none] (13) at (6.125, -1.5) {$\mathsf{BT}^\mathit{n}_\mathit{n}$};
		\node [style=none] (15) at (-3, 0) {$\mathsf{retabulate}$};
		\node [style=none] (16) at (8, 0) {$\mathsf{unTip}$};
	\end{pgfonlayer}
	\begin{pgfonlayer}{edgelayer}
		\draw [style=Wire] (3.center) to (0);
		\draw [style=Wire, bend right=45] (1.center) to (0);
		\draw [style=Wire, bend right=45] (0) to (2.center);
		\draw [style=Wire] (12.center) to (11);
	\end{pgfonlayer}
\end{tikzpicture}

%% file: pics/horizontal-definitions.tikz
\begin{tikzpicture}
	\begin{pgfonlayer}{nodelayer}
		\node [style=Natural transformation] (8) at (-2.5, 0.5) {};
		\node [style=Natural transformation] (9) at (-4, -0.5) {};
		\node [style=none] (10) at (-4, 1) {};
		\node [style=none] (11) at (-2.5, 1) {};
		\node [style=none] (12) at (-2.5, -1) {};
		\node [style=none] (13) at (-4, -1) {};
		\node [style=none] (14) at (-5, -0.5) {$\alpha$};
		\node [style=none] (15) at (-1.5, 0.5) {$\alpha'$};
		\node [style=Natural transformation] (16) at (4, -0.5) {};
		\node [style=Natural transformation] (17) at (2.5, 0.5) {};
		\node [style=none] (18) at (2.5, 1) {};
		\node [style=none] (19) at (4, 1) {};
		\node [style=none] (20) at (4, -1) {};
		\node [style=none] (21) at (2.5, -1) {};
		\node [style=none] (22) at (1.5, 0.5) {$\alpha$};
		\node [style=none] (23) at (5, -0.5) {$\alpha'$};
		\node [style=none] (30) at (-4, -1.5) {$\mathit{F}$};
		\node [style=none] (32) at (-4, 1.5) {$\mathit{G}$};
		\node [style=none] (34) at (2.5, -1.5) {$\mathit{F}$};
		\node [style=none] (36) at (2.5, 1.5) {$\mathit{G}$};
		\node [style=none] (37) at (0, 0) {$=$};
		\node [style=none] (38) at (-5.5, 0) {};
		\node [style=none] (39) at (-1, 0) {};
		\node [style=none] (40) at (1, 0) {};
		\node [style=none] (41) at (5.5, 0) {};
		\node [style=none] (42) at (-2.5, -1.5) {$\mathit{F}\mathrlap{'}$};
		\node [style=none] (43) at (-2.5, 1.5) {$\mathit{G}\mathrlap{'}$};
		\node [style=none] (44) at (4, -1.5) {$\mathit{F}\mathrlap{'}$};
		\node [style=none] (45) at (4, 1.5) {$\mathit{G}\mathrlap{'}$};
	\end{pgfonlayer}
	\begin{pgfonlayer}{edgelayer}
		\draw [style=dashed line] (40.center) to (41.center);
		\draw [style=dashed line] (39.center) to (38.center);
		\draw [style=Wire] (13.center) to (9);
		\draw [style=Wire] (9) to (10.center);
		\draw [style=Wire] (11.center) to (8);
		\draw [style=Wire] (8) to (12.center);
		\draw [style=Wire] (21.center) to (17);
		\draw [style=Wire] (17) to (18.center);
		\draw [style=Wire] (19.center) to (16);
		\draw [style=Wire] (16) to (20.center);
	\end{pgfonlayer}
\end{tikzpicture}

%% file: pics/blank-g-e.tikz
\begin{tikzpicture}
	\begin{pgfonlayer}{nodelayer}
		\node [style=Natural transformation] (0) at (6, 0) {};
		\node [style=none] (1) at (5, -1) {};
		\node [style=none] (2) at (7, -1) {};
		\node [style=none] (3) at (6, 1) {};
		\node [style=none] (4) at (5.25, -1.5) {$\mathsf{BT}^{1+\mathit{k}}_\mathit{k}$};
		\node [style=none] (5) at (7, -1.5) {$\mathit{s}$};
		\node [style=none] (6) at (6, 1.5) {$\mathit{s}$};
		\node [style=Natural transformation] (7) at (0, 0) {};
		\node [style=none] (8) at (0, -1) {};
		\node [style=none] (10) at (7.5, 0) {$\mathit{g}$};
		\node [style=none] (11) at (1, 0) {$\mathit{e}$};
		\node [style=none] (12) at (0, 1) {};
		\node [style=none] (13) at (0, -1.5) {$\top$};
		\node [style=none] (14) at (0, 1.5) {$\mathit{s}$};
		\node [style=Natural transformation] (15) at (-6, 0) {};
		\node [style=none] (16) at (-7, 1) {};
		\node [style=none] (17) at (-5, 1) {};
		\node [style=none] (18) at (-6, -1) {};
		\node [style=none] (19) at (-6, -1.5) {$\top$};
		\node [style=none] (20) at (-5, 1.5) {$\top$};
		\node [style=none] (21) at (-7, 1.5) {$\mathsf{BT}^\mathit{n}_\mathit{k}$};
		\node [style=none] (22) at (-8.5, 0) {$\mathsf{blank}^\mathit{n}_\mathit{k}$};
	\end{pgfonlayer}
	\begin{pgfonlayer}{edgelayer}
		\draw [style=Wire] (3.center) to (0);
		\draw [style=Wire, bend left=45] (1.center) to (0);
		\draw [style=Wire, bend left=45] (0) to (2.center);
		\draw [style=Wire] (8.center) to (7);
		\draw [style=Wire] (12.center) to (7);
		\draw [style=Wire] (18.center) to (15);
		\draw [style=Wire, bend right=45] (16.center) to (15);
		\draw [style=Wire, bend right=45] (15) to (17.center);
	\end{pgfonlayer}
\end{tikzpicture}

%% file: pics/rotation.tikz
\begin{tikzpicture}
	\begin{pgfonlayer}{nodelayer}
		\node [style=none] (0) at (0, 0) {$=$};
		\node [style=Natural transformation] (1) at (-3.5, -2) {};
		\node [style=none] (2) at (-2, -0.5) {};
		\node [style=none] (3) at (-5, -0.5) {};
		\node [style=none] (4) at (-5, 0.5) {};
		\node [style=Natural transformation] (5) at (-5, 1) {};
		\node [style=none] (6) at (-6, 2) {};
		\node [style=none] (7) at (-4, 2) {};
		\node [style=none] (8) at (-3.5, -3) {};
		\node [style=none] (10) at (-5, 0) {$\mathsf{BT}\smash{^n_k}$};
		\node [style=none] (11) at (-3.5, -3.5) {$\top$};
		\node [style=none] (12) at (-2, 0) {$\top$};
		\node [style=none] (13) at (-3.75, 2.5) {$\mathsf{BT}^{\smash{1+k}}_k$};
		\node [style=none] (14) at (-6, 2.5) {$\mathsf{BT}^n_{1+k}$};
		\node [style=none] (17) at (-8, 1) {$\mathsf{retabulate}$};
		\node [style=none] (18) at (-6.5, -2) {$\mathsf{blank}^n_k$};
		\node [style=Natural transformation] (19) at (3.5, -2) {};
		\node [style=none] (20) at (5, -0.5) {};
		\node [style=none] (21) at (2, -0.5) {};
		\node [style=none] (22) at (3.5, -3) {};
		\node [style=none] (23) at (2.25, 0) {$\mathsf{BT}\smash{^n_{1+k}}$};
		\node [style=none] (24) at (3.5, -3.5) {$\top$};
		\node [style=none] (25) at (5, 0) {$\top$};
		\node [style=none] (26) at (6.75, -2) {$\mathsf{blank}^n_{1+k}$};
		\node [style=Natural transformation] (27) at (5, 1) {};
		\node [style=none] (28) at (5, 0.5) {};
		\node [style=none] (29) at (4, 2) {};
		\node [style=none] (30) at (6, 2) {};
		\node [style=none] (32) at (8, 1) {$\mathsf{blank}^{1+k}_k$};
		\node [style=none] (33) at (4.25, 2.5) {$\mathsf{BT}^{\smash{1+k}}_k$};
		\node [style=none] (34) at (6, 2.5) {$\top$};
		\node [style=none] (35) at (-2, 3) {};
		\node [style=none] (36) at (-2, 0.5) {};
		\node [style=none] (37) at (2, 3) {};
		\node [style=none] (38) at (2, 0.5) {};
	\end{pgfonlayer}
	\begin{pgfonlayer}{edgelayer}
		\draw [style=Wire, bend right=45] (6.center) to (5);
		\draw [style=Wire, bend left=45] (7.center) to (5);
		\draw [style=Wire, bend left=45] (2.center) to (1);
		\draw [style=Wire] (5) to (4.center);
		\draw [style=Wire, bend right=45] (3.center) to (1);
		\draw [style=Wire] (1) to (8.center);
		\draw [style=Wire, bend left=45] (20.center) to (19);
		\draw [style=Wire, bend left=45] (19) to (21.center);
		\draw [style=Wire] (19) to (22.center);
		\draw [style=Wire, bend right=45] (29.center) to (27);
		\draw [style=Wire, bend right=45] (27) to (30.center);
		\draw [style=Wire] (27) to (28.center);
		\draw [style=Wire] (36.center) to (35.center);
		\draw [style=Wire] (38.center) to (37.center);
	\end{pgfonlayer}
\end{tikzpicture}